# Nano Superconducting Quantum Interference device: a powerful tool for nanoscale investigations

Carmine Granata and Antonio Vettoliere

*Institute of Applied Sciences and Intelligent Systems "E. Caianiello"- National Research Council, I-80078 Pozzuoli (Napoli), Italy*

**Abstract:**

The magnetic sensing at nanoscale level is a promising and interesting research topic of nanoscience. Indeed, magnetic imaging is a powerful tool for probing biological, chemical and physical systems. The study of small spin cluster, like magnetic molecules and nanoparticles, single electron, cold atom clouds, is one of the most stimulating challenges of applied and basic research of the next years. In particular, the magnetic nanoparticle investigation plays a fundamental role for the modern material science and its relative technological applications like ferrofluids, magnetic refrigeration and biomedical applications, including drug delivery, hyper-thermia cancer treatment and magnetic resonance imaging contrast-agent. Actually, one of the most ambitious goals of the high sensitivity magnetometry is the detection of elementary magnetic moment or spin. In this framework, several efforts have been devoted to the development of a high sensitivity magnetic nanosensor pushing sensing capability to the individual spin level. Among the different magnetic sensors, Superconducting QUantum Interference Devices (SQUIDs) exhibit an ultra high sensitivity and are widely employed in numerous applications. Basically, a SQUID consists of a superconducting ring (sensitive area) interrupted by two Josephson junctions. In the recent years, it has been proved that the magnetic response of nano-objects can be effectively measured by using a SQUID with a very small sensitive area (nanoSQUID). In fact, the sensor noise, expressed in terms of the elementary magnetic moment (spin or Bohr magneton), is linearly dependent on the SQUID loop side length. For this reason, SQUIDs have been progressively miniaturized in order to improve the sensitivity up to few spin per unit of bandwidth. With respect to other techniques, nanoSQUIDs offer the advantage of direct measurement of magnetization changes in small spin systems. In this review, we focus on nanoSQUIDs and its applications. In particular, we will discuss the motivations, the theoretical aspects, the fabrication techniques, the different nanoSQUIDs and the relative nanoscale applications.

**Contents:**

**1. Introduction**

**2. NanoSQUIDs: fundamentals, theoretical elements and spin sensitivity.**

**3. NanoSQUIDs: fabrication and performance**

# 1. Introduction

Direct current Superconducting QUantum Interference Device (dc SQUID) is the most sensitive magnetic flux and field detector known so far [1-4]. Due to the low operating temperature and the quantum working principle, a SQUID exhibits an equivalent energy sensitivity that approaches the quantum limit. The ultra high SQUID sensitivity has also allowed, in the last years, very interesting experiments like the detection of axion dark-matter [5], the dynamical Casimir effect [6], the Majorana fermions [7] investigations, the Josephson heat interferometer [8], the Sunyaev Zeldovich effect [9], effects of the quantum gravity [10] and the detection of the primordial gravitational waves [11]. Thanks to their unique properties, SQUID devices are widely used in several applications like biomagnetism, magnetic microscopy, non-destructive evaluation, geophysics, astrophysics, quantum information, particle physics and, recently, also in nanoscience.

In the last years, one of the most ambitious goals of the high sensitivity magnetometry is the investigation of the magnetic properties of nanoscale objects like nanoparticles, magnet molecules, cold atom clouds, nanowire, single electronic spin [12,13]. These systems have been investigated using a wide variety of methods such as Hall effect in miniaturized semiconductor bars [14], spin-polarized scanning tunnelling microscopy [15], magnetic resonance force microscopy [16], electron holography [17] and, more recently, the promising diamond magnetometer based on nitrogen-vacancy centres [18-21]. Among them, the nano-sized SQUID (nanoSQUID) [22-24] is one of the most promising sensors for nanoscale applications because it exhibits an ultra high magnetic moment sensitivity reaching few Bohr magnetons or spins per unit of bandwidth and allows direct magnetization changes detection in small nano-object systems. Recently, great efforts have been focused to the development of nano-SQUIDs, making such a nanosensor a powerful tool to study the magnetic properties of the nanoparticles at a microscopic level.

In this article, we present a review of nanoSQUIDs and its applications. The first section will be focused on the basic principles of SQUID and the main characteristics of the quantum device. This is very useful to understand the following sections on nanoSQUIDs. The state of the art of nanoSQUIDs will be presented in the second and third section. It will include the motivations, the theoretical aspects, the fabrication techniques and an overview of various nanoSQUIDs. In the last section, the nanoscale applications and the conclusions will be addressed.

**1.1 Fundamentals of a SQUID**

A dc SQUID sensor is a converter of magnetic flux into an electrical current or a voltage having an extremely low magnetic flux noise. Basically, it consists of a superconducting ring interrupted by two Josephson junctions. As the next section will show, the current flowing into the SQUID or the voltage across it is a periodic function of the external magnetic flux treading the SQUID detector with a period equal to a fundamental physical constant $\Phi_0=h/2e=2.07\times10^{-15}$ Wb ($h$ is the Plank constant and $e$ is the charge of the electron).

Why is it so sensitive? The superconductivity is one of the most spectacular manifestations of the macroscopic quantum physics. Being a macroscopic object as a superconductor described by a wave function, there are macroscopic physical quantities related to quantum physical constant. In the case of the SQUID, the voltage across it (few tens of $\mu$V) is related to the quantum flux that is a very small quantity from a macroscopic point of view. Moreover, by using suitable readout electronics, the SQUID can measure a magnetic flux less than $10^{-6}$ $\Phi_0$ per bandwidth unit, resulting in an ultra low noise sensor.

Jaklevic et al [25] gave the first demonstration of quantum interference in a device by using two thin-film Josephson junctions connected in parallel on a superconducting loop. Few years later, Zimmerman and Silver [26] proposed a device consisting of a superconducting ring interrupted by a single Josephson junction (rf SQUID). As no leads are attached to the device, in this case the loop is inductively coupled to a resonant tank circuit that is excited at its resonant frequency by a radio-frequency current. The amplitude of the oscillating voltage across the tank circuit is periodic in the magnetic flux treading the rf SQUID loop with a period of $\Phi_0$.

For about ten year from its discovery, the rf SQUID had a major impact because it required only a single junction which was easier to manufacture than a pair of junctions. The invention of a reliable and reproducible Josephson device technology based on Nb/AlOx/Nb multilayer and the development of a suitable dc SQUID design based on the planar SQUID with an integrated multiturn input coil ensured the dominance of the dc SQUID which offered sensitivity higher than rf SQUID one. Nowadays, most of applications employ dc SQUID based on both low and high critical temperature ($T_c$) superconductors. Only high $T_c$ rf SQUIDs have viable applications prospects as their sensitivity is comparable to high $T_c$ dc SQUID and are almost always single-layer devices.

Beside the magnetic flux, a SQUID can detect with an unprecedented sensitivity also other physical quantities as magnetic field, current, voltage, displacements, etc, if converted in a magnetic flux by using suitable flux transformer circuits [27].

In this section we will provide the fundamentals of SQUIDs (principle of operation, the noise properties, the main applications), while an exhaustive analysis can be found in the references [1-4].

**1.2 Working principle of a SQUID**

The operation principle of a SQUID is based on the Josephson effect [28] and the flux quantization in a superconducting ring [29]. In 1962, Brian Josephson predicted that a supercurrent could tunnel through an insulating barrier separating two superconducting electrodes. A Josephson junction is schematically represented by two superconductor separated by a thin insulation barrier (Fig.1). If the junction is biased with a dc current, the voltage across it remains zero up to a current value called Josephson critical current $I_0$ due to the cooper pair-tunnelling trough the insulation barrier. This so-called dc Josephson effect is due to the overlap of the macroscopic wave functions in the barrier region (Fig.1).

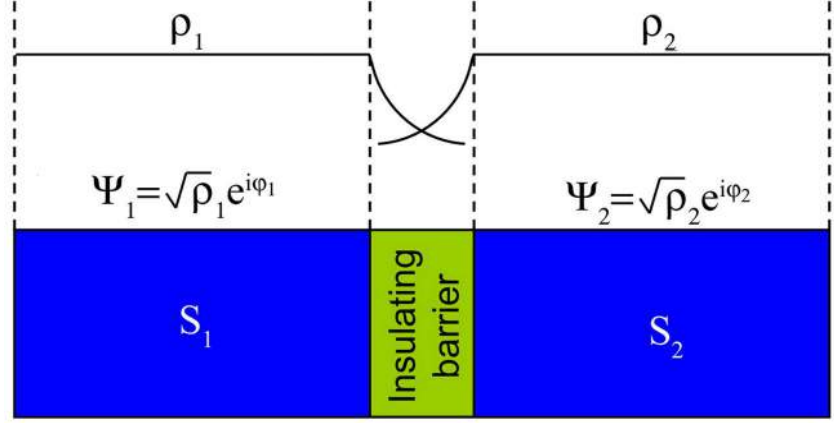

*Fig. 1. Scheme of a Josephson junction. A thin insulator barrier (few nm) separates the two superconductors. The overlap of the macroscopic wave functions allows the tunnelling of the Cooper pair.*

When the bias current is greater than $I_0$, the junction switches in a resistive state where the tunnel is due to the single electrons.

The fundamental equations of the Josephson effect are:

$$I = I_0 \sin \varphi \tag{1}$$

$$\frac{\partial \varphi}{\partial t} = \frac{2e}{\hbar} V \tag{2}$$

Where $I_0$ is the critical current and $\varphi$ the phase difference between the macroscopic wave functions of the cooper pairs relative to the two superconductors and $V$ is the voltage across the junction. If the voltage across the junction is constant, the phase difference is a linear function of the time $\varphi = (2e/h) V \cdot t + \varphi_0$, which substituted in the first Josephson equation gives an oscillating current at non zero voltage: $I = I_0 \sin(2\pi f t + \varphi_0)$ where $f = (2e/h)V/2\pi$ (486.6 MHz/μV). This effect, known as Josephson ac effect [30], has been successfully employed in metrology and several other applications [2-4]. An exhaustive review of the Josephson effect can be found in reference [31]. An overlap of the macroscopic wave functions of the two superconductors which implies the Josephson effect can be also obtained by using different structures such as Dayem bridges (nanoconstrictions of superconductor), point of contacts, by-crystal and step edge grain boundary junctions, normal or semiconductor barrier based junctions etc. [32,33].

The magnetic flux quantization states that the magnetic flux treading a superconducting loop exists only in multiples of the flux quantum *($\Phi=n\Phi_0$)* [29]. Physically, it is due to the Meissner effect [34], a peculiar effect of the superconductivity, which involves the magnetic flux expulsion in a superconductor when it moves from the normal state to a superconducting one in presence of an external magnetic field. In the case of a superconducting ring, the discrete quantities of flux can be trapped into the ring rather than expelled as in a continuous superconductor. This magnetic flux is sustained by a persistent circulating current inside the ring, *$J=\Phi/L$* where *L* is the inductance of the ring and *$\Phi=B\cdot S$* is the applied magnetic flux. The flux quantization within a superconducting ring can be explained also assuming that the macroscopic wave function must be single valued (Sommerfield-Bohr quantization) [26].

The typical SQUID configuration is schematically shown in the Fig. 2. The two Josephson junctions are in a parallel configuration, so the critical current of the SQUID is $I_c=I_1+I_2$ or $I_c=2I_0$ if the Josephson junctions are identical. As will be shown below, in presence of an external magnetic field treading the SQUID loop, $I_c$ oscillates with a period of one flux quantum. This is due to the interference of superconducting wave functions in the two arms of the SQUID and are analogue to the two slit interference in optics. It is the basis of the working principle of a SQUID.

**1.3 SQUID characteristic computation**

Fig. 2 reports the equivalent electrical circuit of a SQUID obtained in the framework of the resistively shunted junction model (RCSJ) [35,36]. In this model, the Josephson junction has a critical current $I_0$ and is in parallel with a capacitance *C* and resistance *R* having a current noise source associated to it. The *R*-value is related to the hysteresis in the current-voltage (*I-V*) characteristic of a junction or a SQUID. In particular if the Stewart-McCumber parameter $\beta_c=2\pi I_c C R^2/\Phi_0 <1$ there is no hysteresis [36].

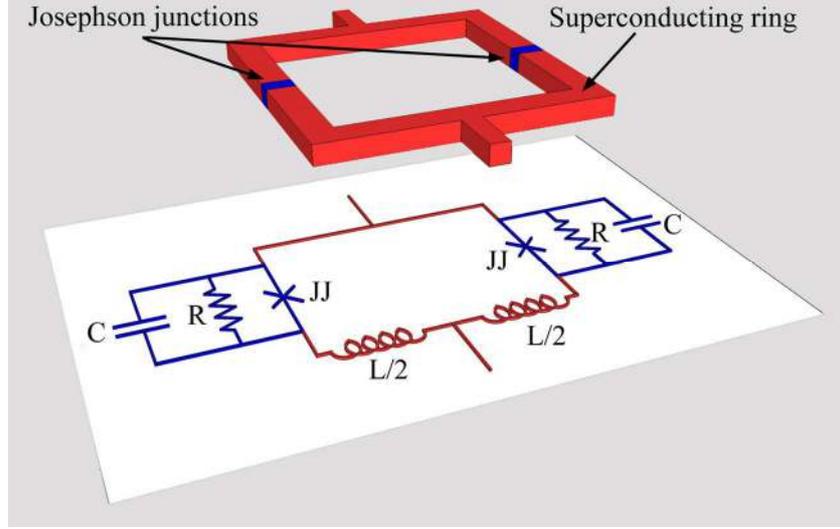

*Fig.2 Equivalent electric circuit of a dc-SQUID in the framework of the resistively and capacitively shunted model of a Josephson junction [35].*

It can be shown that the flux quantization in the presence of a superconducting ring including two Josephson junctions can be write as [31]:

$$\varphi_1 - \varphi_2 = 2\pi \frac{\Phi}{\Phi_0} = 2\pi \frac{\Phi_e + LJ}{\Phi_0} \qquad (3)$$

Where $\varphi_1$ and $\varphi_2$ are the phase differences of the superconducting wave functions across the two junctions. $\Phi = \Phi_e + LJ$ is the total flux threading the SQUID loop given by the external flux $\Phi_e$ and the self flux produced by the screening current circulating into the SQUID loop with an inductance $L$. The circulating current can be expressed as $J=(I_1-I_2)/2$.

First, we consider the case of zero voltage state. Appling the Kirchhoff laws at the circuit of Fig.2, combining the equation 1 and 3 and supposing that the junctions are identical:

$$I_c(\Phi) = I_1 + I_2 = I_0(\sin\varphi_1 + \sin\varphi_2) = 2I_0 \sin\varphi \cos\left(\pi\frac{\Phi}{\Phi_0}\right) \qquad (4)$$

$$\frac{\Phi - \Phi_{ext}}{\Phi_0} = \frac{LJ}{\Phi_0} = \frac{LI_0(\sin\varphi_1 - \sin\varphi_2)}{\Phi_0} = \beta_L \cos\varphi \sin\left(\pi\frac{\Phi}{\Phi_0}\right) \qquad (5)$$

Where $\varphi=(\varphi_1+\varphi_2)/2$ while $\beta_L=2LI_0/\Phi_0$ is the inductance parameter. It is one of the most important parameter, since the SQUID characteristic strongly depends on the $\beta_L$ value.

If the SQUID inductance is very small, $\beta_L \approx 0$, consequently $\Phi \approx \Phi_e$, the SQUID critical current has a simple co-sinusoidal behaviour and the modulation depth, defined as $\Delta I_c = I_c(\Phi_e=0) - I_c(\Phi_e=\Phi_0/2)$, is equal to $2I_0$, that is the SQUID critical current modulates to zero. If the junction critical currents are not equal, the SQUID does not modulate to zero also in the case of $\beta_L = 0$. In this case the $I_c(\Phi)$ is given by:

$$I_c(\Phi) = \sqrt{(I_1 - I_2)^2 + 4I_1 I_2 \cos^2\left(\frac{\pi \Phi}{\Phi_0}\right)} \qquad (6)$$

If $\beta_L$ is not zero, $\Delta I_c$ decreases by increasing the $\beta_L$ value as shown in the Fig. 3a where $I_c$, as a function of the external magnetic flux, is reported for three different $\beta_L$ values. The curves have been obtained by numerically solving the equations 4 and 5. An estimation of the critical current modulation depth is given by: $\Delta I_C/I_C = 1/(1+\beta)$ [37]. For $\beta_L = 1$ the critical current modulates by 50%, and for $\beta_L \gg 1$, $\Delta I_C/I_C$ decreases as $1/\beta_L$.

As the critical current $I_C$ is a regular function of the magnetic flux through the SQUID loop, one can easily calculate the flux change in the SQUID loop by measuring the critical current variations $\Delta I_c$ and dividing it by the magnetic flux to critical current transfer factor ($I_\Phi = \partial I_C/\partial \Phi_e$). In this case, the device is employed as a magnetic flux to current transducer. The signal-to-noise ratio (SNR) and the magnetic flux resolution are:

$$SNR = \frac{\Delta I_c}{I_{c.n}} = \frac{\Delta \Phi \cdot I_\Phi}{I_{c.n}} \qquad \Phi_N = \frac{I_{c.n}}{I_\Phi} \qquad (7)$$

Where $I_{c,n}$ is the measurement error of the critical current and $\Delta I_c = \Delta \Phi \cdot I_\Phi$ is the current corresponding to a magnetic flux variation $\Delta \Phi$. An increase of $I_\Phi$ leads to an increase of both SNR and $\Phi_N$. As the section 3 will show, this configuration can be successfully employed in the case of device showing an hysteretic current-voltage characteristic.

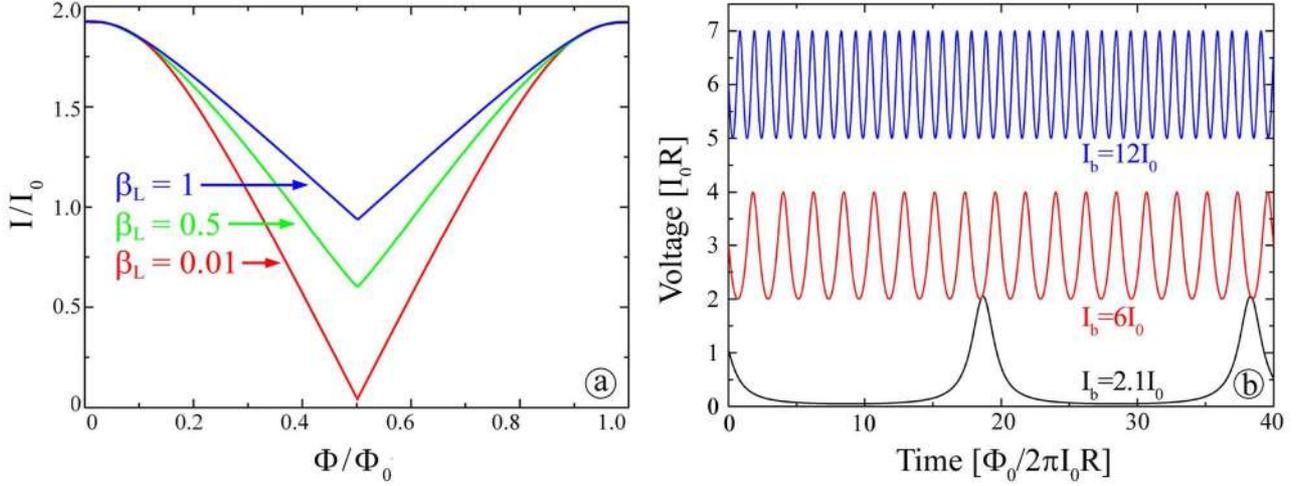

*Fig.3. a) Critical current of a SQUID as function of the external magnetic flux threading the loop for three different $β_L$ values. b)Voltage across a SQUID as a function of the normalized time.*

The voltage state involves the presence of an oscillating current and voltage as predicted by the second Josephson equation (2). In this case, the equations describing the SQUID dynamic are obtained by including in the Kirchhoff law the current terms due to the voltage across the resistance (*V/R*) and the capacitance (*CdV/dt*). Considering the equation 2 and applying the Kirchhoff law to both SQUID arms, we obtain the following two equations:

$$\frac{I_c}{2} + J = I_0 \sin\varphi_1 + \frac{\Phi_0}{2\pi R}\frac{d\varphi_1}{dt} + \frac{\Phi_0 C}{2\pi}\frac{d^2\varphi_1}{dt^2} + I_{N,1}$$
$$\frac{I_c}{2} - J = I_0 \sin\varphi_2 + \frac{\Phi_0}{2\pi R}\frac{d\varphi_2}{dt} + \frac{\Phi_0 C}{2\pi}\frac{d^2\varphi_2}{dt^2} + I_{N,2}$$

(8)

The above Langevin equations, together with the equation 5, provide a complete description of the SQUID characteristics. The voltage is given by: *V=1/2[d(φ₁(t) +φ₂(t))/dt]*. The terms $I_{N,1}$ and $I_{N,2}$ are the Nyquist noise associated to the shunt resistors *R*. In the simplest case, where $I_{N,1}=I_{N,2}=0$ and $β_L$, $β_C<<1$, the equation 8 and 5 can be easily solved, providing the following equation for the SQUID *I-V* characteristic:

$$V(\Phi_e, I) = \frac{R}{2}\sqrt{I^2 - \left(2I_0 \cos\left(\pi\frac{\Phi_e}{\Phi_0}\right)\right)^2}$$

(9)

The voltage swing or peak-to-peak modulation defined as $\Delta V_S = V(\Phi_0/2) - V(\Phi_0)$ is given by:

$$\Delta V_S(I) = \left[\frac{RI}{2} - \frac{R}{2}\sqrt{I^2 - (2I_0)^2}\right] \tag{10}$$

The maximum value is obtained for $I=2I_0$, that is $\Delta V_S=I_0 R$.

In the more general case, the equations 8 and 5 are typically numerically solved [38-40], providing the $\varphi_1(t)$ and $\varphi_2(t)$, which allow to compute all the SQUID characteristics. Neglecting the capacitance and following the normalizations: current $i$ to $I_0$, voltage $v$ to $I_0 \cdot R$, applied magnetic flux $\varphi_e$ to $\Phi_0$, and time $\tau$ to $\Phi_0/(2\pi I_0 R)$, and considering the inductance parameter $\beta_L=2LI_0/\Phi_0$, we obtain the following equation:

$$\frac{i}{2} + j = \sin(\varphi_1) + \frac{d\varphi_1}{d\tau} + i_{n1}$$

$$\frac{i}{2} - j = \sin(\varphi_2) + \frac{d\varphi_2}{d\tau} + i_{n2} \tag{11}$$

$$j = \frac{(\varphi_1 - \varphi_2) - 2\pi\phi_e}{\pi\beta}$$

$$v(\tau) = \frac{1}{2}\left[\frac{d(\varphi_1 + \varphi_2)}{d\tau}\right]$$

Below, we report some simulations neglecting the junction capacitance and the resistor Nyquist noise. In particular, we numerically solved the differential equations (11) by using the Euler method based on the difference equations. We used a time step $\Delta\tau=0.01$ over a total time $T=10000$, corresponding to $10^6$ time units. Such values guarantee a well-defined time averaged voltage. The voltage oscillates as function of the time as shown in the Fig. 3b reporting a $V(\tau)$ for $\beta_L=1$, $\Phi_e=0$ and $I_B/I_0=2.1$, $6$ and $12$. Increasing the bias current, the voltage frequency increases and the amplitude decreases. Supposing an oscillation period of about $\Phi_0/(2\pi I_0 R)$ as in the case of $I_B/I_0=12$ reported in Fig. 3b and considering typical parameters of a dc SQUID ($I_0=10$ μA, $R=3$ Ω) a frequency of about 100 GHz is obtained. In practice, a voltage average is typically measured. Varying the bias current and taking the average voltage $<V>$, the I-V characteristic can be obtained, for a fixed $\Phi_e$ and $\beta_L$ values. The voltage-flux characteristic (V-$\Phi$) can be obtained by varying the external magnetic flux and fixing the $\beta_L$ and $I_B$ value. Fig. 4 reports the *I-V* characteristics computed for $\Phi_e=0$ and $\Phi_e=\Phi_0/2$ and *V-Φ* characteristics for several values of $I_B/I_0$ ratio and $\beta_L=1$, $\beta_c=0$ for all curves. As expected, the *I-V* does not show hysteresis while the *V-Φ* have a periodic behaviour with a period equal to $\Phi_0$. The *V-Φ* amplitude ($V(0)-V(\Phi_0/2)$) depends on the bias current and $\beta_L$ value and reaches its maximum for $I_B=2I_0$.

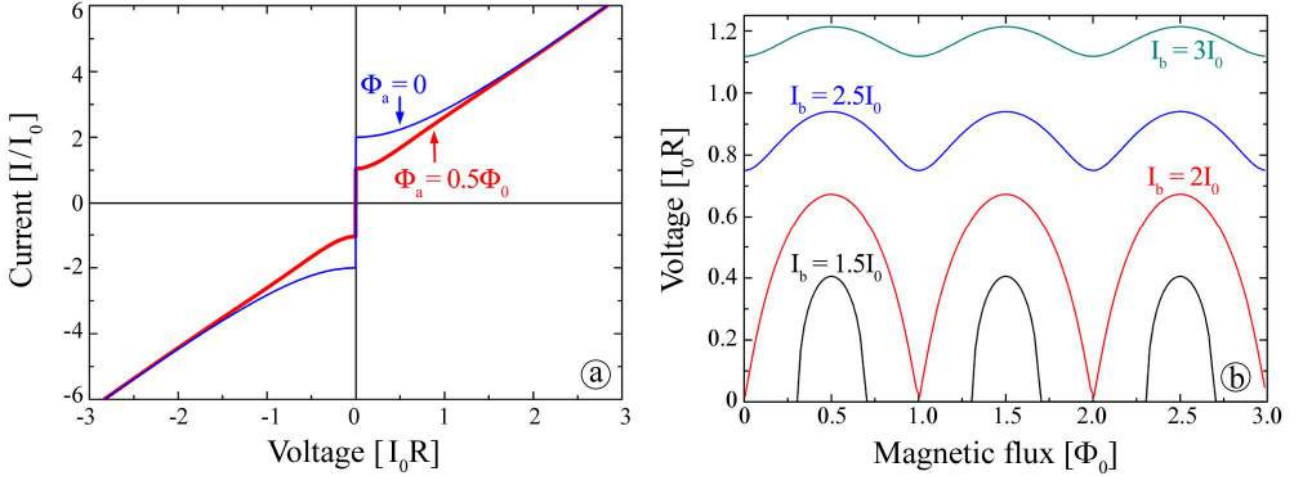

*Fig.4 a) Current-voltage characteristics computed for $\Phi_e=0$ and $\Phi_0/2$. b) Voltage-magnetic flux characteristics computed for $I_B/I_0=1.5, 2.0, 2.5$ and $3.0$. In both figures the $\beta_L = 1$ and $\beta_c=0$.*

Hence, a non-hysteretic SQUID can be considered as a magnetic flux–voltage transducer and can be employed as a magnetic flux detector. In this case, $\Delta\Phi_e=\Delta V/V_\Phi$ where $V_\Phi=\partial V/\partial\Phi_e$ is the voltage responsivity, namely the slope of the *V-Φ* curve in the magnetic bias point. Typically, in this configuration, the SQUID is biased with a constant current close to $I_c$ and an external magnetic flux $\Phi_e=\Phi_0/4$ in order to maximize the $V_\Phi$ and increase the *SNR*.

If the magnetic signals to measure are much smaller than the flux quantum, the SQUID can be work in small signal mode. In fact, as for every continuous signal, in a suitable small range around a point, the response of the SQUID can be considered linear with the external magnetic flux.

If the signals to detect are greater than the flux quantum, the SQUID response has to be linearized. In order to do it, a Flux-Locked-Loop (FLL) configuration is often used [41]. In such scheme, the output voltage is converted into a current by a resistor and fed back into the SQUID as a magnetic flux, via a coil coupled with the sensor, nulling the input magnetic flux (Fig. 5). Therefore, the SQUID works as a null detector of magnetic flux. The output voltage can be read across the feedback resistor being proportional to the magnetic flux input. The FLL linearizes the SQUID output, increasing the linear dynamic range.

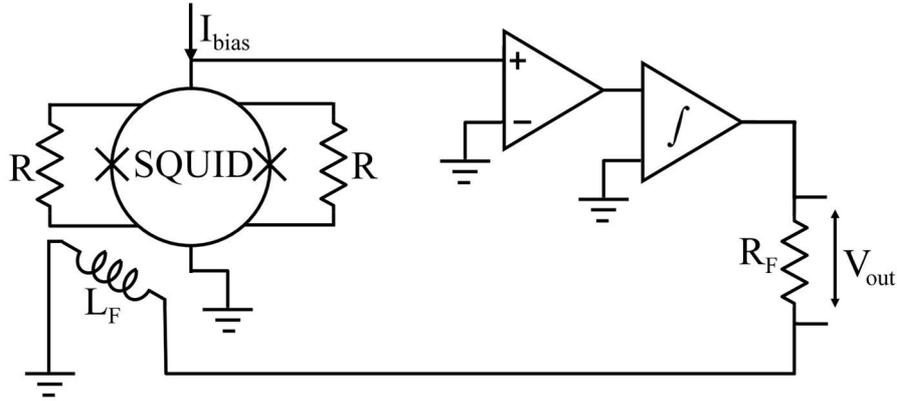

*Fig.5 Flux locked loop circuit employed to increase the linear dynamic range of a dc SQUID.*

Due to the very low output voltage noise of a SQUID, direct voltage readout mode, generally, leads to a reduction of intrinsic SQUID sensitivity. In order to solve this problem, complicated schemes such as ac-flux modulation in combination with an impedance matching were often used [42-43]. In recent years, a second generation of SQUID sensors with a large flux-to-voltage transfer factor and an alternative readout circuit have been developed in order to allow a direct-coupled readout scheme without flux modulation [44]. In comparison with the standard electronics, the direct-coupled readout schemes are simpler, more compact and less expensive. In particular, circuits based on Additional Positive Feedback (APF) are very effective [45]. The APF scheme consist of a RL circuit in parallel to the SQUID which the effect is to render the V-Φ characteristic asymmetric, so, if the SQUID is biased on the steeper side, an effective increase of the SQUID responsivity ($V_\Phi = \partial V/\partial \Phi$) is achieved.

**1.4 Magnetic noise and energy sensitivity**

One of the most important factor of merit of a SQUID device is the magnetic flux noise or more precisely the spectral density of magnetic flux noise. The importance of the noise in Josephson devices has stimulated many theoretical and experimental investigations leading to an exhaustive comprehension of the main mechanisms responsible of the different noise. Theories for voltage, current and magnetic flux noise in resistively shunted junctions [46], rf-SQUID and dc-SQUID have been developed [1, 38-40] as well as quantum charge noise in cooper pair box [47].

In the case of a shunted dc-SQUID, this noise is essentially due to the Nyquist noise associated to the shunt resistor with a current spectral density $S_I=4k_BT/R$ where $k_B$ is the Boltzman constant. In order to preserve the Josepshon coupling, the following condition is required: $I_0\Phi_0/2\pi>>k_BT$. It means that the Josephson energy ($I_0\Phi_0/2\pi$) must be much greater than the thermal energy. In other words there is a thermal current $I_{th}=(2\pi k_BT)/\Phi_0$ so that $I_0>I_{th}$ in order to ensure that the thermal fluctuations do not destroy the Josephson coupling. The value of $I_{th}$ for 4.2 K (liquid helium) and 77 K (liquid nitrogen) are respectively 0.18 µA and 3.2 µA [1]. Such conditions are typically written as $\Gamma=2\pi K_BT/I_0\Phi_0<<1$. The $\Gamma$ parameter plays a fundamental role for both SQUID design and its performances. In dimensionless units the spectral density of the Nyquist current noise relative to the shunt resistance can be written as $S_I=4\Gamma$. Another mandatory condition is that the magnetic energy associated to the SQUID inductance must be much greater than the thermal energy: $\Phi_0^2/2L>>2\pi k_BT$. Also in this case it is possible to define an inductance (fluctuation threshold inductance) $L_F=\Phi_0^2/(4\pi^2 k_BT)$ so that $L<L_F$. At T=4.2 K and 77 K the values of $L_F$ are 1.9nH and 102 pH respectively [1].

What are the effects of resistor Nyquist noise and how we can simulate its effect on SQUID characteristics? The current noise in the shunt resistors induces a voltage noise $V_N(f)$ across the SQUIDs, moreover it rounds the *I-V* characteristic and reduces the critical current. Consequently, the *I-Φ, V-Φ* characteristics and the corresponding responsivities ($I_\Phi$ and $V_\Phi$) are degraded too. In order to simulate the SQUID characteristics in presence of noise, we need to insert the term $I_{N,1}$ and $I_{N,2}$ in the equations 5. It can be shown that the thermal noise current has no time correlation. It means that the correlation function is a delta function:

$$\langle I(t)I(t+t_i)\rangle = \frac{2K_B T}{R}\delta(t_i) \tag{12}$$

So the current noise results in a frequency independent noise (white noise) with a zero average $<I(t)> =0$ and a Gaussian distribution of the amplitudes. Hence, in the equation 5, $I_{N,1}$ and $I_{N,2}$ are Gaussian distributions depending on the noise parameter $\Gamma$. In the following simulations, the discrete noise source $i_n$ is a sequence of random numbers, so it is a Gaussian random variable with mean square deviation $<\Delta i^2_n>=2\Gamma/\Delta\tau$ and average $<i_n>=0$.

Regarding the calculation procedure, it is the same of that describe above: the two phase allow to calculate $V(t)$, after that a fast Fourier transformer of $V(t)$ provide the power spectral density (PSD) $S_V(f)$ (the normalized unit of $S_V$ is $I_0 R \Phi_0/2\pi$). The PSD of the flux noise will be given by: $S_\Phi = S_V/V_\Phi^2$, where the voltage to flux transfer factor $V_\Phi$ is obtained by taking the derivative of $V$-$\Phi$ characteristic at $\Phi=\Phi_0/4$ (maximum responsivity). However, the values of practical interest are the spectral densities given by the root square of the PSD: $S_V^{1/2}$ (V/Hz$^{1/2}$) and $S_\Phi^{1/2}$ ($\Phi_0$/Hz$^{1/2}$). Fig. 6a reports the spectra of the voltage noise relative to two different $\Gamma$ values (0.01 and 0.05). They have been computed for an external flux bias of $\Phi_a=\Phi_0/4$ and a bias current of $I_B=2.1 I_0$.

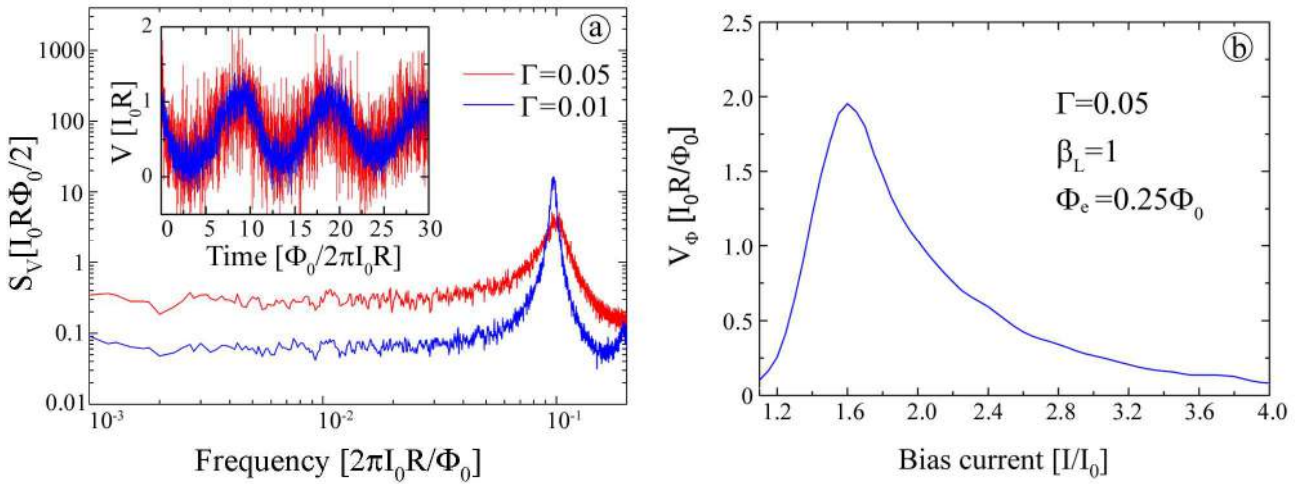

*Fig. 6 a) Power spectral densities of the voltage noise for two different $\Gamma$ values. The inset shows the voltage as a function of the time for the same $\Gamma$ values. The flux and current bias are $\Phi_e=\Phi_0/4$ and $I_B=2.1 I_0$ respectively. b) Voltage responsivity as a function of the bias current. The maximum value is obtained for a bias lower than the SQUID critical current ($2I_0$).*

It is evident the presence of the peaks related to the fundamental Josephson frequency $f_J=<V>/\Phi_0$ (in normalized unit $f_J=<v>/2\pi$).

The $S_V$ values are typically taken in the white region of the spectrum that is at a frequency smaller than the peak relative to the fundamental Josephson frequency. The inset of the Fig. 6a shows the $v(\tau)$ relative to $\Gamma=0.01$ (blue color) and $\Gamma=0.05$ (red color). As expected the peak frequency in the spectrum corresponds to the inverse of the period of the $v(\tau)$. The PSD of $v(\tau)$ have been obtained via Welch's method by using a sampling frequency $\omega_s=1/\Delta\tau$. The spectra have been obtained by averaging 60 spectra of time traces ($v(\tau)$). In the figure 6b, the voltage responsivity $V_\Phi$ as a function of the bias current $I_B$ is reported.

The power spectral density of the voltage $S_V$ and the spectral density of magnetic flux $S_\Phi^{1/2}$ as a function of the bias current $I_B$ are reported in the Fig. 7 for $\beta_L=1$, $\Phi=0.25\,\Phi_0$ and $\Gamma=0.05$. From the figure, we observe that the minimum of $S_\Phi^{1/2}$ corresponds to about $1.6\,I_c$ where the values of the $V_\Phi$ and $S_V$ are:

$$V_\Phi \cong \frac{2I_o R}{\Phi_0} = \frac{R}{L} \quad S_V \cong \frac{0.8 I_0 R \Phi_0}{2\pi} = 16\,K_B T R$$
$$S_\Phi^{1/2} \cong 4\sqrt{\frac{K_B T}{R}}\,L \tag{13}$$

In order to compare SQUID with different inductance, SQUID noise is often presented as the noise energy for unit bandwidth:

$$\varepsilon \cong \frac{S_\Phi}{2L} \cong \frac{8 K_B T L}{R} \tag{14}$$

It is expressed in unit of $\hbar$. A SQUID can reach an energy resolution as low as few $\hbar$ [48,49], in other words it is limited by quantum mechanics uncertainty principle.

It has been proved that the condition $\beta_L=1$ and $\Phi=0.25\,\Phi_0$, optimize the SQUID performances [38]. Hence in order to reduce the flux noise of a SQUID, we have to reduce the inductance of the loop, increase the shunt resistance value preserving the conditions $\beta_C<<1$, $\Gamma<<1$, $\beta_L=1$.

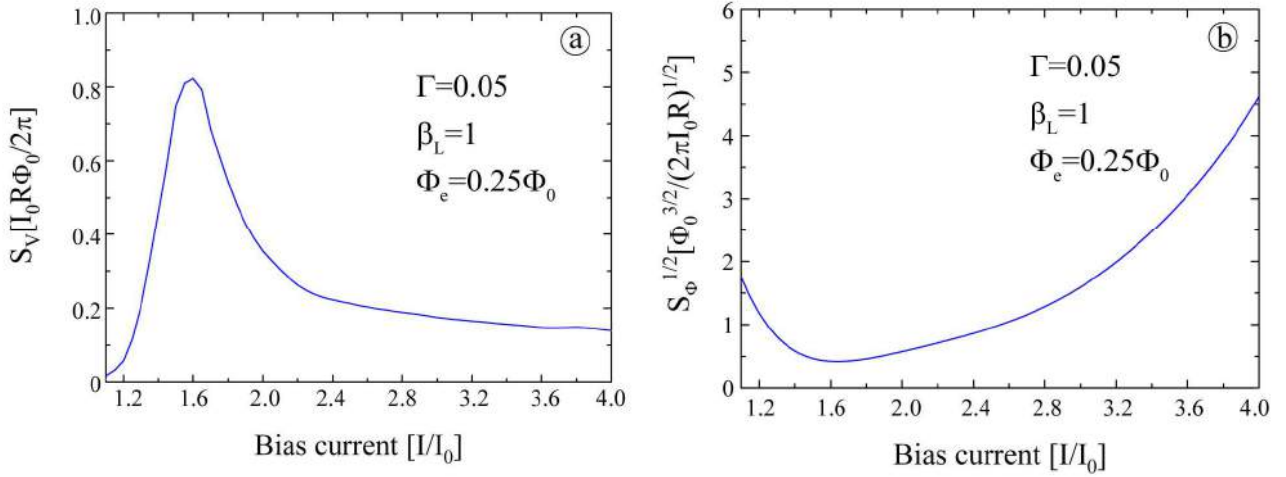

*Fig.7 a) Power spectral density value in the white region of the voltage noise as a function of the bias current. For bias current much greater than the SQUID critical current, the voltage noise tends to the Nyquist noise of the normal resistance ($R_s/2$). b) Spectral density of the magnetic flux noise as a function of the bias current. It has been obtained by taking the root square of the the ratio $S_V/V_\Phi^2$.*

In the practical case, the $\beta_C$ value is not zero, because the Josephson junction have a non negligible capacitance, however simulations taking into account also the second derivative in the equations 8, show that the results for bias current less than the critical currents of SQUID are not far from those for very small $\beta_C$ values.

It is worth to note that for overdamped SQUID ($\beta_C << 1$) operating at large thermal fluctuations, an analytical theory has been developed [50]. It is based on the equivalency between the coupled Langevin equations and the single two-dimensional Fokker-Plank equations.

The total voltage or magnetic flux noise is given by $V_n = [<V(t)^2>-<V(t)>^2]^{1/2}=\sigma_v$, $\Phi_n=[<\Phi(t)^2>-<\Phi(t)>^2]^{1/2}=\sigma_\Phi$, where $\sigma$ is the full width at half height (FWHM) of the voltage or flux distribution (in the Gaussian case $\sigma=FWHM/2.35$). $V_n$ and $\Phi_n$ are respectively related to the PSD of the voltage $S_V(f)$ and of the magnetic flux $S_\Phi(f)$ by :

$$V_n = \sigma_V = \sqrt{\int_0^{f_s/2} S_v(f)\,df} \quad ; \quad \Phi_n = \sigma_\Phi = \sqrt{\int_0^{f_s/2} S_\Phi(f)\,df} \tag{15}$$

The integration span ranges from zero to half of the sampling frequency (Shannon-Nyquist theorem).

In presence of hysteresis, the SQUID can operates as magnetic flux to current transducer (see section 1.2) and the current noise is given by $I_n = [<I(t)^2> - <I(t)>^2]^{1/2} = \sigma_I$ which is the width at half maximum of the critical current distribution P(I). In analogy to the equation 15, $I_n$ is related to the PSD of the critical current fluctuation $S_{\Delta Ic}(f)$ by:

$$I_n = \sigma_I = \sqrt{\int_0^{f_B/2} S_{\Delta I_c}(f)\,df} \tag{16}$$

Where $f_B$ is the physical bandwidth of the system. A reasonable assumption is that the physical bandwidth is equal to the plasma frequency of the junction $\omega_p = 2\pi f_p = (2\pi I_c/C\Phi_0)^{1/2}$. If the noise is white, from the equation 16, the PSD white level of the critical current noise is given by $2\pi\sigma_I^2/\omega_p$. So, the spectral density of magnetic flux noise of a hysteretic device can be written as [51]:

$$S_\Phi^{1/2} = \frac{\sigma_I}{I_\Phi}\sqrt{\frac{2\pi}{\omega_p}} \tag{17}$$

In practice, the bandwidth is limited by the readout electronic or the speed of the measurements, therefore the (17) can be written as:

$$S_\Phi^{1/2} = \frac{\sigma_I}{I_\Phi\left(\frac{\Delta t}{I_c}\frac{dI}{dt}\right)} \tag{18}$$

Where $\Delta t$ is the measurement time interval and $dI/dt$ is the current sweep rate.

Before to close this short compendium about the basic principle and the noise of a dc SQUID, it is mandatory to mention the low frequency noise (1/f or flicker noise [52]) in a SQUID device. The theoretical and experimental investigation of such a noise is very important for many applications such as biomagnetism, geophysics and quantum computing. In particular, concerning the latter topic, it seems that the low frequency noise is related to the decoherence time in Josephson qubits [53,54]. This has stimulated many theoretical [55-57] and experimental [58-62] investigations of low frequency noise in Josephson device.

As reported in a pioneering work of Koch et al [63], there are two main sources of low frequency noise in a dc SQUID: the fluctuations of the critical current of the Josephson junctions (critical current noise) and the motion of the magnetic vortex trapped in the SQUID body (flux noise). Being a SQUID biased with a constant current, a critical current fluctuation produces voltage fluctuations and so a magnetic flux noise. The critical current fluctuations may occur in the process of tunnelling of electrons through the barrier that can be trapped in a defect and subsequently released. Occupation of the trap induces a local change in the height of the tunnel barrier and hence in the critical current density of that region. Hence, a single trap leads to a randomly switching critical current $I_0$ of the junction between two values. It can be shown that this process leads to a 1/f spectrum [41,64]. Fortunately, this noise can be reduced by using suitable readout based on reversing the bias current at some frequency above the corner of 1/f noise [41,44].

The second main source of 1/f noise in SQUIDs is a flux noise and cannot be reduced by any bias reversal scheme. Due to the presence of defects in the body of SQUID or in any superconductive circuital elements connected to it, during the cooling process a magnetic vortex can be trapped in a defect acting as a pinning state. If the thermal energy is sufficiently large, the vortex may overcome the pinning energy and hop back and forth between two or more adjacent pinning sites inducing a change in the flux coupled to the SQUID. In order to reduce this noise, it is needed to decrease the number of defects by employing very high quality superconducting film or to design the superconducting component so that the vortices do not enter, by using for example narrow linewidth structures. In particular, the linewidth should be reduced below to $(\Phi_0/B)^{1/2}$ [65] where B is the magnetic field in which the device is cooled. Another way to reduce this kind of noise is to prevent the motion of the vortex by improving its pinning inside superconducting structures. To this aim, suitable structures inserted in the SQUID ring such as moats and slits can be used [66-67]. They act as a reliable traps preventing the motion of the vortices and reducing the flicker noise. In the case of SQUID device moving in an ambient magnetic field, it has been shown the effectiveness of the flux Dam that is a weak link in the superconducting structure, in reducing the 1/f noise [68]. In fact, the flux Dam limits the circulating supercurrents, which induce the generation of the vortices.

However, it is worth to stress that the origin of 1/f flux noise in $LT_c$ dc-SQUID is a controversial issue and, recently, several models have been proposed to explain it in Josephson devices. In 2007, Koch et al [58] assumed that the low frequency flux noise was generated by the magnetic moments of electrons in defect states, which they occupy for a wide distribution of time before escaping. Two years later, Sendelback et al [61] showed that, due to spin-spin interactions on the surface, the SQUID inductance, fluctuated with a 1/f power spectrum generating inductance noise highly related to conventional 1/f flux noise. In 2013, an experimental study of 1/f noise in SQUIDs with systematically varied geometry evidenced that their results were incompatible with a model based on the random reversal of independent surface spins [69]. Very recently, a theory of spin diffusion was proposed to explain the low frequency flux noise in the SQUIDs [70].

### 1.5 Detection of physical quantities

A SQUID is essentially a magnetic flux detector. However, it can detect with a ultra-high sensitivity any physical quantities that can be converted in a magnetic flux trading the SQUID loop. Depending on the quantity to measure, it is needed to employ a suitable SQUID sensor design. In this section, we report the basic principle of the main SQUID configurations.

### 1.5.1 Magnetic field measurements

In the case of a bare SQUID, the rms magnetic field noise $S_B^{1/2}(f)$ is simply given by $S_\Phi^{1/2}/A_\ell$ where $A_\ell$ is the geometrical area of the superconducting loop. Since the flux noise increases with the ring inductance (equation 13), it is not possible increase the magnetic field sensitivity by increasing the geometrical area of the SQUID ring. Moreover, an increase of the loop inductance leads also to a decrease of voltage responsivity.

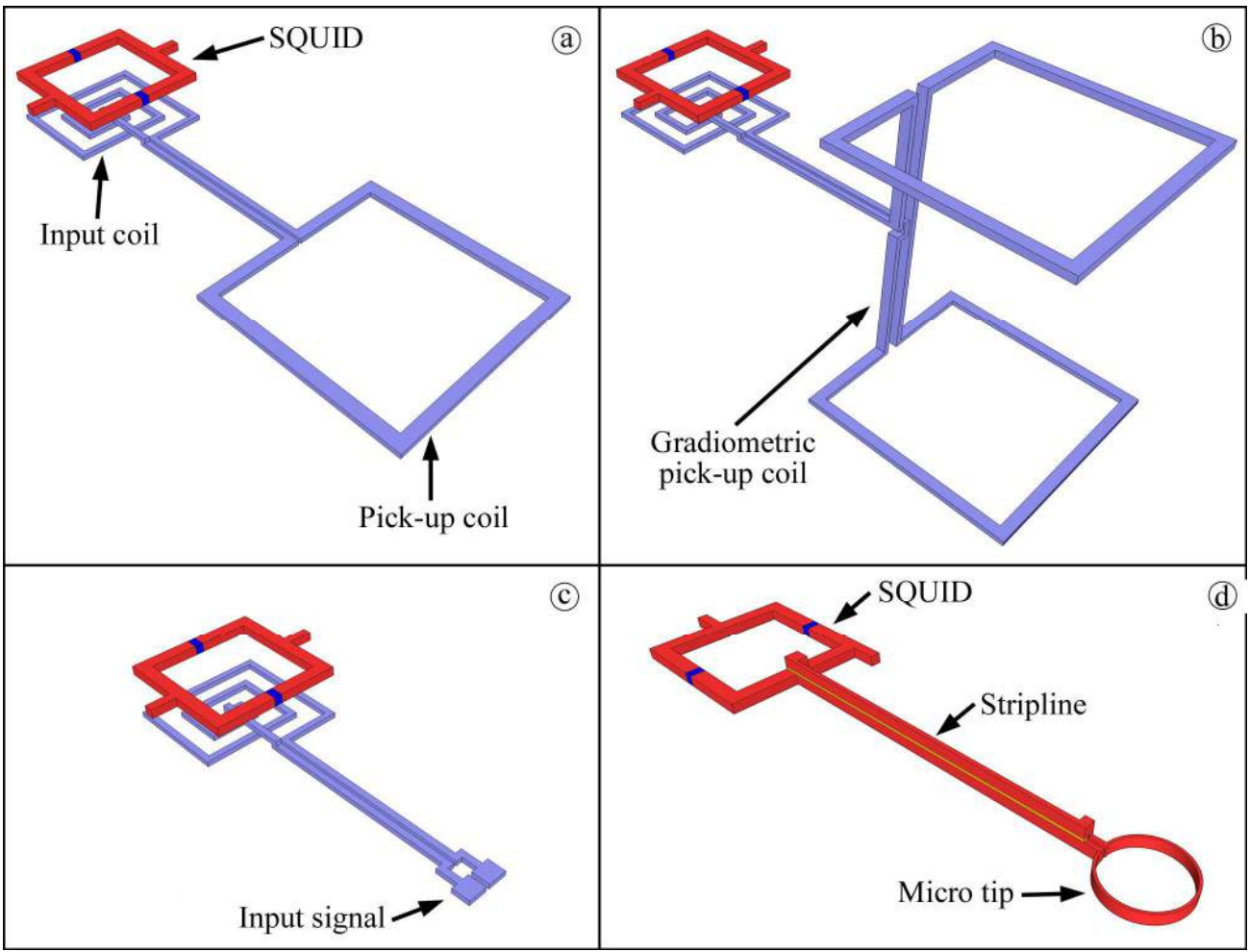

*Fig. 8 Schemes of the main SQUID configurations: a) and b) magnetometer and gradiometer configuration, c) voltage or current sensor, d) SQUID for magnetic field detection with a high spatial resolution.*

Therefore, in order to increase the magnetic field sensitivity, a superconducting flux transformer is employed. It consists of a superconducting primary coil working as a magnetic flux pick-up (pick-up coil) connected in series with a superconducting secondary coil magnetically coupled to the SQUID (input coil) (Fig.8a). When a magnetic flux $\Phi_p$ is applied, due to Meissner effect, a screening current flows into pick-up coil to nullify the total magnetic flux. Such screening current flows also in the input coil inducing a magnetic flux $\Phi_S$ into the SQUID loop [27,41]:

$$\Phi_S = \frac{M_i}{L_i + L_P}\Phi_p = \frac{k_i\sqrt{LL_i}}{L_i + L_p}\Phi_p \tag{19}$$

Where $L_i$ e $L_p$ are the inductances of the input coil and the pickup coil respectively, $k_i$ is a coupling factor, and $M_i$ is the mutual inductance between the SQUID loop and input coil.

By inverting the (19) and considering that $B_p=\Phi_p/A_p$ ($A_p$ is the pick-up coil area), it is possible to obtain the spectral density of the magnetic field noise $S_B^{1/2}(f)$ of the SQUID magnetometer [27, 41,71]:

$$S_B^{1/2} = \frac{S_{\Phi,p}^{1/2}}{A_p} = \frac{L_p + L_i}{M_i A_p}S_\Phi^{1/2} = B_\Phi S_\Phi^{1/2}; \quad B_\Phi = \frac{L_p + L_i}{M_i A_p} \tag{20}$$

$B_\Phi$ is the magnetic flux to magnetic field conversion efficiency or SQUID sensitivity. It is a fundamental parameter for a SQUID magnetometer and assumes the minimum value when $L_p = L_i$.

It is worth to note that, by using the superconductive flux transformer, an increase of field sensitivity of two orders of magnitude can be easily obtained with respect to the bare SQUID. Typically, a SQUID magnetometer with a square pick-up coil of about 1 cm$^2$ exhibits a magnetic flux noise as low as 1fT/Hz$^{1/2}$. An alternative to the superconducting flux transformer is the multiloop or cartwheel SQUID. In this configurations, the SQUID loop consists of several large loops forming a cartwheel [45,72]. The loops are connected in parallel across the same junctions in order to reduce the total inductance of the SQUID and the effective area is that of a single loop. Usually, a multiloop magnetometer exhibits a magnetic field sensitivity $B_\Phi$ lower than a flux transformer based magnetometer with the same size. Using such a design and sub-micrometer cross type Josephson tunnel junctions, a SQUID magnetometer exhibiting a magnetic field noise as low as 0.3 fT/Hz$^{1/2}$ has been developed few years ago [73].

It worth noting that promising superconductive magnetometers not based on the quantum interference have been also developed in the last years. They include magnetometers based on flux-flow in long Josephson junctions [74-76] and kinetic inductance magnetometers [77].

**1.5.2 Magnetic gradient measurements**

The magnetometer configuration is able to detect both uniform and non-uniform magnetic fields. To reject the uniform magnetic field, the two pick up loops are wound in opposite directions and balanced. So, a change in a uniform magnetic field does not induces supercurrent in the input coil and subsequently there is no magnetic flux treading the SQUID, whereas a magnetic filed gradient generates a supercurrent (proportional to the gradient) that is detected by the SQUID (Fig. 8b). Being sensitive to the magnetic field gradient, these devices are called first derivative gradiometer. To measure fields gradient in z direction ($\partial B_z/\partial z$), the axial configuration is used while the planar configuration detects those in x-y directions ($\partial B_x/\partial y$ or $\partial B_y/\partial x$).

Considering the axial configuration (Fig. 8b), the magnetic flux coupled into the SQUID via mutual inductance $M_i$ is [78]:

$$\Phi_s = M_i J_s = \frac{2d^3 M_i}{2L_p + L_i} \left( \frac{\partial B_z}{\partial z} \right) \tag{21}$$

Where $d$ is the distance between the pickup coil centres (baseline); thus the spectral density of the noise gradient is expressed by:

$$\left. \frac{\partial B_z}{\partial z} \right|_n = \frac{2L_p + L_i}{M_i} \frac{S_\Phi^{1/2}}{2d^3} = B_\Phi \frac{S_\Phi^{1/2}}{2d^3}; \qquad B_\Phi = \frac{2L_p + L_i}{M_i} \tag{22}$$

Balances of 1 part in 100 to 1 part in 1000 are typically obtained in axial gradiometers. Analogue expression can be found for SQUID gradiometers in planar configuration that offers the advantage of both a very high intrinsic balance, limited by the precision of the photolithographic techniques (1 part in 10000), and a good matching between the input inductance of the SQUID and the detection superconducting circuit inductance ensuring an optimal signal transfer. Moreover, a planar SQUID gradiometer avoids the unreliable superconducting soldering, guarantying a better reliability during the thermal cycles.

It is also possible to realize second derivative gradiometers by arranging the detections coils in a suitable way. If well balanced, they are insensitive to the uniform magnetic fields and its first spatial derivative. The SQUID gradiometers in both configurations are widely employed in multichannel systems for biomagnetism where a high (ambient field) rejection factor is required.

### 1.5.3 Current and voltage measurements

A simple way to measure an electrical current by a SQUID is to send it in a coil coupled to the SQUID loop [1, 41] (Fig. 8c). In this case, the magnetic flux coupled to the SQUID loop is $\Phi = M\,i$ and the spectral density of the current noise will be:

$$S_I^{1/2} = \frac{S_\Phi^{1/2}}{M} = I_\Phi S_\Phi^{1/2}; \quad I_\Phi = \frac{1}{M} \tag{23}$$

Where $M$ is the mutual inductance between the input coil and the SQUID loop. A suitable way to obtain a practical and reliable SQUID amperometer is to use a flux transformer that converts easily the electrical current into a magnetic flux treading the SQUID loop and allows to obtain fully integrated sensors. An efficient way to couple the signal coil to the SQUID is also obtained by using a double transformer coupling [79-81] by inserting a matching (intermediary) transformer between the signal coil and the input terminals of the classic configuration. With respect to the single transformer, it allows to efficiently match the low-inductance SQUID with a very high signal coil inductance (tens of µH), obtaining an ultra low electric current noise. In the last years, superconducting current sensors based on the SQUID including an integrated signal coil (single transformer), two stage sensor including double-transformer coupling exhibiting a spectral current noise of few tens of fA/Hz$^{1/2}$ at T=4.2 K, have been developed [82-83].

A SQUID voltmeter can be simply obtained by connecting the signal source in series with the input coil of the SQUID, via a resistance. In this configuration, the feedback current is sent into a known resistance to obtain a null balancing measurement of voltage. The noise is essentially limited by the Nyquist noise of the resistance which can vary from $10^{-6}$ to 100 $\Omega$ giving a spectral density of the voltage noise ranging from $10^{-14}$ to $10^{-10}$ V/Hz$^{1/2}$.

**1.5.4 Superconducting flux transformer design**

As seen, the magnetometer, gradiometer and amplifier designs can be optimized by minimizing the value of $B_\Phi$ or $I_\Phi$ in equations 20-23. If the flux-transformer is integrated, an excellent coupling to the SQUID is obtained using a Ketchen-type design [84,85]. The SQUID loop, in a square washer configuration, is coupled to a multi-turn thin film input coil, which is connected to a square single-turn pick-up loop (Fig.9). In such a way, the coupling between the washer and the input coil is very effective and the input coil inductance is proportional to the turn numbers and hole inductance. In this configuration, the SQUID inductance does not depend on the outer dimension of the washer but only on the hole dimension. Hence, the input coil inductance can be adjusted to match a suitable load by varying the outer dimension of the washer to accommodate the required number of input coil turns. The Josephson junctions are located on the outer edge of the square loop, away from the higher field region at the centre square hole. Consequently, a slit through the conductor loop is used, introducing a parasitic inductance. Such additional inductance is only partially coupled to the coil turns, reducing the overall coupling efficiency. So, it is preferable to avoid very long slit.

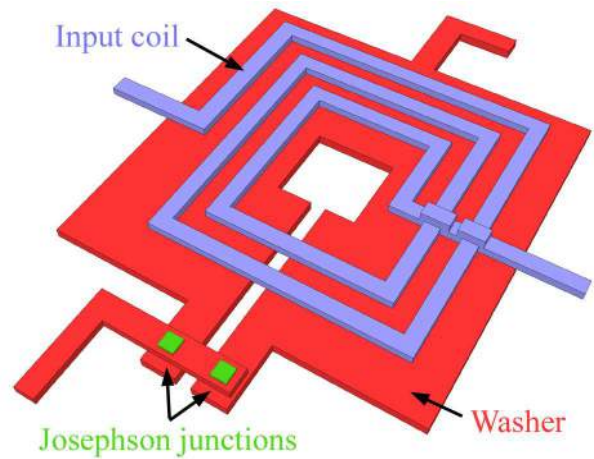

*Fig. 9 Ketchen-Jaycox flux transformer scheme: a multi turn input coil is magnetically coupled to a SQUID ring in a washer configuration.*

Moreover, the washer structure tends to focus magnetic flux into the central hole by an amount proportional to the product of outer dimension *l* and the hole dimension *d* increasing the flux capture area by a factor $(A_w/A_h)^{1/2}$ ($A_w$ and $A_h$ are the geometrical areas of the washer and hole respectively) [86]. The flux focusing effect is largely employed to fabricate SQUID sensors for many applications.

**1.5.5 High spatial resolution measurements**

In order to increase the magnetic field sensitivity, it is needed to increase the detection coil or the SQUID loop dimensions. However, in this way the spatial resolution decreases and so, the capability of the SQUID to distinguish two or more magnetic sources close together. Hence, if a high spatial resolution is required, as in the magnetic microscopy, the pick-up coil should be as small as possible [87-89]. If the area of the detection coil is comparable or lower than the SQUID loop area, the flux transformer is not effective and the SQUID loop itself can be employed as detection coil. However, it is preferable to keep the Josephson junctions away from the magnetic source under investigation. Typically, a circular pick-up loop is a part of the SQUID self-inductance and it is connected by a stripline to Josephson junctions (Fig. 8c). The stripline consists of two superconducting layers separated by a double insulation layer and it is insensitive to normal fields, avoiding additional parasitic detection area [90]. Between the end of the stripline structure and the Josephson junctions a washer structure can be inserted in order to modulate the sensor and to operate in flux locked loop (FLL) mode. In such a way, direct coupling of the modulation magnetic flux to the measurement volume is avoided. SQUID having a pick coil area of the order of few $\mu m^2$ are called microSQUIDs and are mainly employed in superconducting scanning microscopes. As expected, microSQUIDs exhibit a poor magnetic field sensitivity. In fact for a micropick-up coil area ranging from 4 to 25 $\mu m^2$ we have a $B_\Phi=1/A_p=80\text{-}500$ $\mu T/\Phi_0$ and so a high value of magnetic field noise spectral density $S_B^{1/2}$ 80-500 $pT/Hz^{1/2}$ (here a magnetic flux noise of $1\mu\Phi_0/Hz^{1/2}$ has been assumed).

**1.6 Main SQUID applications**

Due to their ultra-high sensitivity, SQUIDs are widely employed in many applications. Here, we briefly mention the main SQUID applications. An exhaustive review of most of the SQUID applications, expect those related to nanoSQUIDs, can be found in the references [2-4]. In the Fig. 10, we report the typical magnetic field sensitivity and the bandwidth of the different applications. The sensitivities of the LTs and HTs SQUID sensors are also indicated.

*Biomagnetism,* namely the study of magnetic field associated to the electric activity in the human body, is one of the most important applications of SQUIDs and in particular of SQUID magnetometers and gradiometers. Since from the first SQUID measurements of biomagnetic signal arising from the heart and the brain in the early 70', many efforts have been devoted to the development of biomagnetic instrumentations. Nowadays, suitable multichannel systems, up to some hundreds channels [91-94], are available for the study of the magnetic activity of the brain (magnetoencephalography, MEG) and the heart (magnetocardiography, MCG). MEG is a non-invasive, functional imaging technique that measures magnetic fields generated by the neuronal activity of the brain using ultra high sensitivity SQUID sensors. Among the available brain functional imaging methods, MEG uniquely features both a good spatial and an excellent temporal resolution, allowing the investigation of many key questions in neuroscience and neurophysiology. MEG measurements reflect intracellular electric current flowing in the brain and provide direct information about the dynamics of evoked and spontaneous neural activity. In addition, MEG measurements are not subject to interferences due to the tissues and fluids lying between the cortex and the scalp, and magnetic fields are not distorted by the different conduction of the skull, unlike with electroencephalograms (EEGs). These features make MEG an excellent tool to localize the subcortical sources of brain activity and to investigate dynamic neuronal processes, as well as to study cognitive processes, such as language perception, memory encoding and retrieval and higher-level tasks. As regard as the clinical applications, it has been proven that MEG is a useful diagnostic tool in the identification, prevention and treatment of numerous diseases and illnesses.

The MCG is a non-invasive electrophysiological mapping technique that provides unprecedented insight into the generation, localization, and dynamic behaviour of electric current in the heart. The aim of the MCG measurement is to determine the spatio-temporal magnetic field distribution produced by the cardiac electric activity in a measurement plane just above the thorax. MCG signals, unlike ECG, are not attenuated by surrounding anatomical structures, tissues, and body fluids, thereby providing more accurate information. Moreover, compared with surface potential recordings, multichannel MCG mapping is a faster and contactless method for 3D imaging and localization of cardiac electrophysiologic phenomena with higher spatial and temporal resolution.

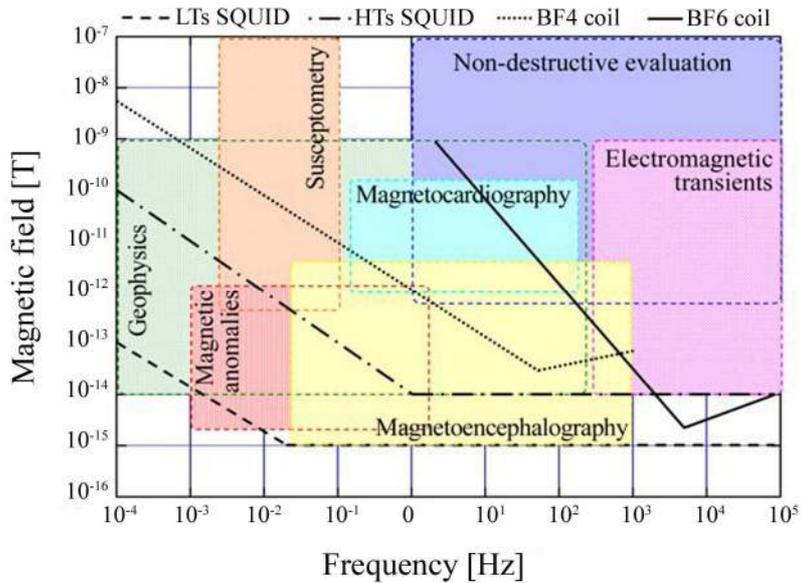

*Fig.10 Magnetic field sensitivity and the bandwidth of the main SQUID applications. The lines indicate the sensitivities of four different magnetic sensors (LTs and HTs SQUIDs, BF-4 and BF-6 coil). Note that only the LTs SQUID is enough sensitive to cover all application fields.*

As documented by numerous papers, MCG can also be useful to improve diagnostic accuracy. Recently, the development of standardized instrumentations for unshielded MCG, its ease of use and reliability even in emergency rooms has triggered a new interest from clinicians for magnetocardiography, leading to several new installations of unshielded systems worldwide. In addition to brain and heart, the non-invasive measurement techniques based on SQUIDs have been applied also to other human organs such as stomach and intestines (magnetoenterography), lung (magnetopneumography) [95], muscles (magnetomyography) and peripheral nerves (magnetoneurography) [96]. Other useful biomedical applications include liver iron concentration detection which allows a completely non invasive measurement of iron overload in patient affect by thalessemia or hemochromatosis [97,98]. SQUID Magnetic immunoassay employs functionalised molecules bound to an antibody or antigen. By measuring the remnant magnetic field or the magnetic relaxation is possible to determine the bound or unbounded molecules [99].

An interesting field of application of SQUID sensors is *nondestructive evaluation (NDE)* [2-4, 100,101]. NDE is the non-invasive identification of structural or material flaws in a specimen. Examples are the imaging of surface and subsurface cracks or pits due to corrosion or fatigue in aging aircraft and reinforcing rods in concrete structures. There are several competing methods for NDE such as acoustic, thermal, and electromagnetic techniques. The advantages of the SQUID for NDE include high sensitivity (about 10–100 fT·Hz$^{-1/2}$), wide bandwidth (from dc to 10 kHz), broad dynamic range (>80 dB). Moreover, the ability of SQUIDs to operate down to zero frequency allows them to sense much deeper flaws than traditional techniques, to detect and monitor the flow of steady state corrosion currents, and to image the static magnetization of paramagnetic materials. An important application of SQUIDs in NDE is the detection of subsurface damage in metallic structures such as aircraft parts by eddy current techniques. Here, an alternating magnetic field produced by a drive coil is applied and the fields generated by the induced eddy currents in the structure are lock-in detected. The structural flaws diverts the eddy currents distorting the magnetic field. Since the eddy currents flow over a skin depth, which is inversely proportional to the square root of the frequency, deep defects require correspondingly low frequencies. Here, the flat frequency response of SQUIDs is a distinct advantage over the response of currently used coil systems, which fall off with decreasing frequency.

*Scanning SQUID microscopy (SSM)* is a technique capable of imaging the magnetic field distribution in close proximity across the surface of a sample under investigation with high sensitivity and modest spatial resolution [2-4, 88, 102]. Most often, the sample is moved over the SQUID in a two-dimensional scanning process and the magnetic signal is plotted versus the coordinates to produce an image. The frequency at which the image is obtained ranges from near zero, where simply the static magnetic field produced by the sample is measured, to beyond 1 GHz.

Today, SQUID microscopes with cold samples have a spatial resolution of about 5µm, while those with room temperature samples have a resolution ranging between 30 and 50µm. The advantage of the scanning SQUID microscope is its very high sensitivity. In fact, the scanning SQUID microscope is orders of magnitude more sensitive to magnetic fields than the other techniques. A disadvantage of SSM is its relatively poor spatial resolution. Whereas for scanning SQUID microscope a resolution of only 5µm has been demonstrated, scanning electron microscopy with polarization analysis, for example, has a spatial resolution of 30-50 nm. Nevertheless, there are many possible applications of SSM, which do not require submicron spatial resolution.

SQUID systems are also employed in determining the magnetic properties of the earth [2-4]. This concerns both the characterization of specific earth samples (rock magnetometry) and the mapping of the earth magnetic field as well as its electromagnetic impedance. In this framework, an important application of SQUID is in *magnetotellurics*, involving the simultaneously measurements of the fluctuating horizontal components of the electric and magnetic fields at the earth's surface originated in the magnetosphere and ionosphere. From these frequency-dependent fields, the impedance tensor of the ground can be calculated estimating the spatial variation of the resistivity of the ground. The interesting frequency range is about $10^{-3}$ to $10^2$ Hz corresponding to a skin depth between about 50 km and 150 m (assuming a resistivity of 10Ω·m). The sensitivity required for magnetotellurics is about 20-30 fT/√Hz in the white noise regime and a 1/f knee of 1 Hz. Currently, magnetic measurements in geophysics are mostly made with induction coils. Below about 1 Hz, the spectral density of the noise in coils increases as *$1/f^3$*, whereas that of SQUIDs increases only as 1/f, giving the latter magnetometer a substantial advantage at low frequencies. Furthermore, coils for use below 1 Hz can be as long as 1.5 m, and the deployment of three such coils orthogonally, buried in the ground for stability, is a tedious undertaking. So, a three-axis SQUID magnetometer in a compact Dewar with a long hold time becomes competitive.

SQUIDs play a key role in *Metrology* and in particular in the development of quantum electrical standards [2-4]. Significant applications includes the cryogenic current comparator that enables the accurate measurement of the ratio of two direct currents. Promising and interesting results have been also obtained in other domains: ionising radiation, X-ray and Γ-ray spectrometry, thermometry, etc [2-4].

Moreover, as mentioned in the introduction SQUID devices have been successfully employed in several experiments of *basic Physic* including cosmology, astrophysics, general relativity, particle physics, quantum optic, and quantum computing.

## 2. NanoSQUIDs: fundamentals, theoretical elements and spin sensitivity

In the previous section, we have seen that a SQUID is able to detect, with an ultra high sensitivity, a magnetic flux or any physical quantity that can be converted in a magnetic flux threading the SQUID loop. The device design strongly depends on the measurement to be performed. What can we say about the capability of SQUIDs to detect the magnetic moment associated to a nano-object? How to design and fabricate an appropriate SQUID sensor? How we can evaluate the magnetic moment sensitivity of a SQUID?

In this section, we will provide an answer to all previous questions. We will start with an introduction to nanoSQUIDs stressing the motivation, the principle of measurement of magnetic nano-objects and their potentiality in view of investigations of small spin populations.

### 2.1 NanoSQUID: origin and motivation

As seen in the section 1, the energy sensitivity of a SQUID is proportional to the loop inductance $L$ (equation 13, 14). Hence, to increase the sensitivity, a small inductance has to be employed. Following this criteria, in the 1980 Voss et al [37,103], developed a niobium dc-SQUID to reach an energy resolution close to the limit of the uncertainty principle. By using electron beam lithography they fabricated a SQUID having a loop area less than 1 $\mu m^2$ and two micro Dayem bridges and obtained an an energy sensitivity of 19 $\hbar$ at T=4.2 K. Few years later, Awschalom et al [49] and Ketchen et al [104] showed that the SQUID capability to detect a magnetic moment is proportional to the diameter of its loop and introduced a new figure of merit: the spin sensitivity, where the spins are counted in unit of Bohr magnetons ($\mu_B = e\hbar/2m_e = 9,274 \times 10^{-24}$ J·T$^{-1}$). Their devices were based on two micro pick-up coils in a gradiometric configuration connected to an input coil magnetically coupled to a SQUID loop. An energy sensitivity as low as 1.7 $\hbar$ was obtained at T=290 mK [49]. They proposed to employ these SQUID micro-gradiometers to investigate the magnetic properties of micro-sized samples at low temperature. At same time, Gallop et al. [105], proposed a prototype SQUID magnetometer based on the measurement of the free precessing nuclear magnetization of a sample of $^3$He. They calculated the minimum number of nuclear spins detectable, founding the same Ketchen's result. In nineties of the last century, Wernsdorfer et al [106,107] successfully employed micro-sized SQUIDs in several experiments to investigate the magnetization of nanoparticles and nanowires. The interesting experimental study performed by using such microSQUIDs, stimulated the researcher to develop even smaller SQUID in order to study very small clusters of magnetic nano-objects, with the aim to detect the single spin. Thanks to the progress in nanotechnology, in the 2003 Lam and Tilbrook [108] developed the first nanoSQUID for the detection of small spin populations, having a hole side length of 200 nm. Thereafter, several nanoSQUIDs with an adequate sensitivity to detect the magnetic response of few spins have been developed. As will be shown in the following, high magnetic moment sensitivity can also be obtained by sitting the nanomagnet to be investigated on the SQUID Josephson junctions with a very small cross section. Thus, the nanoSQUIDs can have either submicron loop or nanometric Josephson junctions.

To investigate magnetic properties, the nano-object should be positioned within the pick-up coil or SQUID loop. By applying an in-plane magnetic field, a magnetization of the sample is induced resulting in a magnetic flux variation proportional to magnetization change, trough a coupling factor that depends on SQUID and sample geometries [106, 107, 109].

**2.2 Magnetic moment and spin sensitivity**

In order to intuitively understanding why the capability of a SQUID to detect a small magnetic moment increases by decreasing the loop size, consider the sketch reported in the Fig. 11. A representation of the magnetic flux lines related to a magnetic moment oriented in the z direction and SQUID having three different coil sizes are reported. In the largest coil case (a), we can see that only few field lines contribute to the total flux threading the loop, the others return within the loop and will give no net contribution to the magnetic flux. By decreasing the detection coil size, the number of flux lines which return within the loop decreases and the net magnetic flux increases. In the smallest coil (c), all field lines give a contribution to the magnetic flux. On the other side, if the size of coil tends to infinite, there is no net magnetic flux linkage into the loop because all flux lines return within the loop.

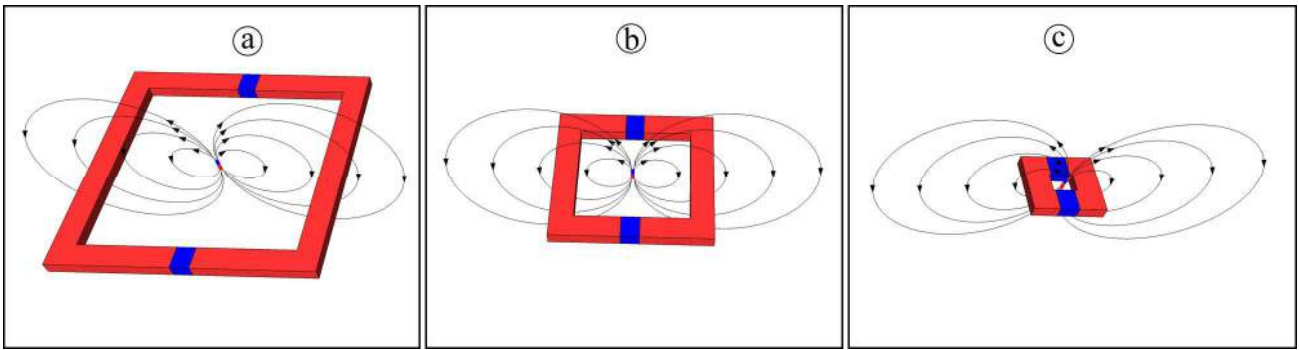

*Fig.11 Schematic representation of magnetic flux lines of a magnetic moment located in centre of a SQUID flux capture area for three different loop sizes.*

Beyond this intuitive explanation, we will provide a formal demonstration of the above statement for a simple case and will give the definition of the spin or magnetic moment sensitivity.

Let us consider the elementary magnetic moment ($\mu_B$) or a single spin positioned over a square coil with a side length *L* and oriented along the z-axis (Fig.12a). We also suppose that the SQUID coil is composed of an idealized filamentary square loop and that the spin lies in the same plane of the SQUID loop and it is located in its centre, in other words $x'=y'=z'=0$. To obtain the magnetic flux through the SQUID loop, we have to calculate the surface integral on the loop area of the magnetic field produced by the magnetic moment $\mu_B$. It is equivalent to calculate the line integral of the vector potential associated to $\mu_B$ across the contour ring. Under the aforementioned assumptions, the Cartesian components of the magnetic vector potential A(**r**) at generic position r(x,y,z), are:

$$A_x = -\frac{\mu_0 \mu_B}{4\pi} \frac{y}{r^3}; \qquad A_y = \frac{\mu_0 \mu_B}{4\pi} \frac{x}{r^3} \qquad (24)$$

Here, $\mu_0$ is the magnetic vacuum permeability, and $r=[x^2+y^2+z^2]^{1/2}$. The magnetic flux due to the elementary magnetic moment is:

$$\Phi_\mu = \oint \vec{A} \cdot d\vec{s} \qquad (25)$$

Where the integral is considered along the closed line of the SQUID loop. The contribution of the four sides of the square loop can be calculated as [110]:

$$\Phi_\mu = \frac{\mu_0 \mu_B}{4\pi}\left(2\int_{-L/2}^{L/2} dx \frac{L/2}{[(L/2)^2+x^2]^{3/2}} + 2\int_{-L/2}^{L/2} dy \frac{L/2}{[(L/2)^2+y^2]^{3/2}}\right) = \frac{2\sqrt{2}\,\mu_0\mu_B}{\pi L} \qquad [26]$$

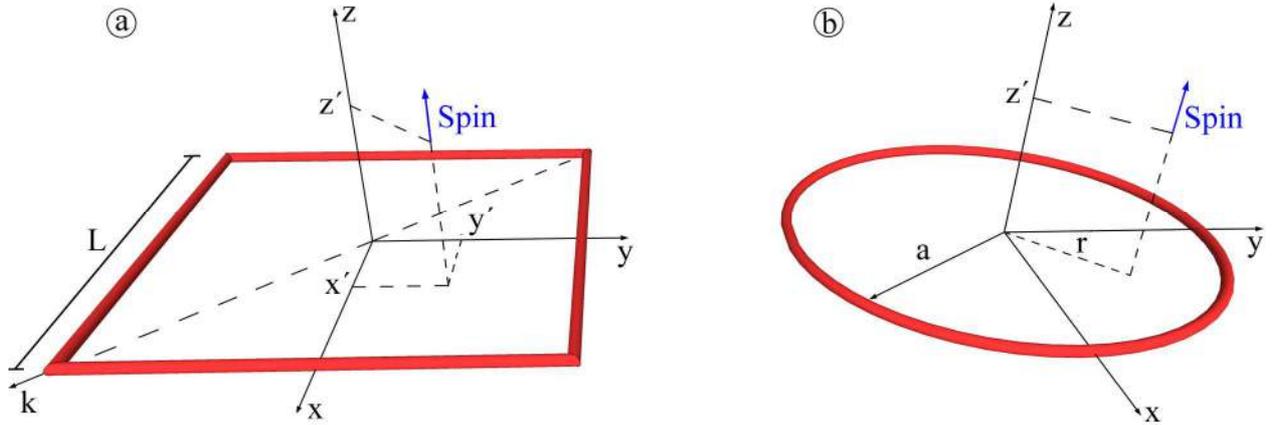

*Fig.12 Scheme of a square (a) and circular (b) SQUID detection coil including an elementary magnetic moment (Bohr magneton or spin) located in a generic position $P=(x',y',z')$.*

The same result is obtained by integrating the magnetic field of the dipole over the SQUID loop area [111]. In the case of circular loop (Fig.12b), $\Phi_\mu = \mu_0 \mu_B /2a$, where $a$ is the coil radius and imposing conditions $r'=\theta'=z'=0$ [104,108]. A relevant figure of merit of a nano-SQUID is the minimum detectable spin number per band unit or spectral density of the spin noise $S_n^{1/2}$. It can be obtained by taking the ratio of the spectral density of the flux noise $S_\Phi^{1/2}$ and the net magnetic flux due to the single spin:

$$S_n^{1/2} = \frac{S_\Phi^{1/2}}{\Phi_\mu} = \frac{S_\Phi^{1/2} \pi L}{2\sqrt{2}\,\mu_0\mu_B} \qquad \text{(square loop)} \qquad (27)$$

$$S_n^{1/2} = \frac{S_\Phi^{1/2}}{\Phi_\mu} = \frac{S_\Phi^{1/2} 2a}{\mu_0 \mu_B} \qquad \text{(circular coil)} \qquad (28)$$

In both case, the spin sensitivity increases by decreasing the side length $L$ or the diameter $2a$ of the SQUID loop. As for the SQUID magnetometer, gradiometer or amperometer, even in this case we can define a transfer factor $M_\Phi$, equal to the inverse of $\Phi_\mu$ so that the spin sensitivity can be expressed as: $S_n^{1/2} = M_\Phi S_\Phi^{1/2}$. As will be shown in the section 2.6, the spin sensitivity strongly depends on the position of the magnetic moment within the SQUID loop and on the distance from the SQUID plane.

**2.3 A simple way to make a nanoSQUID**

As shown in the previous section, the capability of a SQUID to measure small magnetic nano-objects depends on the detection coil size. In particular, if we consider a flux noise of 0.1-1.0 $\mu\Phi_0/Hz^{1/2}$, we can estimate by the formulas (25, 26) that a sensitivity of few spin per bandwidth unit requires a loop side length of 100 nm or a radius of 50 nm in the case of circular loop.

SQUIDs having an effective area much smaller than 1 $\mu m^2$, require deep sub-micron Josephson junctions in order to maintain the detection size as designed. However, a typical SQUID employs two Josephson tunnel junctions that are limited by photolithographic process to about one micron. Apart from the technological limitations, there are major difficulties in using conventional tri-layer Josephson tunnel junctions at deep sub-micron scale, since the junctions should have a critical current density as high as $10^7$-$10^8$ A/cm$^2$ in order to guarantee a suitable noise parameter of the SQUID device (see section 1.4). This requirement adds further complications to the fabrication process. As will be show in the next section, very recently, most of the above obstacles have been overcome.

In any case, good alternatives to tunnel junctions are superconducting nanobridges [112] (or Dayem nanobridges) which consist of nano-constrictions in a superconducting film with length and width less than one micrometer (Fig. 13). The Josephson effect into the nanobridge junction were predicted in 1964 by Anderson and Dayem [113]. In particular, they first demonstrated that superconducting bridges display microwave-induced steps in their current-voltage characteristics, similar to those observed in Josephson tunnel junctions. Nano-bridges junctions show a much higher critical current and they can be made by the same superconducting material as the rest of the SQUID in a single layer superconducting thin film. Thus, the fabrication of nanoSQUIDs based on nanobridges is relatively simple being obtained by a single nanopatternig step, avoiding the alignment of several layers on top of each other.

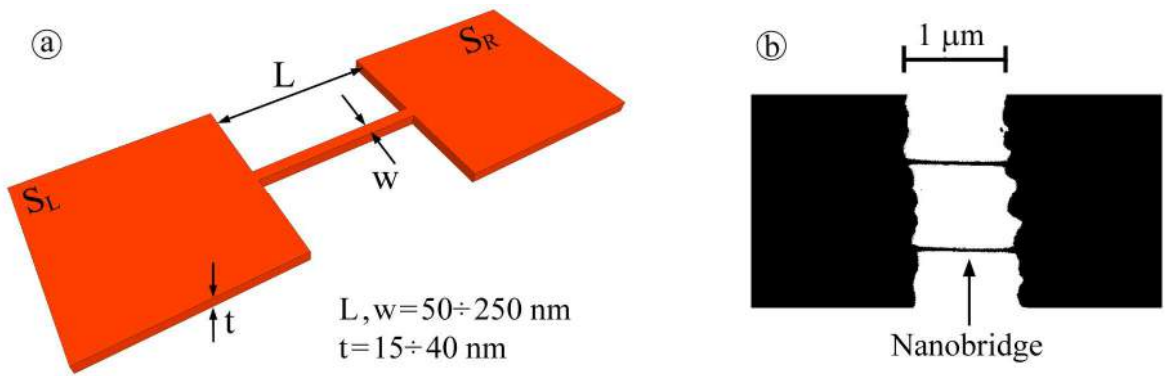

*Fig.13 a) Schematic view of a nano-constriction Josephson junction. If the length and the width of the bridge are comparable with the coherence length of the superconductor, a Josephson tunnelling occurs. b) microSQUID based on Dayem bridges fabricated by electron beam lithography. Image adapted from [37].*

Contrary to the tunnel junctions, Dayem nanobridges exhibit a high current density and are resilient to the magnetic field applied in the plane containing the SQUID loop. Moreover, the use of very thin film (10÷20 nm) to fabricate SQUID based on nanobridges, prevent flux trapping also for applied magnetic field up to several T. The insensitivity to high magnetic fields applied in the SQUID plane is a necessary condition for the measurement of the nanoparticles magnetization. On the other hand, unlike an over-damped SQUID based on tunnel Josephson junctions, the current-voltage (*I-V*) characteristics of SQUIDs based on Dayem nanobridges are typically hysteretic at T≤4.2 K. This prevents the use of the SQUID as a magnetic flux to voltage transducer, resulting in a loss of device sensitivity. Figure 13b, shows the nanobridge based SQUID reported in the pioneering work of Voss et al [103].

Note that, if the width and the thickness of a superconducting line is small (less than 1 μm), the kinetic inductance can no longer be neglected. For a nanowire, the kinetic inductance is [114]:

$$L_K = \frac{\mu_0 \lambda^2 \ell}{wt} \quad (29)$$

Where $\lambda$ is the London penetration depth, $l$ and $w$ are the length and the width of the bridge respectively, and $t$ is its thickness. If we consider a niobium nanobridge having a length of 100 nm, a width of 80 nm and a thickness of 20 nm, the kinetic inductance is 1.5 pH, which could be comparable or greater than the SQUID geometrical inductance. Therefore, we can expect that total inductance is dominated by the contributions of the kinetic inductances of the bridges.

Despite their simple structure, we would like to point out that the superconducting nanobridges have been proposed also in other interesting applications as photon detectors, hot electron bolometers as detectors in astrophysical observations at terahertz frequencies and in flux qubits [115-118].

**2.4 Theoretical aspects of Dayem bridges**

There are several exhaustive reviews describing the theoretical aspects of Josephon junction based on nanobridges [112, 119-120], so below we will provide just the main peculiar aspects useful to understanding the predictions of the main features of the SQUID based on nanobridges.

The most important issue of a Josephson structure (tunnel junction, Dayem bridge, point contact, etc.) is the relationship between the supercurrent and the phase difference across the superconducting structure. The first Josephson equation [1.1] states that the Current Phase Relationship (CPR) is sinusoidal; however, it is not generally true. If the length of the nanobridge (L) is smaller than both the coherence length (ξ) and the magnetic field penetration depth (λ) of the superconducting film, the classical sinusoidal dependence of the supercurrent $I_0$ on the phase difference φ between the two superconductor electrodes is preserved.

$$I_0(\varphi) = \frac{\pi \Delta_L \Delta_R}{4eR_N T} \sin(\varphi) \quad (30)$$

Where $\Delta_L$ and $\Delta_R$ are the superconducting order parameters of the left and right superconductors, $T$ is the temperature and $R_N$ is normal state resistance (Aslamazov and Larkin model) [121].

Likharev and Yakobson [122] studied the effects of an increasing nanobridge length on both critical current and CPR. They founded that by increasing the nanobridge lengths, the critical current initially diminishes tending to a constant value when $L > 8\,\xi$. The CPR also depends on the ratio between the bridge effective length $L$ and the coherence length $\xi$ of superconducting film. In particular, for $L/\xi > 1$, a deformation of the CPR from sinusoidal shape proportional to the $L/\xi$ ratio itself occurs. Pei et al. [123] observed such CPR deformations in indium microbridges.

For $L > 3.5\,\xi$, the CPR becomes multivalued with a shape close to that given by the curve:

$$I_S(\varphi) = I_c \frac{\varphi\,\xi(T)}{L}\left[1 - \left(\frac{\varphi\,\xi(T)}{L}\right)^2\right] = \frac{\Phi_0}{2\pi L_K}\left[\varphi - \left(\frac{\xi(T)}{L}\right)^2 \varphi^3\right] \qquad (31)$$

Where $L_k$ is the kinetic inductance given by the (29). The transition from single to multivalued CPR, determines the onset of a hysteresis in the I-V characteristic.

Note that, both Aslamazov-Larkin and Likharev-Yakobson predictions are valid only close to the critical temperature of the superconductors being based on Gizburg-Landau theory.

Recent numerical calculations of CPR as a function of $L/\xi$ ratio based on the analysis of Likharev and Yacobson are reported in the references [124-126]. They numerically solved the equations (32) and (33) derived by the reduced Ginzburg–Landau differential equations:

$$\frac{L}{\xi} = 2\int_{a_0}^{1} \frac{da}{\left|\sqrt{a_0^2 - \frac{a_0^4}{2} + \frac{j_0^2}{a_0^2} - a(x)^2 + \frac{a(x)^4}{2} - \frac{j_0^2}{a(x)^2}}\right|} \qquad (32)$$

$$\varphi = 2 j_0 \xi \int_{a_0}^{1} \frac{da}{a_0^2 \left|\sqrt{a_0^2 - \frac{a_0^4}{2} + \frac{j_0^2}{a_0^2} - a(x)^2 + \frac{a(x)^4}{2} - \frac{j_0^2}{a(x)^2}}\right|} \qquad (33)$$

Where $a(x)$ is a function so that $a(0)=a(L)=1$; in the region of the nanobridge $a(x)<1$ and has its minimum $a_0$ at the centre of the bridge ($a(L/2)=a_0$) [125]. $j_0$ and $\varphi$ are the scaled current density and the phase difference across the Dayem bridge respectively. For a given L/ξ

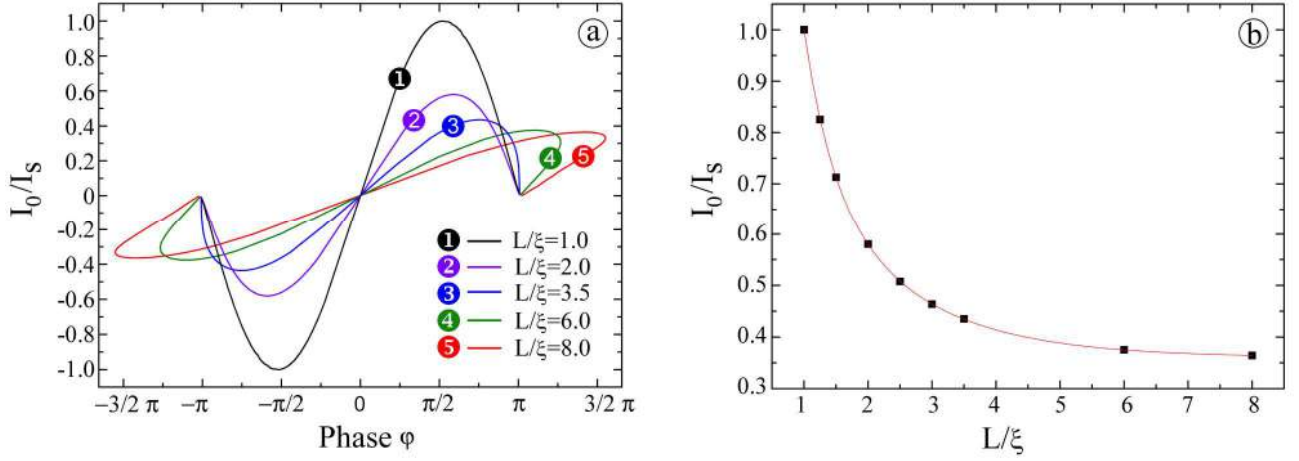

*Fig.14 a) Current-phase relationship for different L/ξ ratio calculated by numerically solving the equation (32) and (33). For L/ξ greater than 3.5, the curve becomes double valued. b) Maximum critical current as a function of the L/ξ ratio (adapted from [126]).*

ratio, by varying the $a_0$ parameter in (32), the corresponding $j_0$ values is found by using a scalar nonlinear zero finding function. The pairs ($a_0$, $j_0$) are substituted in the (33) to find the corresponding phase values. The pairs ($\varphi$, $j_0$) are used to build a current-phase curve $P_{L/\xi}(\varphi)$. Fig. 14a reports the CPRs for *L/ξ* ranging from *1* and *8*; we can see that the CPR is sinusoidal for *L/ξ=1* and becomes double value for *L/ξ>3.5* giving rise the occurrence of hysteresis in the current-voltage characteristic. The Fig. 14b shows the maximum critical current $I_0$ as a function of the *L/ξ* ratio, showing an exponential-like decay of the $I_0$. The saturation value, reached for *L≈8ξ*, corresponds to the depairing critical current density of an infinitely long superconducting wire [122], in this case $I_S/I_C=(2\sqrt{3})/9$.

We would like to emphasize that the hysteresis in the Josephson weak link like the nanobridge is a controversial issue. In fact, many researches address such behaviour to a thermal effect that is, when the nanobridge switch in the resistive state, the Joule dissipation leads to a local temperature increase causing a decrease of the Ic and a consequent hysteresis in the I-V characteristic. Skocpol et al [127], carried out a detailed experimental and theoretical study about the thermal effects in superconducting nanobridge. They provided a model based on the self-heating hotspot to describe the nanobridge I-V characteristics exhibiting hysteresis. As will be shown later, to avoid the thermal hysteresis, some nanoSQUIDs are fabricated by depositing on the superconducting film a normal film that act as a thermal shunt.

As regard as the behaviour of the CPR as a function of the temperature, the first predictions were made by Kulik and Omelyanchuk (KO-1 theory) [128] who considered a one-dimensional wire connecting two superconductors. The wire had a length $L<<(\xi l)^{1/2}$ (dirty limit) and a width $w<<L$, where $l$ is the electronic mean free path. In the dirty limit case is possible to use the Usadel equations [129], to simplify the complicated general equations of the stationary superconductivity based on the quantum field theory [130,131]. For a symmetric junction formed by identical superconductors, they derived the following formula, which allows to calculate the CPR at arbitrary temperature:

$$I_S = \frac{4\pi T}{e R_N} \sum_{\omega>0} \frac{\Delta \cos(\varphi/2)}{\sqrt{\Delta^2 \cos^2(\varphi/2) + \omega^2}} \arctan \frac{\Delta \sin(\varphi/2)}{\sqrt{\Delta^2 \cos^2(\varphi/2) + \omega^2}} \qquad (34)$$

Where $\omega = \pi T(2n+1)$ are the Matsubara frequency. The curves $I_S(\varphi)$ are non sinusoidal at low temperatures, reducing to the Aslamazov-Larkin result for $T \to T_c$. On the other hand, we note that the $\xi(T)$ increases with the temperature, so, for a fixed bridge length, an increase of the temperature leads to a decrease of the ratio $L/\xi(T)$. Therefore, we expect to observe a greater deformation of the CPR by decreasing the temperature.

Direct measurement of the temperature dependent CPR on niobium nanobridge were performed by Troeman et al. [132]. They observed a deformation from Josephson-like sinusoidal characteristics at high temperatures to saw-tooth shaped curves at intermediate and multivalued relationships at low temperatures.

Note that, the conditions of the K-O1 theory are not easily feasible in the practice. In fact, if we consider an Al nanobridge (10-30 nm thick), 1 nm long, the resulting coherence length is about 30 nm, so that $(\xi(0)l)^{1/2} \approx 5.5$ nm making the short limit difficult to achieve with the available nano-lithographic tools. Few years ago, Vijay et al computed the CPR [133] for Al nanobridges having a length comparable to the coherence length at *T=0* and inserted between two-dimensional (2D) and three-dimensional (3D) superconducting electrodes. They used the general theory of Gor'kov [130, 131] in the diffusive limit and numerically solved the Usadel equations for both cases. They considered a bridge width of *45 nm*, a length *L* ranging from *15* to *240 nm* and a temperature of *0.15 K*. In the 2D case, the CPR had a nearly linear behaviour also for the smallest bridge length (*L=15 nm*), while in the three dimensional case the CPR resembles a slight distorted sinusoid and becomes more linear as the bridge length increases. They computed also a two dimensional map of the phase within the superconducting structure (nanobridge and banks). Their calculations showed a logarithmic phase evolution in the bridge and the banks in the 2D case, whereas in the 3D case the phase varies mostly in the bridge region and quickly reached the imposed value in the banks within a few ξ(0). In the latter case, the electrodes act as a phase reservoir, leading to an almost sinusoidal CPR. On the contrary, in the 2D case, there is no phase reservoir and the structure resembles a superconducting wire with a linear CPR.

Since it is very difficult to fabricate niobium nanobridges having dimensions smaller or equal to the coherence length, most of the theoretical predictions based on standard dc-SQUID theory are usually unreliable and these devices exhibit characteristics that do not resemble those of standard SQUIDs.

## 2.5 NanoSQUID performance predictions

To describe the operation of a SQUID based on nanobridges, the equations 11 have to be solved by using the CPR of the nanobridges instead the sinusoidal one.

$$\frac{i}{2} + j = P_{L/\xi}(\varphi_1) + \frac{d\varphi_1}{d\tau} + i_{n1}$$

$$\frac{i}{2} - j = P_{L/\xi}(\varphi_2) + \frac{d\varphi_2}{d\tau} + i_{n2} \qquad (35)$$

$$j = \frac{(\varphi_1 - \varphi_2) - 2\pi\phi_e}{\pi\beta}$$

$$v(\tau) = \frac{1}{2}\left[\frac{d(\varphi_1 + \varphi_2)}{d\tau}\right]$$

$P_{L/\xi}(\varphi)$ strongly depends on the ratio $L/\xi$ as shown in Fig. 14a. Again, the noise term is given by the Nyquist noise in the normal resistance of the SQUID.

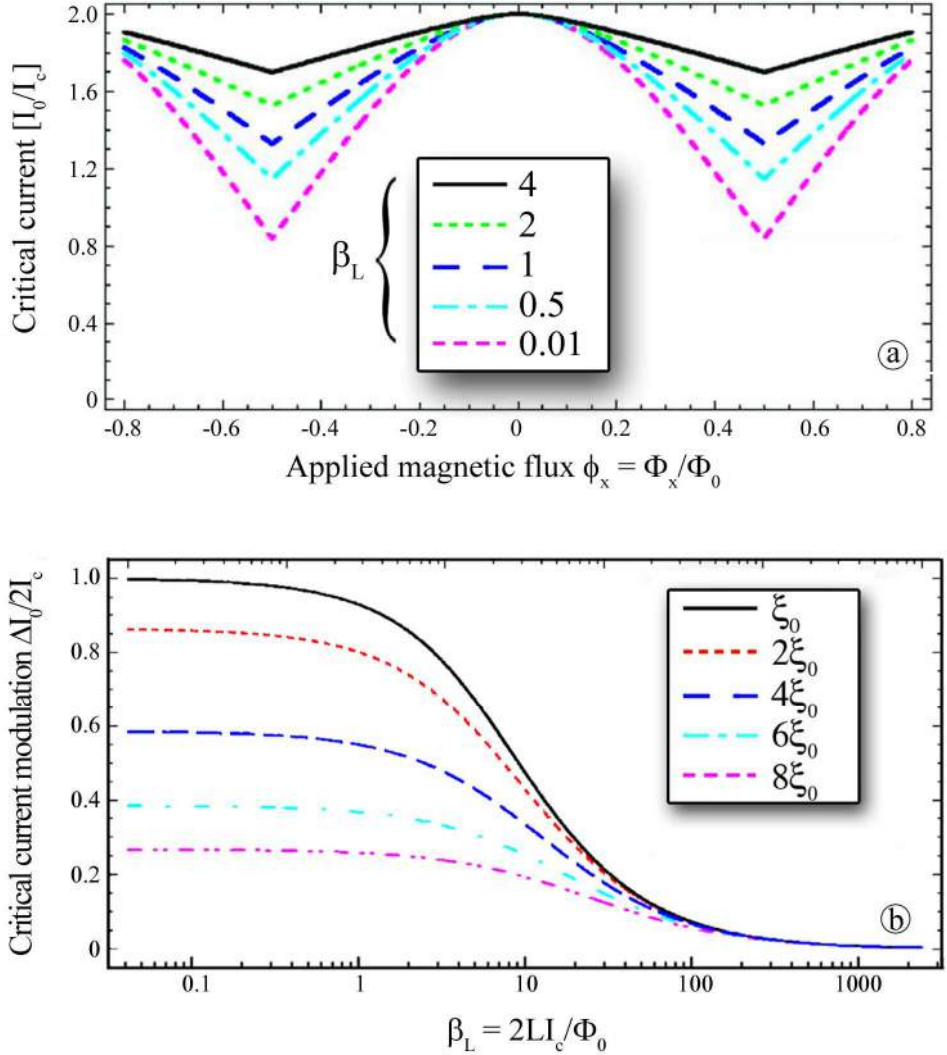

*Fig.15 a) Critical current as a function of the external magnetic flux for $L/\xi=4$ and different $\beta_L$ values. b) Critical current dependence on $\beta_L$ for different $L/\xi$ values (adapted from [126]).*

The static characteristics of the SQUID were studied by K. Hasselbach et al [124] and by Podd at al [125]. They numerically solved the equations (35) for V=0 (v=0, $i_{n1}=i_{n2}=0$) and found the critical current as a function of the external magnetic flux for different $L/\xi$ values.

The results of their simulations are reported in figures 15. The $I_c$-$\Phi_e$ for $L/\xi=4$ are reported in Fig. 15a for different $\beta_L$ values. The curves are triangular-like for high $\beta_L$ values and becomes smoother by decreasing the $\beta_L$ value. It is worth to note that, also for very low $\beta_L$ values, the critical current does not reaches the zero value as in a conventional SQUID. This is more evident in the figure 15b, showing the modulation depth $\Delta I_c/I_c$ as a function of $\beta_L$ value for different $L/\xi$ ratios. By decreasing the $L/\xi$ values, the modulation depth increases approaching the value 1 for $L/\xi=1$ corresponding to the standard sinusoidal case. Therefore, the highly non-sinusoidal CPR leads to both a reduced modulation depth and a triangular shape dependence of the critical current on the applied magnetic flux. Hasselbach et al. [124] studied also the temperature dependence of the critical current in both niobium and aluminium based microSQUIDs. In both case, they found that the increase of the $I_c$ by decreasing the temperature is appreciably greater than the sinusoidal CPR case. The theoretical predictions were in good agreement with the experimental data.

Granata et al. [126] solved numerically equation (35) for $V \neq 0$. In particular, their analysis was focused on the non-hysteretic regime that is $L/\xi$ ranged from 1 to 3.5. As concern as other parameters, since their analysis regarded nano-SQUIDs having very small loop inductance a reasonable value, $\beta_L = 0.05$, was assumed. It corresponds to a SQUID critical current less than 100 μA, which is a reasonable upper limit for the non-hysteretic nano-SQUIDs [132]. A noise parameter $\Gamma$ of *0.02* was chosen in correspondence to $L/\xi = 1$, and it was progressively increased up to *0.046* for $L/\xi = 3.5$, taking into account the decrease of the critical current for increasing $L/\xi$ ratios (see Fig. 14b). In such a way, it is possible to compare the device with different Dayem bridge lengths. In the computational scheme, the $P_{L/\xi}(\varphi)$ is an external function which is recalled to each computation step. In figure 16a, the voltage-flux characteristics corresponding to three different $L/\xi$ value are reported. The computation was performed considering an incremental step of the external magnetic flux of 0.005 $\Phi_0$. The two curve sets correspond to two bias current values. An evident decrease of the voltage amplitude is observed by increasing $L/\xi$ values. Figure 16b reports the $V_\Phi$ as a function of the bias current for different L/ξ ratios. They have been computed by evaluating the derivative of the *V-Φ* characteristics at *Φₑ=Φ₀/4*.

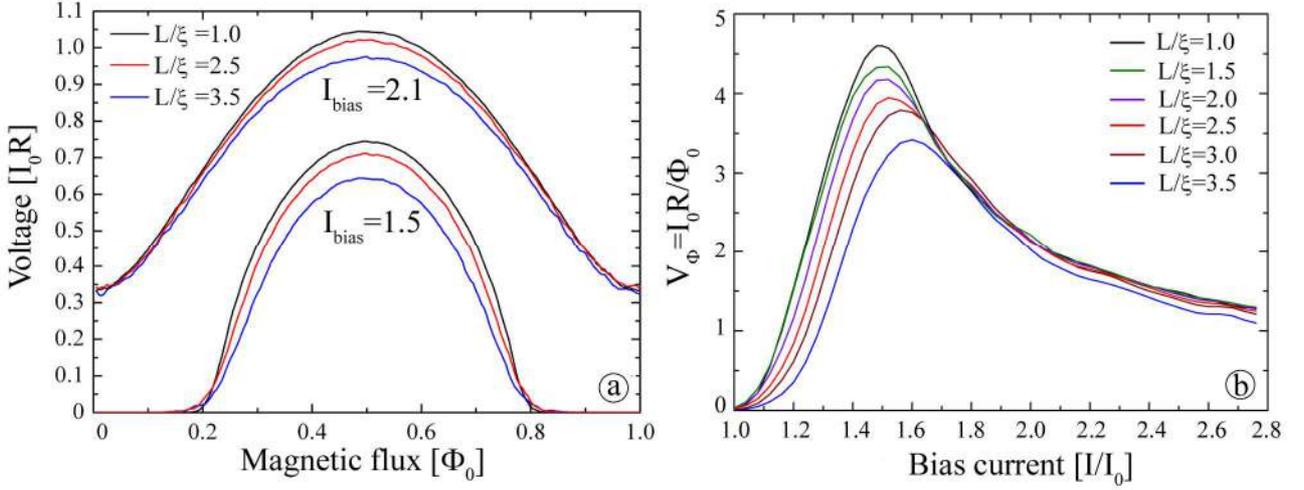

*Fig.16 a) Voltage vs. magnetic flux characteristics corresponding to three different L/ξ values, computed for two different values of bias current.* b) *Flux-to-voltage transfer factor $V_\Phi$ as a function of the bias current for different L/ξ ratios obtained by evaluating the derivative of the V-Φ characteristics at Φₑ= Φ₀ /4 (adapted from [126]).*

As well as the *V-Φ* curves, also the $V_\Phi$ values decrease by increasing the L/ξ ratio. Similarly to conventional SQUIDs, the $V_\Phi$ curves show a steep slope for low bias current, a maximum for a bias current lower than the critical current (about 75% of $I_c$), and a smooth slope for high bias current. It is also possible to observe a shift of the maxima toward high bias currents for increasing L/ξ ratios, due to the different critical current modulation depths. The spectral densities of the voltage white noise as a function of the bias current for *L/ξ* ratio ranging from *1* to *3.5* are reported in figure 17a. As for the $V_\Phi$ computation, also in this case, the external magnetic flux was set to a value of $\Phi_0/4$. As shown in Fig.17a, the noise increases by increasing the bridge length. Also in this case, it is possible to observe a shift of the maxima due to the same reason given above for the $V_\Phi$ curves. Figure 17b reports the power spectral density of the flux noise $S_\Phi$ as a function of the bias current for the same range of L/ξ values. It was obtained by taking the ratio between the voltage noise and the responsivity square at $\Phi_a=\Phi_0/4$. As expected, the noise increases by increasing the ratio L/ξ.

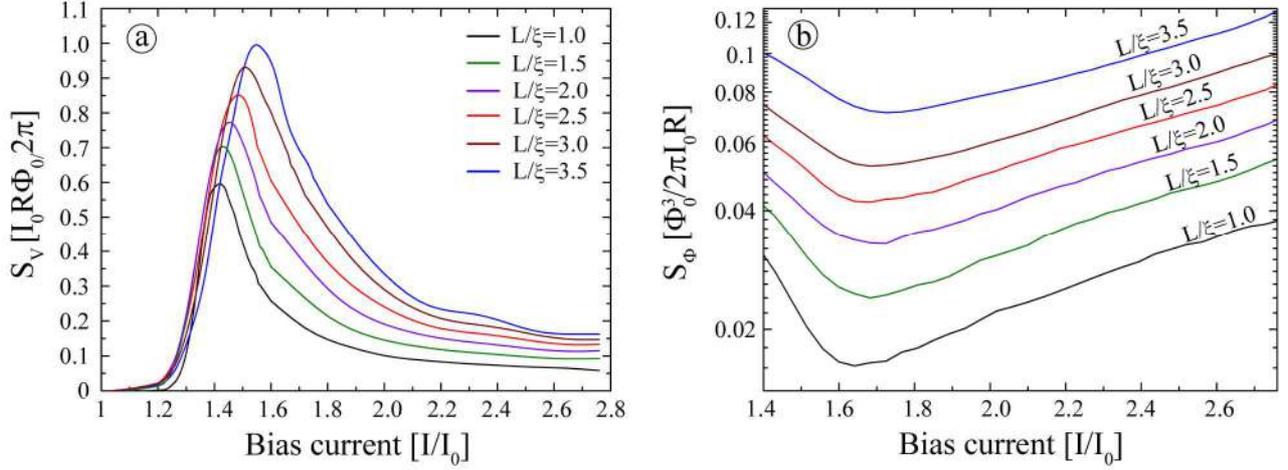

*Fig.17 a) Power spectral densities of the voltage noise as a function of the bias current. The $S_V$ values have been taken in the white region of the spectrum. b) Power spectral densities of the magnetic flux noise ($S_\Phi$) as a function of the bias. It has been obtained by computing the ratio between the voltage noise and the square of responsivity. For both figures, L/ξ ratio ranges from 1 to 3.5 and the external magnetic flux bias is $\Phi_e = \Phi_0/4$ (adapted from [126]).*

An empirical expression of the white flux noise was derived from the above simulations [126]:

$$S_\Phi = \frac{\Phi_0^3}{2\pi I_0 R}\left[ A(\Gamma)\left(\frac{L}{\xi}\right)^{4/3} + B(\Gamma) \right] \qquad (36)$$

where $A(\Gamma)$ and $B(\Gamma)$ are numerical coefficient which depend on $\Gamma$ value. It was obtained by fitting the behaviour of the flux noise as a function of L/ξ value. The predictions of the described noise theory were successfully employed to fit some experimental noises data found in the literature.

## 2.6 Flux coupling and spin sensitivity computation

As seen in the section 2.2, the transfer or coupling factor between the magnetic dipole and SQUID is needed to evaluate the spin sensitivity of the nanosensor. In section 2.2 we have considered the simple case of the magnetic dipole co-planar with the SQUID and lying in the loop centre. In this section, we will take into account a more general case where the position of the magnetic dipole can vary within the loop and the distance from the SQUID plane in not zero. This issue has been extensively studied [106-107, 110-111, 133-140] and useful results have been obtained. First, we will discuss about the filamentary model already mentioned in the section 2.2; namely, square [110] or circular [137] loop of the nanoSQUID is assumed as a dimensionless wire and the magnetic flux is obtained by calculating the line integral of the vector potential along the SQUID loop line. We will start to consider the square loop case; the results obtained in the case of circular loop are very similar. Let us consider a magnetic moment $m_z$ oriented along the z-direction and positioned in a generic position (x',y'z') (Fig.12), the Cartesian components of the magnetic vector potential A(**r**) at position **r**(x,y,z), are:

$$A_x = -\frac{\mu_0 m_z}{4\pi} \frac{(y-y')}{|\vec{r}-\vec{r}'|^3}; \qquad A_y = \frac{\mu_0 m_z}{4\pi} \frac{(x-x')}{|\vec{r}-\vec{r}'|^3} \tag{37}$$

$|\mathbf{r}\text{-}\mathbf{r'}| = [(x\text{-}x')^2+(y\text{-}y')^2+(z\text{-}z')^2]^{1/2}$. By executing the line integral (equation 25), taking into account the contributions of the four sides of the square loop, we obtain the magnetic flux threading the loop produced by a magnetic moment $m_z$ as a function of its position within the loop:

$$\Phi_\mu(x',y',z') = \frac{\mu_0 m_z}{4\pi} \left\{ -\int_{L/2}^{-L/2} dx \frac{(L/2-y')}{\left[(x-x')^2+(L/2-y')^2+z'^2\right]^{3/2}} - \int_{L/2}^{-L/2} dy \frac{(L/2+x')}{\left[(L/2+x')^2+(y-y')^2+z'^2\right]^{3/2}} \right.$$

$$\left. + \int_{-L/2}^{L/2} dx \frac{(L/2+y')}{\left[(x-x')^2+(L/2+y')^2+z'^2\right]^{3/2}} + \int_{-L/2}^{L/2} dy \frac{(L/2-x')}{\left[(L/2-x')^2+(y-y')^2+z'^2\right]^{3/2}} \right\}$$

(38)

Note that for $x'=y'=z'=0$, the equation (38) reduces to equation (26).

In the Fig.18, the normalized magnetic flux distribution valued over the SQUID loop and relative to a magnetic moment $m_z$ at height $z'=0.05\,L$, is reported. Since the z' value is much smaller than the loop side length $L$, the magnetic particle can be reasonably considered in the plane of the loop and no field divergence arises in correspondence of the loop edges. As shown in the figure, the magnetic flux is strongly dependent on the spin position within the loop; in particular, the higher values are obtained when the spin is close to the sensor edges reaching the maximum values in the corners of the ring. In order to have an estimation of the flux coupling as a function of its position within the nanoSQUID sensitive area for different distances from the SQUID plane, the normalized magnetic flux coupling over the diagonal of the square loop for different z' values is reported in Fig.19a

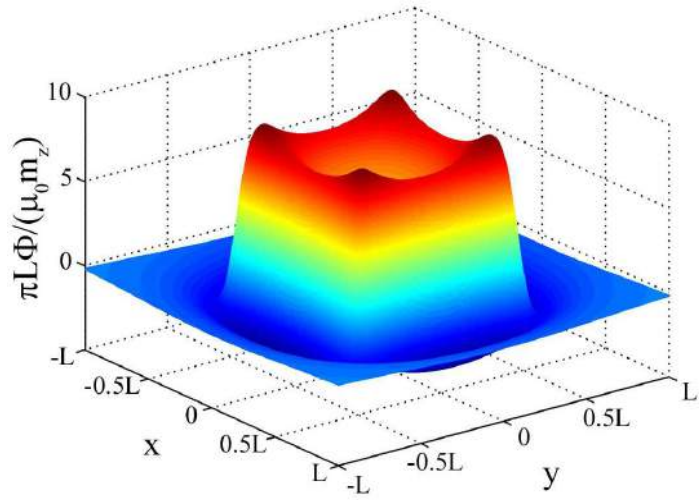

*Fig.18 Normalized magnetic flux distribution threading the sensor square loop produced by a magnetic moment $m_z$ oriented along the z-axis (arranged as in fig.12), as a function of its position within the sensor capture area for z'=0.05 L.*

From a practical point of view, a distance *z'* ranging from 10 nm to 40 nm takes reasonably into account the thickness of both passivation layer and the nano-objects size. Figure 19b reports the spin noise as function of the spin position along the loop diagonal for different z' values, computed by using the equation (27) and assuming a magnetic flux noise of 1 $\mu\Phi_0/Hz^{1/2}$. As shown in the figure, there is an appreciable enhancement of the sensitivity near the perimeter of the loop. For a large *z'/L* values, the magnetic flux coupling and spin noise tend to a constant value, indicating that they no longer depend on the spin position as also shown in simulations of scanning SQUID microscopy [66, 133, 141-143].

Similar calculations, for square loop having an area ranging from (400x400) nm$^2$ to (25x25) nm$^2$, were made by Volarick and Lam [144]. They considered a magnetic moment of 300 $\mu_B$ and a distance above the nanoSQUID plane in the range (0-100) nm.

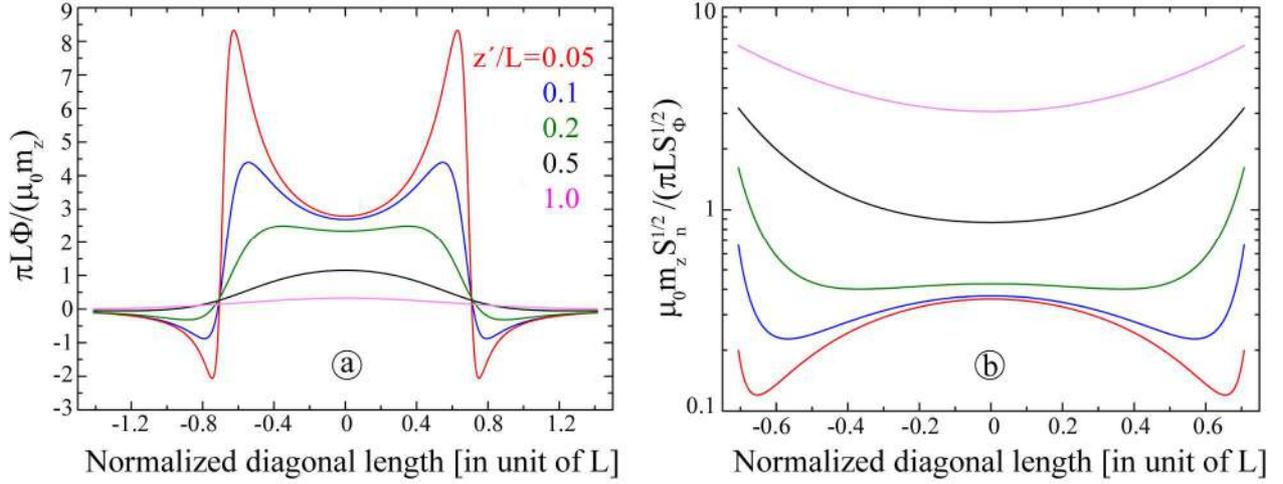

*Fig.19 Normalized magnetic flux coupling (a) and normalized spin noise spectral density $S_n^{1/2}$ (b) as a function of the position within the loop computed along a diagonal of the nanoSQUID loop for different z'/L values.*

Tilbrok performed the calculations in the case of a circular loop [137] following the same procedure reported above but using spherical coordinates. The total flux due to a magnetic moment oriented along the z-axis, is given by:

$$\Phi_\mu = \frac{\mu_0}{\pi} \frac{a}{(z'^2+r^2+a^2)^{3/2}} \frac{1}{(1+k)\sqrt{1-k}} \left\{ \frac{r\,m_z}{k}\left[(1+k)K\left(\frac{2k}{k-1}\right) - E\left(\frac{2k}{k-1}\right)\right] + a\,E\left(\frac{2k}{k-1}\right)m_z \right\}$$

(39)

Where *a* is the loop radius, *r* and *z'* are the x-ordinate and the height of the dipole above the x-y plane respectively, $k = 2ar/(z'^2+r^2+a^2)$ and finally, $K[2k/(k-1)]$, $E[2k/(k-1)]$ are elliptic integrals of the first and second kinds respectively. For $r=0$ (dipole on the axis of symmetry), $k=0$ and the (39) assumes the simplified form:

$$\Phi_\mu = \frac{a^2 \pi_0 m_z}{2(a^2+z'^2)^{3/2}} \quad (40)$$

Therefore, for any given height *z'*, there is an optimum nanoSQUID loop radius $a = z'\sqrt{2}$ that maximize the (40). This is useful information because in many situations the height will be constrained by experimental limitations. He considered also the case of a magnetic moment oriented in the x direction finding that the magnetic coupling is very low, except for magnetic moment near the perimeter of the loop.

It is worth to stress that the filamentary model used in the above calculation is based on equation (27) that involves a line integral over a loop, which is a closed mathematical line of zero thickness. On the other hand, the SQUID is a physical device with a finite thickness, so, more detailed considerations about the magnetic field-SQUID interaction should be taken into account. Therefore, these simulations are mainly suitable for very thin film nanoSQUIDs.

Moreover, the filamentary model does not consider flux focusing due to finite width of the superconducting tracks [110,145]. For a circular loop, such increasing can be approximately evaluated as $A_{eff}=A_g\ [(1+\Lambda(T)/a)]^2$ where $\Lambda(T)=\lambda^2(T)/d$ is the two-dimensional effective penetration depths [146], $\lambda(T)$ is the London penetration depth at the temperature $T$, $d$ is the superconducting film thickness and $A_g$ is the geometrical area of the SQUID loop. With typical values of loop radius and film thickness employed in nano-SQUID fabrication, the correction factor may not be negligible. However, this effect does not appreciably modify the behaviour of the flux lines within the sensor loop (except close to the sensor edges), so we can consider the area of the sensor under investigation as an effective area, which includes the flux capture area increasing.

The coupling factor calculation of Bouchiat [138] was based on a different approach: the Lorentz reciprocity theorem [147]. It stands for electromagnetism in a linear isotropic medium: the sources and the created fields can be interchanged. In this case, it is equivalent to consider the case of evaluation of the magnetic field created by the SQUID loop at the dipole position **r** instead of calculation of the flux generated by the dipole and threaded by the loop.

Bouchiat [138] explored in a quantitatively way also other coupling geometries, including the case where the assembly of spins sits directly on the nanobridge (Fig. 20), as already proposed by Wernsdorfer et al some years before [107]. His simulations show that the greatest coupling efficiency is obtained when a nanobridge acts as active detecting element. In this case (near field regime, z<<L) the sensitive part is no longer the loop but the weak link itself and the original Ketchen's formula (28) has to be replaced by:

$$S_n^{1/2} = \frac{S_\Phi^{1/2}}{\Phi_\mu} = \frac{S_\Phi^{1/2} 4 r}{\alpha \mu_0 \mu_B} \qquad (41)$$

Where $r^2$ is the section of the nanobridge and $\alpha$ is a geometrical factor which depends on the nanobridge size and the distance from it. For r = 25 nm (nanobridge) and r = 1 nm (nanotube), the geometrical factor $\alpha$ assumes the values of 0.06 and 0.03 respectively (Fig. 20c) [138]. In the ideal case of a magnetic dipole in direct contact to a very small nanobridge section (r →0), $\alpha$ tends to 1/2. It corresponds to the case where half of the magnetic flux lines are threaded by the nanobridge (Fig 20b).

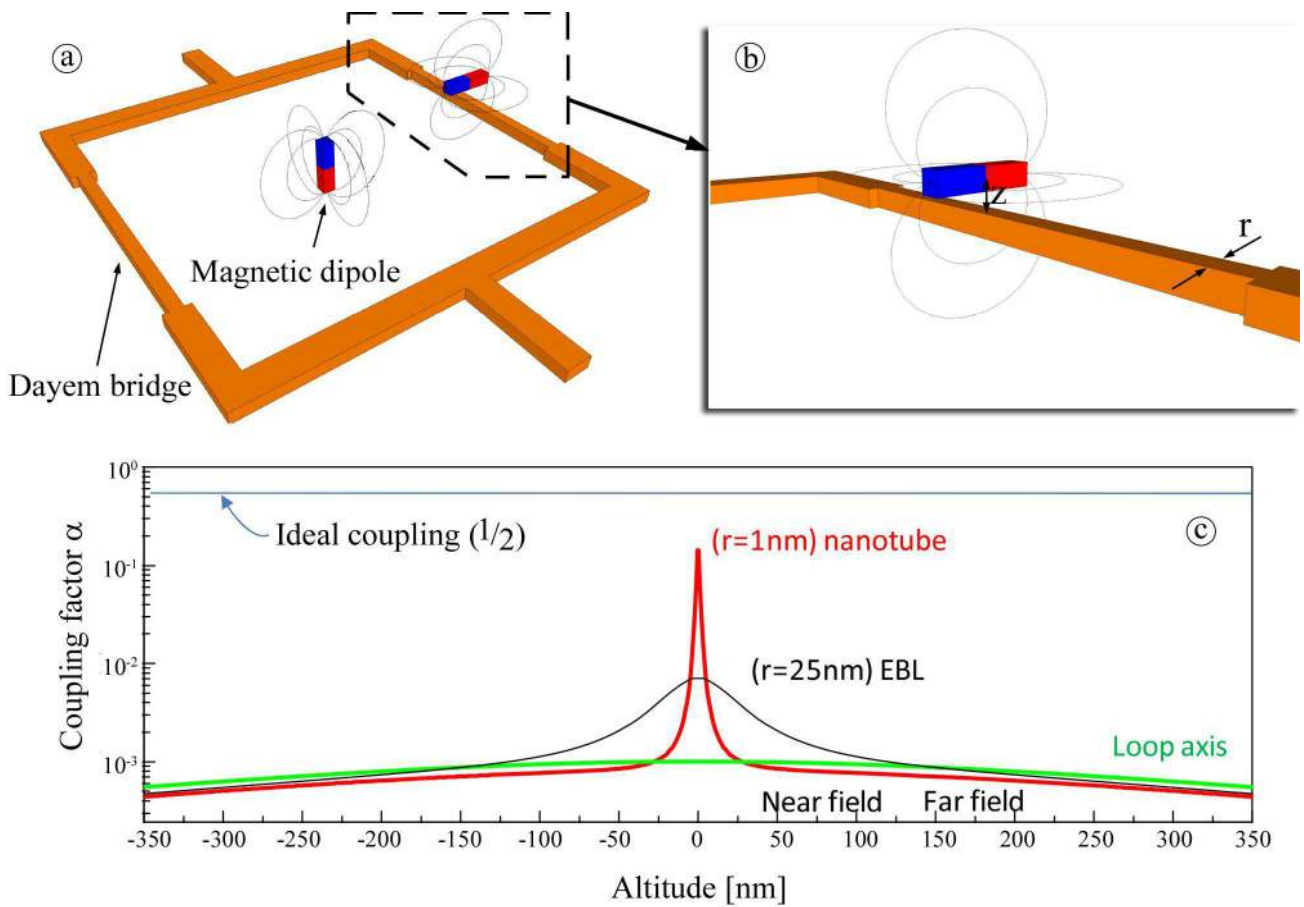

*Fig.20 a) and b) Scheme of two different way to couple a magnetic moment to a SQUID: in the centre of nanoSQUID (standard Ketchen's scheme) or directly on the top of nanobridges. c) Coupling factor $\alpha$ as a function of the distance z' from the SQUID plane for a magnetic moment $m_z$*

*traveling along a trajectory tangent to the nanobridge. The red and the black curve are relative to two different nanobridge cross section. The green curve refers to a moment $m_z$ traveling along the z symmetry axis (adapted from [138]).*

Starting from energetic considerations, Nagel et at [139] reached the same Bouchiat's results. They showed that the calculation of the magnetic flux per magnetic moment ($\Phi_\mu/\mu$) coupled into the SQUID loop is equivalent to calculate the three-dimensional magnetic field distribution produced by the SQUID loop treaded by an arbitrary current. Using an upgraded version of such calculation method, a detailed study of spin sensitivity for high critical temperature nanoSQUIDs based on YBCO grain boundary junctions was recently carried out by Wolbing et al. [140]. They calculated both the magnetic flux noise and the coupling factor and derived an explicit expression for the spin sensitivity as a function of the geometrical and electrical parameters of their nanodevices. They showed that a spin sensitivity of few $\mu_B/Hz^{1/2}$ could be reached if the nanoparticle was placed 10 nm above the centre of the nanoconstriction with its magnetic moment perpendicular to the grain boundary and oriented in the plane of the SQUID loop.

Based on the above results, we can say that a suitable spin sensitivity can be obtained by either decreasing the SQUID loop or using constrictions as small as possible. As will be shown in the next section, both criteria has taken into consideration to develop high sensitive nanoSQUID.

**3. NanoSQUIDs: fabrication and performance**

In this section, we will describe the main nanoSQUIDs based on nanobridges reporting the design criteria, the main characteristic and the performance. We will also describe other nanoSQUID types stressing the advantages and the disadvantage with respect to nanobridge SQUIDs. In particular, we will deal with nanoSQUID based on: sandwich type nanojunctions, carbon nanotubes, high critical temperature (YBCO) and $MgB_2$ superconductors. In addition, we will describe nanoSQUID built on an apex of nanometric tip and those based on graphene and diamond materials.

## 3.1 Nanofabrication techniques

There are two different approaches to realize nanosized objects. The first is to scale down a bulk material (top-down approach) the second is to assemble the nanoscale elements. Mainly, a nanoSQUID is fabricated by using the first approach exploiting the abilities of the Electron Beam Lithography (EBL) [149] or the Focused Ion Beam (FIB) [150] sculpting to realize nanostructures. The aforementioned techniques are used together with thin film deposition process, optical photolithography and other techniques usually used in standard fabrication of micro device. However, these aspects will be no addressed in the following, limiting the analysis to the fundamental aspects of EBL and FIB techniques. However, even if most of nanoSQUIDs are fabricated by using the above-cited techniques, it is worth to mention that a technique based on Atomic Force Microscopy (AFM) has been successfully employed to fabricate SQUID based on nanobridges [151, 152]. This technique involves local anodization of superconductor films under the voltage-biased tip of the AFM.

### 3.1.1 Electron Beam Lithography (EBL)

The Electron Beam Lithography (EBL) is a technique allowing to create pattern having a resolution down to 10-20 nm or less [153] by using a beam of electrons to define a nanometric structure in an electron sensitive resist deposited on a sample. The exposed pattern becomes soluble in a specific removal leaving the sample uncovered along the same pattern. The purpose, as with photolithography, is to create very small structures in the resist that subsequently can be transferred to the substrate material, by lift-off process or etching.

The electron beam is generated in the column by an electron source and focused by two or more lenses. Under the column, a vacuum chamber houses high-resolution mechanical stages to move the sample under the beam and to expose step-by-step the substrates under the relatively narrow field of focus of the electron beam.

The emission of electrons can be made by thermionic way but for higher resolution beam a field emission sources is required. The size of the source determines the amount of demagnification that has to be applied to have small spot at the target. The electrons beam are focused by using electrostatic force to obtain the convergence of the beam. However, electrostatic lenses usually produce higher aberrations, therefore magnetic lens are preferred to focus the beam. The spherical and chromatic aberrations can be minimized by reducing the convergence angle with the consequent confinement of the electrons into the centre of the lenses even if it leads to a reduction of beam current.

By using a magnetic lens, the divergence of the magnetic flux can be exploited in the focusing action causing a rotation of the electrons (and the image) along Z-axis that does not affect the performance of the lens, but the design, alignment and operation of the system.

Other optical elements include aperture, deflection systems, alignments coils, blanking plates and stigmators. Deflection of the electron beam is used to scan the beam across the surface of the sample. Typically, for very small beam deflections electrostatic lenses are used, larger beam deflections require electromagnetic scanning. Deflecting the beam off axis introduces additional aberrations like the spot shape distortion and deviations from linearity. These effects are less evident if a magnetic deflection is used. A stigmator is used to balance the imperfections in the creation and alignment of the electron beam.

Typically, the writing field is of the order of 100 μm – 1 mm while for larger patterns a movable stage is required. In this case, the accuracy of the stage is a critical point in order to avoid stitching phenomena and to realize exact overlaying in aligning different patterns.

When the beam interacts with the resist, some electrons can scatter along small angles leading to broaden the beam diameter. During the penetration toward the substrate, the electrons can scatter by a large angle (backscattering) causing the proximity effect in which the specific pattern area can be influenced by the electrons scattered in the close areas. In the meantime, the electrons are slowed down progressively, producing a cascade of low voltage electrons (secondary electrons). Forward scattering can be reduced by using a thin film of resist and the high accelerating voltage. The secondary electrons, having a low energy, contribute, as the forward scattering, to enlarge the beam diameter resulting in a loss of resolution with respect to the expected one. Some secondary electrons can preserve higher energy and contribute to the proximity effect. Note that there are several techniques to fix the proximity effect like, just to cite, the dose modulation, pattern biasing, GHOST or software methods.

### 3.1.2 Focused Ion Beam (FIB)

With respect to EBL technique, the Focused Ion Beam (FIB) uses a beam of ions instead electrons to induce a solubility change along a nanometric pattern of a resist distributed on a sample [150]. More frequently, the FIB is used as a sculpting technique exploiting the milling process due to the collisions of the ions with atoms of the structure, thus avoiding the use of any kind of resist. In fact, since the mass of the atom is comparable with those of the incident ion, a large amount of ion energy is transferred to the atom that acquire enough energy and speed to leave the lattice sites. Furthermore, the FIB can be used also for ion beam-induced deposition and chemical reaction of surface species. Since the Ion Beam Lithography (IBL) is very similar to EBL technique, in the following, only the sculpting technique will be briefly addressed, leaving aside the deposition and chemical reaction techniques.

The obtainable spatial resolution is of the order of the beam diameter, typically of about one hundred of nanometres. The milling rate increases as the beam current increase, while for a more precise milling, a small beam spot is required that is obtained using a small beam current. In addition, the rate depends on the mass of the atom and its binding energy to the lattice.

Similarly to the EBL, a suitable source generates the ion beam that is then focused in the optics column. Such a source can be gaseous, in which the gas molecules irradiated by an electron beam becomes ionized. In such a way, the ions are emitted from a relatively large area. For a point source and a limited angle of emission, a field ionization source has to be used. In this case, the sources operate by desorption of ions from a sharp tip in a strong electric field. A high efficiency can be reached by using liquid metal ion source in which a thin needle or a capillary is wetted by a thin film of liquid source metal, which has been heated to the liquid state. Typically, the gallium is employed due to its low melting temperature and high brightness of obtained ion beam. Moreover, the Gallium has a mass large enough to sputter a wide range of materials.

The focusing of ion beam is obtained by using electrostatic lens consisting of electrodes operated at high potentials. The electric fields generated by these lenses are used to deflect and accelerate the ion beam whose diameter is influenced by the chromatic aberration, due to the energy spread between the emitted ions. The same diameter, although to a lesser extent, is influenced also by the spherical aberration due to the different path length of the ions from the source to the lens. In particular, the beam diameter results larger than the expected one.

Despite the beam diameter is of the order of one hundreds of nanometres, exploiting the non-homogeneity of the ion beam and overlapping the beam profiles, structures having dimensions less than the beam diameter can be realized.

The collisions between ions and atoms of the substrate or resist generate both forward and backward scattering that are responsible of beam broadening leading to a loss of resolution in the resist exposure. Furthermore, the collisions cause a trajectory deflection displacing the atom from the lattice sites. It leads to a formation of an amorphous layer or causes the implantation of the primary ions in the interstitial lattice. This is an important issue during the milling usage of FIB because the damaged layer may have different mechanical or electric properties that have to be taken into account in the realized nanostructures. As example, in the case of superconductors, the damaged layer can lose the superconducting state.

Nevertheless, the absence of any kind of resist and relative lift-off or etching processes together with the reducing of number of fabrication steps represents an important advantage with respect to EBL technique, especially if the effects of damaged layer can be neglected or eliminated, for instance, by an anodization process.

**3.2 NanoSQUID based on Dayem nanobridges**

In this section, most of nanoSQUID types fabricated by EBL or FIB and relative performance are discussed, following the same chronological order of their publishing.

As seen in the section 2.1, the first SQUID having an effective flux capture area much smaller than 1 μm$^2$ (Fig.21), was developed by Lam and Tilbrook [108]. By using EBL technique, they fabricated a niobium based nanoSQUID having an hole size of 200 nm and nanobridges with a width and a length in the range of 70-200 nm. The nanoSQUID was fabricated starting from a bilayer of Niobium/Gold having a thickness of 20 and 25 nm respectively.

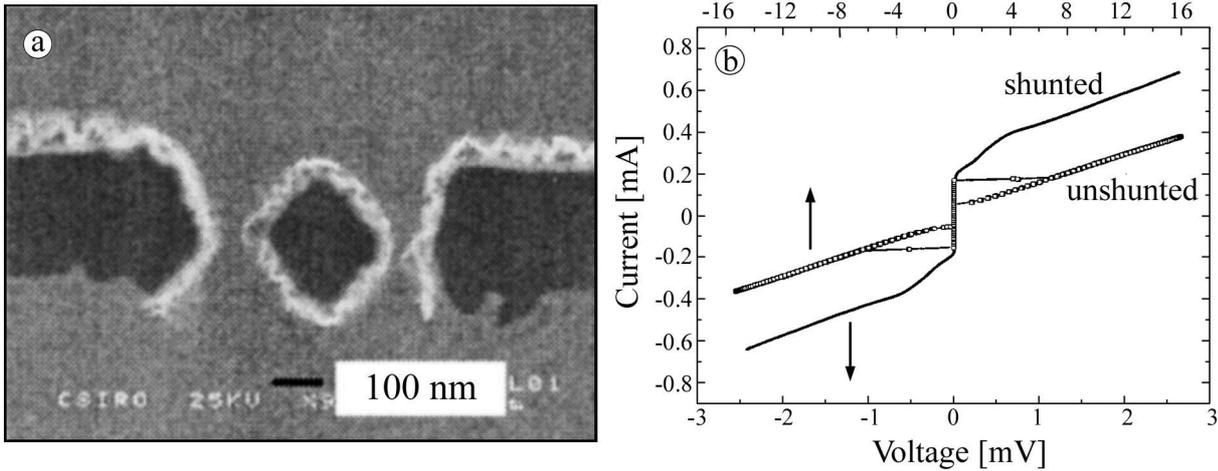

*Fig. 21 a) SEM image of First nanoSQUID having an effective flux capture area much smaller than 1 μm$^2$, b) Current-voltage characteristics measured at T=4.2 K with and without gold thermal shunt (adapted from [108]).*

The Au layer was used as both a mask for etching the Nb film and as a resistive shunt in the completed devices. The nano-SQUID had a washer shape in order to improve the heat dissipation occurring when operated in resistive mode. Two Dayem nano-bridges constituted the lateral arms of the washer, so it was different with respect to the typical washer loop and, therefore, the flux focusing effect [86] was less effective. Dayem bridges had a hyperbolic shape; therefore, the effective length can be significantly larger than the geometrical one. Fig. 21b shows the I-V characteristics with and without the gold layer, showing that the hysteresis disappears in the presence of shunt. The measured magnetic flux noise was $7\times10^{-6}$ μ$\Phi_0$/Hz$^{1/2}$ corresponding to a spin sensitivity of 250 μ$_B$/Hz$^{1/2}$ calculated by using the equation (28). In a later article [153], a slight noise improvement on the same nanoSQUID type was obtained. A magnetic flux noise of $5\times10^{-6}$ $\mu\Phi_0/Hz^{1/2}$ was reported and there was no significant increase in the noise level by applying a static magnetic field up to 2 mT.

A year later, Troeman et al [154], fabricated niobium nanoSQUIDs having an effective area ranging from of $3.6\times10^{-2}$ $\mu m^2$ to 900 $\mu m^2$ by using FIB milling technique (Fig. 22a). They obtained critical currents of 4-25 $\mu A$ and flux sensitivities of 40-200 $\mu V/\Phi_0$ (Fig. 22 b and c) for sensors based on 80 nm wide, 50 nm thick, and 150 nm long nano-bridges. The inductance of the smallest device, due mainly to the kinetic inductance of the bridges, was 150 pH. It was the first nanoSQUID fabricated by FIB technique and even though, the resolution of the used FIB system was modest, the small sizes of the realized device and its quality proved the potential of the technique. As mentioned in the previous section, the gallium implantation during the FIB process leads to a suppression of the superconducting properties. However, as shown by Troeman et al., the damage of the superconductive materials in not crucial if at least a 50 nm thick niobium is employed. At same time, the slight damage leads to a non-hysteretic device. The reported magnetic flux noise was 1.5 $\mu\Phi_0/Hz^{1/2}$ for a device with an area of 900 $\mu m^2$.

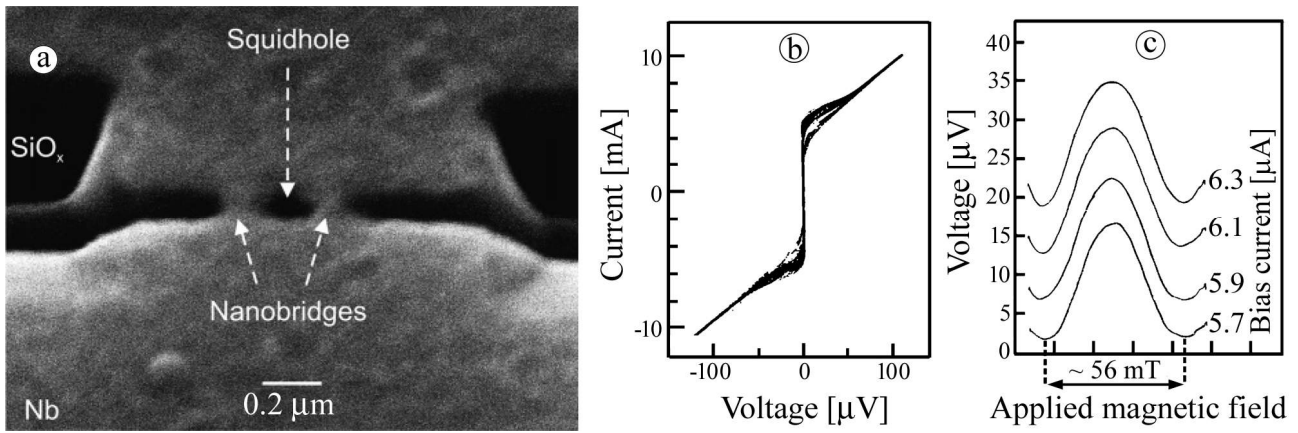

*Fig. 22 a) Picture of niobium based nanoSQUID having an effective area of 0.036 $\mu m^2$, fabricated by focused ion beam technique. b) Current-voltage characteristics measured by applying different values of magnetic field. c) Voltage – magnetic field curves corresponding to different bias current values. The measurements were performed at T=4.2 K (adapted from [154]).*

Combining conventional optical lithography and ion etching with a subsequent FIB milling step, Hao et al. [155-157], Romans et al. [158], Rozhko et al. [159] and recently, Patel et al. [160] fabricated ultra low noise nanoSQUIDs based on a Niobium/Tungsten (*Nb/W*) bilayer. The design and fabrication parameters employed were: loop diameter ranging from 200 nm to 800 nm, niobium thickness of 70-200 nm, tungsten thickness of 150 nm and nanobribridge size of 65-80 nm in width and 60-80 nm in length. A SEM image of one of their nanoSQUIDs is reported in the Fig. 22a [155]. As in the Lam's device, the *W* layer acts as both thermal shunt and resistive shunt eliminating the hysteresis in the I-V characteristic (Fig. 22b) and provides, in addition, protection against the ion beam damages.

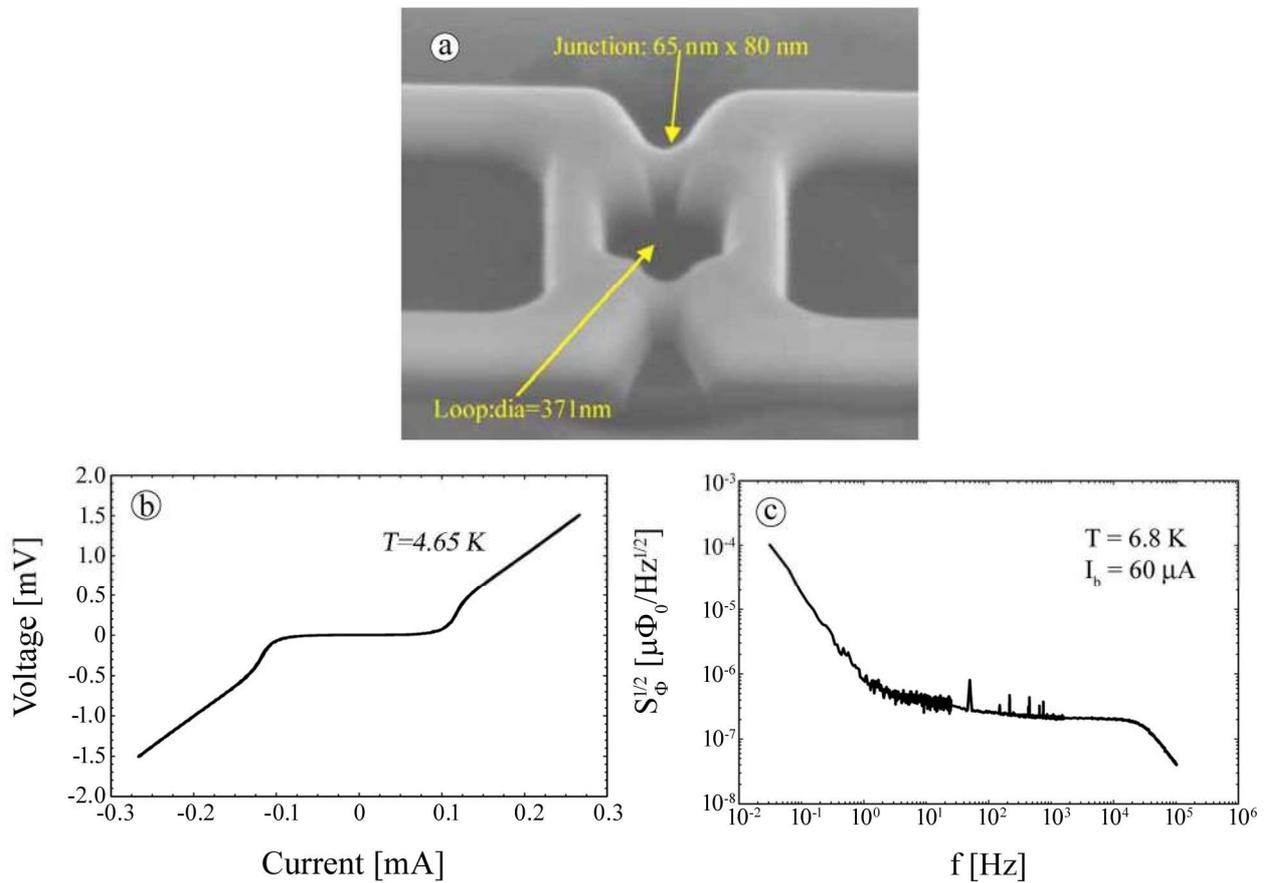

*Fig. 23 a) Scanning electron micrograph of a Dayem bridge based nanoSQUID made of niobium/tungsten bilayer. b) Current-voltage characteristic of the nanoSQUID measured at T=4.65 K. c) Magnetic flux noise spectral density measured at T=6.8 K (adapted from [155]).*

With respect to *Au* film, the *W* film exhibits a high resistivity and it is temperature independent. At T=6.8 K, the nanosensor showed a magnetic flux noise as low as 0.2 $\mu\Phi_0/Hz^{1/2}$ (Fig. 22c). It was measured in a flux-locked loop mode using a 16 SQUID series array [44, 161,162] to pre-amplify the nanoSQUID output signal.

NanoSQUIDs have a flux capture area less than 1 $\mu m^2$, so high magnetic fields (10-50 mT) are needed to measure the voltage–flux characteristics and to tune the device at its optimal working point. Typically, large coils wound outside the tail of the Dewar and permanent magnets are used to provide the magnetic fields for the device calibration and to identify the flux state of the sensor. Nevertheless, the magnetic field values provided by a standard set-up sometimes cannot be larger enough to fulfil the characterization requirements of the device, due to the occurrence of Joule heating of the coils. In order to solve this problem, Granata et al [163] developed an integrated nanoSQUIDs including two micrometric integrated niobium coils located very close to the nanosensor (Fig. 24a) in order to modulate, tune and operate the SQUID in flux locked loop configuration. With respect to external coils, the higher mutual inductance allows the effective measurement of the voltage-flux characteristics as well as a suitable flux biasing of the SQUID to obtain the maximum value of the flux to voltage transfer factor $V_\Phi$.

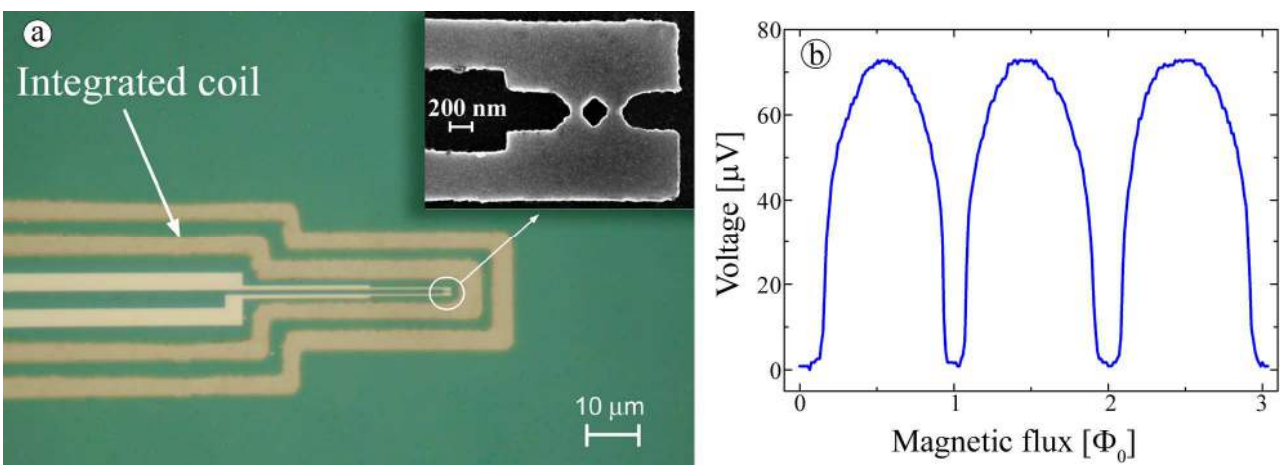

*Fig. 24 a) Picture of the nanoSQUID device including the integrated coils for the sensor modulation and tuning. In the inset, a SEM image of the SQUID loop and nanobridges are shown*

*(Adapted from [163]). b) Voltage-magnetic flux characteristic measured at T=4.2K (adapted from [164]).*

The device design is similar to that reported in ref. [108] and includes a niobium loop with a hole of *200 nm* interrupted by two bridges having a length and a width of 100 nm and 80 nm respectively. They employed a bilayer *Nb(30 nm)/Al (30 nm)* patterned by EBL and shaped by lift-off process and reactive ion etching (RIE) which is a selective chemical dry etch. Also in this case, the Al layer acts as both a resistive and thermal shunt and a self-aligned mask for the RIE process. They obtained non-hysteretic characteristic at T=4.2 K, with a critical current of about *50 μA* and a modulation depth $\Delta I_c/I_c$ as high as 35%. Thanks to the integrated coils, they measured up to three-voltage oscillation as a function of the external magnetic flux (Fig. 24b). Due to high voltage responsivity (about *1.5 mV/Φ_0*), a magnetic flux noise of *1.5 μΦ_0/Hz^{1/2}* was obtained by using a standard low noise direct-coupled amplifier. The corresponding spin noise evaluated at the centre of the loop was of *60 μ_B/Hz^{1/2}* [164].

A higher critical current modulation depth was obtained by Vijay et al [165] by using an aluminium nanoSQUID with three-dimensional contact banks (3D nanobridge SQUID). In such a configuration, nanoSQUID was contacted with banks ticker than the bridges (Fig.25b).

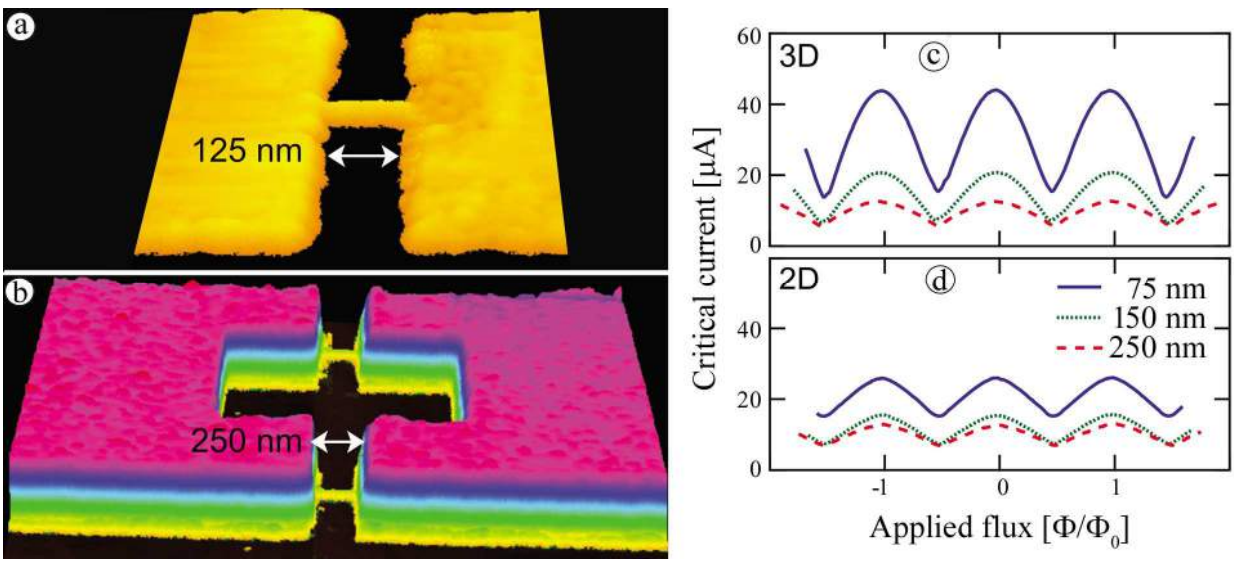

*Fig. 25 a, b) Atomic force micrograph of aluminium nanobridges between two contact pads having a 2D or 3D structure. c, d) Critical current as a function of applied magnetic flux relative to the*

*nanoSQUIDs in 2D and in 3D structure respectively, measured at about 300 mK (adapted from [165]).*

The three-dimensional banks enhance nonlinearity by acting as good phase reservoirs and confining the variation of the superconducting order parameter to the weak link [133] (see section 2.4.1). So, in the case of short nanobridge ($L<\xi$) the magnetic patterns can be described in term of the analytical results of KO-1 theory [120, 128] for a diffusive weak link at zero temperature with a phase variation constrained to the bridge region. They compared such a design with a standard one where the contacted banks have the same thickness of the bridge (Fig.25a). The SQUID loop had an area of *1.5 μm²* and a geometric inductance of *L =5 pH*. The bridges were *8 nm* thick, *30 nm* wide and had a variable length ranging from *75 to 400 nm*. At about *300 mK,* the nanoSQUID, fabricated by using EBL and lift-off process, exhibited a modulation depth as high as 70% approaching to an ideal weak links behaviour (Fig.25c). As observed in the figure, the $I_c$-$\Phi$ characteristics of nanoSQUID with three-dimensional banks is deeper and has more non-linearity with respect the two dimensional device which exhibits a quite linear $I_c$-$\Phi$ pattern (Fig 25d). It is indicative of a non-linear and almost linear CPR for 3D and 2D device respectively. Note that, the presence of thick banks could prevent the operation of such device in large parallel magnetic field, since the critical magnetic field decrease by increasing the superconducting film thickness. However, Antler et al [166] demonstrate the successful operation of a similar nanoSQUID in an applied in-plane magnetic fields up to 60 mT. Since this nanodevice shows hysteretic I-V, it cannot be employed as a conventional magnetic flux to voltage transducer. Nevertheless, an ultra low magnetic flux noise (17 n$\Phi_0$/Hz$^{1/2}$) in the white region was obtained by operating the SQUID in the dispersive mode [166,167]. The nanoSQUID was shunted by one chip capacitor to realize a flux tunable resonator (Fig. 26).

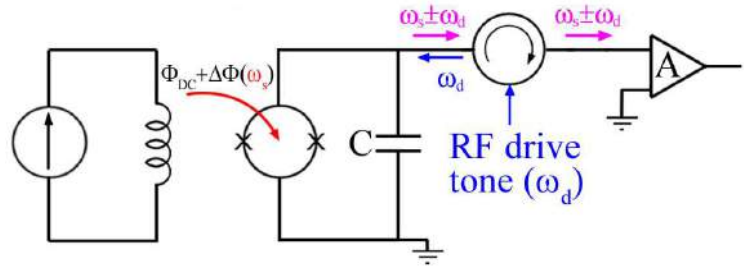

*Fig. 26 Scheme of the measurement circuit including the SQUID and a circulator with a RF drive tone input, employed by Adler et al [166] to operate the nanoSQUID in a dispersive mode [167].*

The input magnetic flux signal to measure induces a change in resonant frequency, which is read out by microwave reflectometry [168]. Such a configuration allows a high bandwidth (about 100 MHz) while avoiding the dissipation associated with conventional nanoSQUID devices as typically operated.

Most of nanomagnetism measurements require high magnetic fields (up to several T). Therefore, a nanoSQUID for nanomagnetism applications should be able to operate in very high magnetic field. The use of very thin superconducting film and a precise alignment system allows the use of magnetic field large enough to perform nanoparticle magnetization measurements. However, the vortex penetration at magnetic field much lower than the critical magnetic field of superconductors, causes additional flux coupling into the SQUID and so a device performance degradation. In order to overcome these problems, Lam et al. [169] fabricated a nanoSQUID with a very small capture area ($70 \times 70\ nm^2$). In their design, the washer was removed in order to eliminate vortex penetration and trapping in the film near the nanoSQUID hole (Fig.27a). As shown in figure, the nanoSQUID was obtained by making a small hole in superconducting submicron track having a width of *250 nm*. The width of nanobridge was approximately *90 nm*, while the thicknesses of bilayer *Nb/Au* films were *20* and *25 nm* respectively. Measurements of *I-V* and *I-Φ* characteristic were performed in magnetic field applied both perpendicularly and parallel to the SQUID plane. They found that, for a perpendicular magnetic field up to *150 mT*, there is no flux shift periodicity indicating the absence of the magnetic flux entry into the SQUID (Fig 27b).

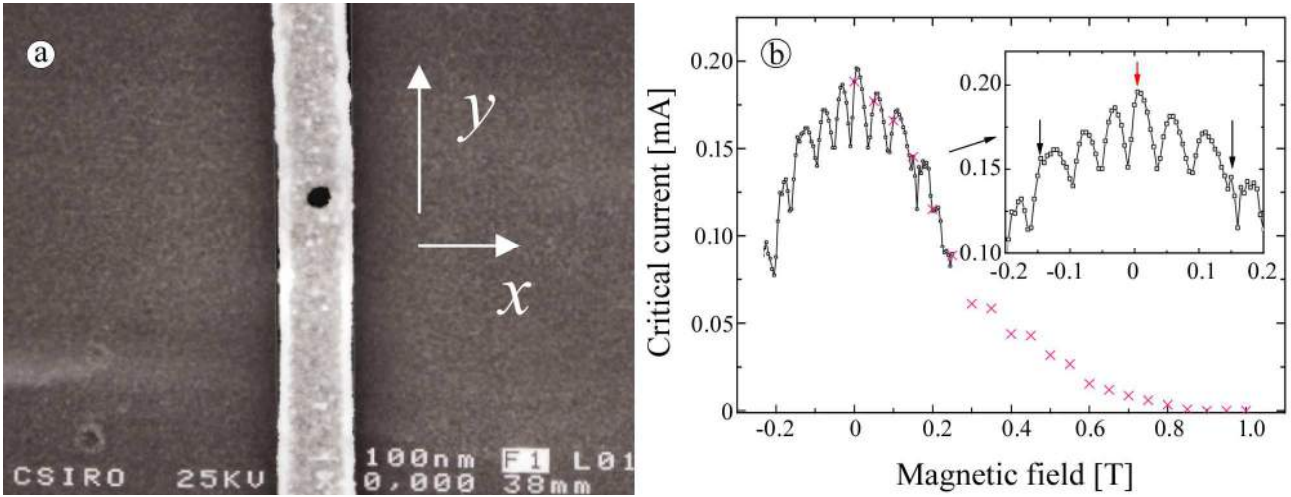

*Fig. 27 a) Scanning electron micrograph of the niobium based nanoSQUID coated with a gold layer. b) Critical current as a function of magnetic field applied perpendicularly to the nanoSQUID. The black arrows indicate the flux shift periodicity due to the magnetic flux entry into the SQUID body. The measurement was performed at T=5K (adapted from [169]).*

So, considering an alignment uncertainty of 5°, a parallel magnetic field up to *150mT/sen(5°)≅1.7 T* can be at least employed for nanomagnetism measurements. The magnetic noise of the nanosensor was *5 $\mu\Phi_0/Hz^{1/2}$* corresponding to *70 $\mu_B/Hz^{1/2}$* for a magnetic moment located at the centre of the hole.

A further step forward was made by Chen et al. [170] that extended the possibility to perform magnetization measurements up to 7 T. They employed a *Nb* nanoSQUID having a very thin thickness (5.5 nm) ensuring a superconducting state even for 7 T in-plane applied magnetic field. In addition, they employed a feedback circuit to increase the linear dynamic range (see section 4.1.2) and a fine alignment of magnetic field in the plane of the SQUID [170].

Since the fabrication of nanoSQUIDs based on nanobridges without hysteresis in the *I-V* characteristic at helium liquid temperature is not very reliable, the nanoSQUIDs in many case have to employed as magnetic flux to current transducer as explained in the section 1.2. In this case, it is appreciable to have a high magnetic flux to current transfer factor $I_\Phi$ in order to improve both the SNR and the magnetic flux resolution (see equation 7). In this framework, high sensitive niobium nanoSQUIDs exhibiting a high current transfer factor were developed [171-176]. High $I_\Phi$ values were obtained by using a strong asymmetric current bias as proposed in a pioneering work of Clarke and Paterson [177]. This is obtained by designing the SQUID with a strong loop inductance asymmetry (Fig. 28a) [174].

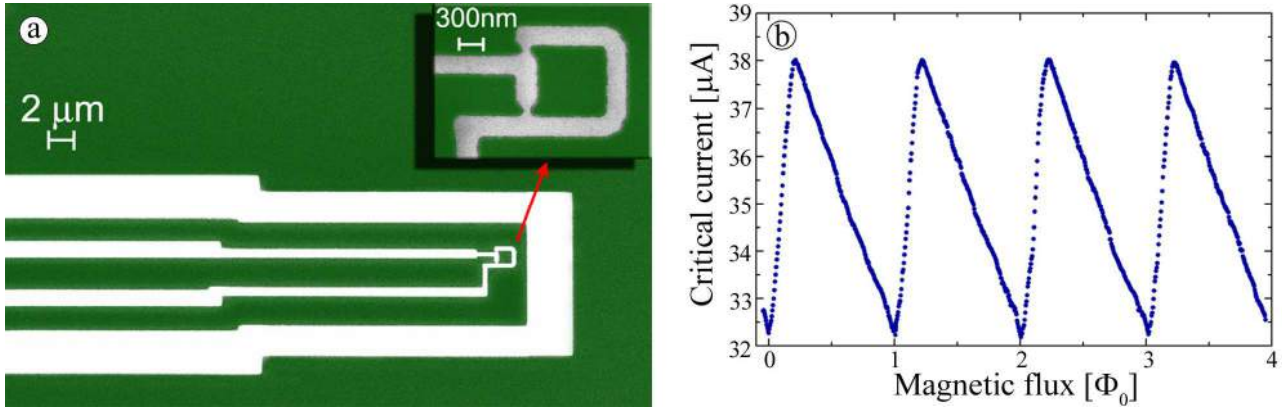

*Fig. 28 a) Scanning electron micrograph of a niobium based nanoSQUID surrounded by an integrated coil for the modulation, calibration or feedback operations. In the inset, a zoom of the SQUID loop and Dayem nanobridges is shown. b) Critical current as a function of the applied magnetic flux, measured at T = 4.2 K. The curve tilting is due to the inductance loop asymmetry (adapted from [174]).*

This asymmetry produces a deformation of the *I-Φ* pattern (Fig.28b), in which one of two branches becomes steeper yielding an increased slope. Since the $I_\Phi$ is the slope of the I-Φ curve, such design allows to increase the current responsivity improving SNR and the magnetic flux resolution. Moreover, the inductance asymmetry causes an offset of the critical current peak (Fig.28b), which could provide an advantage if the shift falls within linear region of the $I_c$-Φ characteristic. In this case, it is not necessary to apply an additional magnetic flux to bias the SQUID at the maximum responsivity point. The nanoSQUIDs reported in [174] consisted in a hole geometry with a flux capture area of *0.5 μm²* and nanobridges with a width *w* of *60 nm* and a length *l* of *140 nm*. The thickness of niobium film was *20 nm*. The nanodevice included a micrometric integrated niobium coils located very close to the nanosensor in order to modulate, to bias and to operate the nanoSQUID in a feedback mode. The maximum current responsivity was *60 μA/Φ₀*, which is about five times greater than the value of a symmetric device having the same modulation depth. Considering a current measurement error $I_{cn}$ of about *0.01 µA,* corresponding to about *1* part to *10^4* of the critical current, an overall flux noise *($\Phi_n=I_{c,n}/I_\Phi$)* of *0.16 m$\Phi_0$* was obtained. Other hysteretic nanoSQUIDs were developed by the same group and reported in refs. [175,176].

An effective way to employ hysteretic nanoSQUIDs based on nanoconstrictions as flux to voltage transducer was proposed by Hazra et al. [178]. Combining a suitable SQUID design with a readout based on bias current modulation and a lock-in voltage amplifier, they obtained regular voltage oscillations up to *20 mT* within the temperature range from *230 mK* to *1.5 K*. The nanoSQUIDs were based on *20 nm* aluminium/*80 nm* niobium/*25 nm* tungsten trilayer and fabricated by EBL. The top *Al* layer acts as an etch mask for lithographic definition of the SQUID structure, as well as a protective layer against oxidation and as mentioned above has the role of limiting the power dissipation. The tungsten layer induces a proximity effect in the Nb film lowering the critical temperature that is a useful issue at very low temperature.

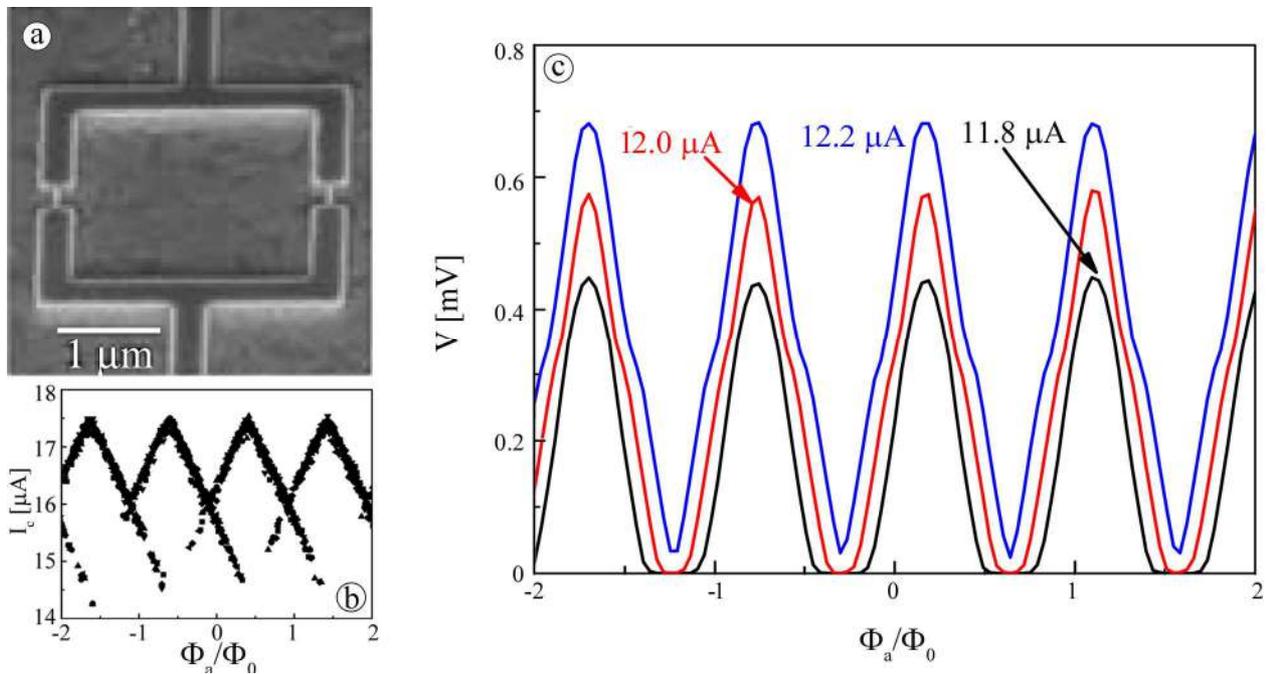

*Fig. 29a) (a) Picture of a nano-SQUID made of aluminium-niobium-tungsten trilayer fabricated on a Si wafer. b) Critical current as a function of applied magnetic flux. c) Voltage-magnetic flux*

*characteristics measured at T= 250 mK and relative to different bias current amplitudes (adapted from [178]).*

The nanodevice, reported in Fig 29a, had a loop area of *6.25 μm$^2$*, a length and width of the nanobridges of 180 nm and 40 nm respectively. The *I-V* characteristics were hysteretic and the *I-Φ* characteristics showed a typical triangular shape with crossing of the two branches of $I_c$ (Fig.29b), indicating that the superconducting property of the nanoSQUID was governed by the short coherence length of *Nb* and the current-phase relationship was non sinusoidal. Biasing the device with a sinusoidal current having amplitude a little bit greater than the critical current, regular *V-Φ* characteristics were achieved (Fig. 29c). The voltage modulation and the voltage responsivity were as high as *500 μV* and *2 mV/Φ$_0$* respectively while the magnetic flux noise was *5×10$^{-5}$ μΦ$_0$/Hz$^{1/2}$*.

To overcome the thermal hysteresis at ultra-low temperature, Blois et al developed a nanoSQUID exploiting the proximity effect of a titanium-gold bilayer [179].

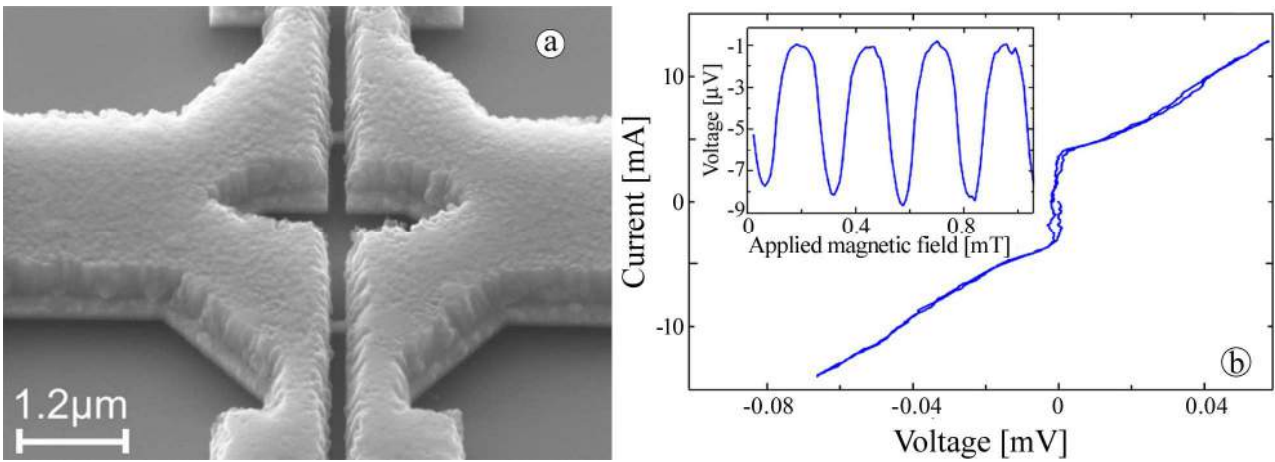

*Fig. 30 a) SEM images of Ti/Au bilayer nanoSQUID with cooling fins. (b) Current–Voltage characteristic measured at T = 100mK. The inset shows the voltage modulation as a function of applied magnetic field (adapted from [179]).*

Even if, the use of thin *Au* or *Al* shunt layers can reduce the hysteresis close to $T_c$ thanks their higher thermal conductivity, hysteresis still occurs below a significant fraction of $T_c$. In this nanosensor, besides the thermal shunt, the gold layer depresses $T_c$ of the nanobridge region via the proximity effect so it can be tuned to match the desired operating temperature. The nanoSQUID (Fig.30a) was fabricated by EBL and the nanobridges had a width and a length of *40* and *120 nm* respectively. They used a three-dimensional contact banks as in the nanoSQUID developed by Vijay et al. [165]. Even in this case, the ticker titanium layer of the banks, increases the heat extraction from the weak link regions and also reduces the kinetic inductance contribution of the loop. In addition, they added cooling fins to the side of the loop (partly visible at top and bottom of the image, Fig.30a) to further improve the usable bias range. The authors performed measurements of *I-V* characteristics at ultra low temperature for different *Au* thickness and found that the nanodevice were non-hysteretic down to *60 mK* if the thickness of the normal metal was greater than 23 nm. In the Fig.30b and 30c, the *I-V* and *V-Φ* characteristics measured at *T=100 mK*, are reported. There is no hysteresis and the voltage oscillations are sinusoidal like. Due to the flux focusing effect, the authors measured an effective flux capture area, of about *7.5 μm$^2$*, which is greater than the geometrical area. The magnetic flux noise spectral density was *1.1 μΦ$_0$/Hz$^{1/2}$* corresponding to a spin sensitivity of about *60 μB/Hz$^{1/2}$* for spins located at the centre of a circular SQUID loop having the same effective area. This nanodevice provides a sensitive tool for studying phase transitions and quantum phenomena at the nanoscale in coupled magnetic or superconducting systems at ultra-low temperatures. It is well established that a nanoSQUID based on nanobridge should have a bridge length smaller than the coherence length of the superconductor and a critical magnetic field as high as possible. Hazra et al have addressed these issues in the development of their nanodevices [180]. They developed a suspended-bridge nanoSQUID by using two superconductors (*Al* and *Nb*). This has the advantage that the bridge is composed of *Al* with a long coherence length, while the bulk of the SQUID consists of *Nb/Al* bilayer, with a relatively higher critical temperature and critical magnetic field (Fig.31). The bilayer consisted of *30 nm* thick *Nb* and *25 nm* thick *Al*. The suspended *Al* nanobridge is clearly shown in the Fig.31c where a high contrast is used to evidence the suspended structure. It was obtained by using EBL and RIE in Sulfur hexafluoride (SF6) atmosphere.

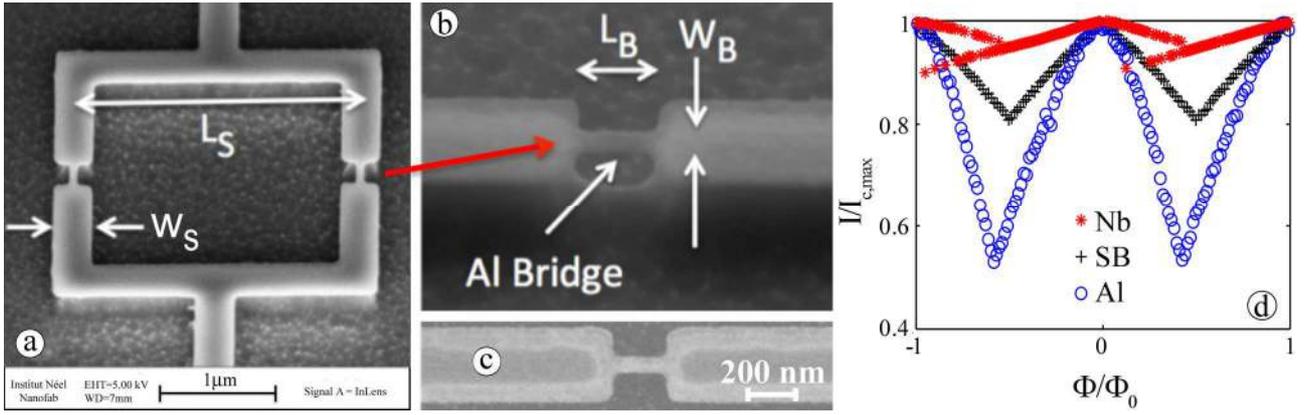

*Fig. 31 a) Image of a nanoSQUID having suspended bridges. b, c) Particular of a suspended-bridge shown at different angle of viewing. d) Critical current as a function of applied magnetic flux for different nanoSQUIDs made of Nb (red/\*), Al (blue/o) and Nb/Al bilayer with the bridge of Al only (black/+) (adapted from [180]).*

They performed a considerable over-etch so that the *Nb* under the *Al* was completely removed. The loop area of the SQUID was 6.25 µm$^2$ while the bridge width and length were *60* and *230 nm* respectively. As expected, these nanoSQUID exhibited intermediate performance between the *Al* and the *Nb* device as shown in the Fig.31d where a comparison of the *I-Φ* of three different device are reported at *T=250 mK*. Being the *I-V* hysteretic, the ultimate noise performance is limited by thermally activated switching out from zero voltage state. From the switching current distribution, measured at *250 mK*, a magnetic flux noise of *400 µΦ$_0$/Hz$^{1/2}$* was estimated by using the equation (18) with *dI/dt =0.048 A/s*, *I$_c$=42 µA*, and *I$_Φ$=17 µA/Φ$_0$*.

Before to conclude this section about the nanoSQUID based on Dayem nanobridges, we would mention to a post fabrication mechanism to modify the critical current of such nanodevices. In fact, the junction critical current can be controlled by hot photon injection by using Dayem nanobridge with integrated heaters [125, 181-183]. In these devices, a current is injected into the heater causing a local increase of the temperature of the nanobridge and, consequently, a decrease of the critical current. The heater consists in a normal metal thin film separated from the superconductor by an insulating layer.

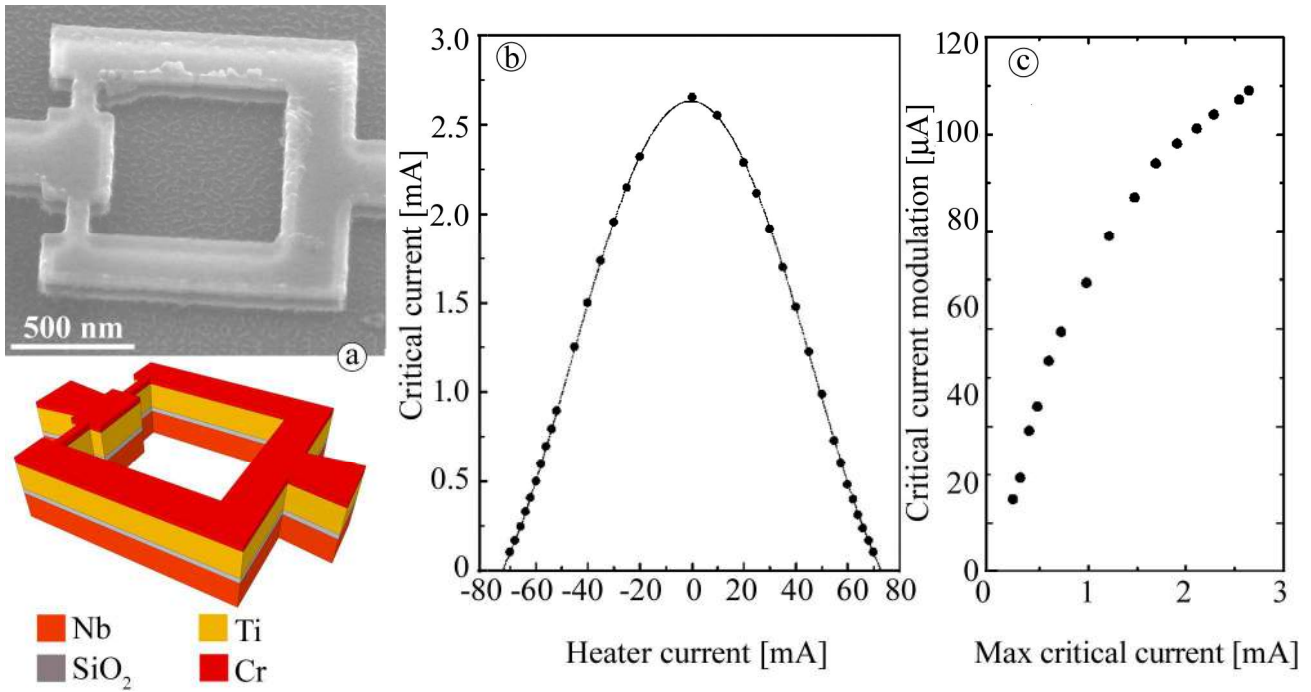

*Fig. 32 a) Scanning electron micrograph of dc-nanoSQUID with a superimposed Ti control layer. The multilayered structure is depicted in the lower sketch. b) Maximum critical current value as a function of the heater current. c) Modulation depth of critical current on varying of the maximum critical current (adapted from [182]).*

The nanoSQUID depicted in the Fig.32a has a loop area of *0.6 μm$^2$* and two nanobridges *200 nm* long and *120 nm* wide. It was fabricated by using EBL to pattern sputtered thin films of *50 nm* thick of *Nb*, *30 nm* thick of *SiO$_2$* (insulating layer) and *50 nm* thick of *Ti* (heater). The chromium film is used as a protective mask, during the RIE etching. The Fig.31b reports the critical current as function of the control current flowing in the heater layer while the Fig.31c shows the modulation depth as a function of the critical current. As shown, this design allows to obtain a fine control of the critical current and a consequent reduction of thermal hysteresis.

Finally, it is worth to mention that several experimental studies about the noise properties of nanoSQUIDs based on nanobridges have been carried out [153, 158-160, 184]. These investigations included measurements of both white and low frequency magnetic noise as a function of the applied magnetic field, the temperature and the nanoSQUID design.

## 3.3 NanoSQUID based on sandwich type nanojunctions

As seen in the previous section, the current-voltage characteristics of nanoSQUIDs based on nanobridges are typically hysteretic, unless be very close to their transition temperature. The stochastic nature of the switching to the voltage state forbids the use of standard SQUID readout schemes, compromising the devices sensitivity. Moreover, due to thermal hysteresis, it is difficult to operate at much lower temperatures that hinder their application to many quantum measurements. In fact, excessive heat dissipation in the SQUID loop is clearly undesirable for studying samples of interest in close proximity since they may be directly affected by the heating or indirectly via other mechanisms that can lead to quantum decoherence. In addition, the reliability and robustness against the thermal cycles of nano-bridges is poor. In fact, in nanoSQUID based on single niobium thin film (20-25 nm), a degradation of performance after some thermal cycles can occurs. In particular, the critical current and its modulation in an external magnetic field decrease. Furthermore, the critical current values and the modulation depth strongly depend on the thickness and the size of nano-bridges. However, as seen in the previous section, some of the drawbacks above reported can be partly overcome by using a further metal layer, which acts as a shunt and a protection layer.

These motivations have stimulated, in the recent years, the development of nanoSQUIDs based on niobium sandwich type Josephson nano-junctions (JJs). Obviously, the fabrication of overdamped JJs, as required for dc-SQUIDs, with submicron dimensions is more challenging with respect the nanobridge junctions. Compared to Dayem bridge, the main advantages of the nanoSQUIDs based on JJs are a better control of the critical current, a high modulation depth, an ultra-low noise, being based on a fully reliable niobium technology. Furthermore, they exhibit a non-hysteretic behaviour, allowing the flux locked loop operation and the use of more effective readout scheme.

In this framework, Nagel et al. [185] developed nanoSQUIDs in both magnetometer and gradiometer configurations based on Superconductor/Normal metal/superconductor (SNS) nano-junctions. In particular, they employed niobium JJs with HfTi barrier without resistive shunt [186, 187]. The outer loop size of gradiometer was ($1.5 \times 1.5$) $\mu m^2$ and the area of the SQUID magnetometer hole was ($500 \times 500$) $nm^2$. The area of the junctions was ($200 \times 200$) $nm^2$ while the barrier thickness was *24 nm*. Fig. 33a and 33b show scanning electron micrographs of the device in gradiometer and magnetometer configurations; in the inset of Fig. 33b a washer configuration is shown. At *T=4.2 K*, the nanodevices did not exhibit hysteresis, the critical current was of *178 µA* whit a modulation depth of about 80%. Comparing with the theory, a $\beta_L$ equal to *0.18* and a corresponding SQUID inductance of *2.1 pH* were estimated. The spectral density of magnetic flux noise as low as *250 $n\Phi_0/Hz^{1/2}$* was measured by using a SQUID amplifier readout (Fig.33c).

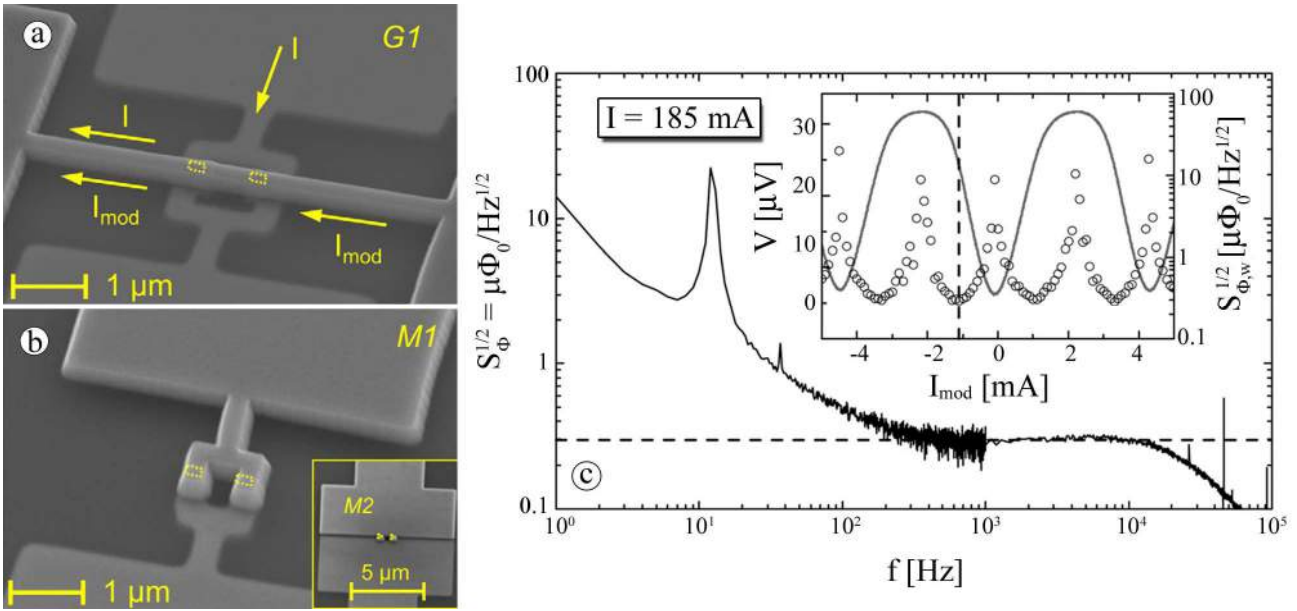

*Fig. 33 SEM images of the SQUID in a gradiometer (a) and a magnetometer (b) configuration. In the inset, a washer-type magnetometer is also displayed. The Josephson junctions with HfTi barrier are indicated in the dotted box. (c) Magnetic flux noise spectral density of SQUID gradiometer measured at optimal working point indicated by the dashed line in the inset. The latter displays the voltage as a function of modulation current (solid line) and the magnetic flux spectral density value in the white region (open circles) (adapted from [185]).*

Following the calculation procedure reported in section 2.6 [140], a spin sensitivity of *29 $\mu_B/Hz^{1/2}$* was obtained for a spin placed directly on the top of the SQUID loop. A few years later, Wolbing et al. [188] demonstrated that a modified version of this nanoSQUID could operate in a magnetic field up to *0.5 T*. In particular, they developed a vertical design where the SQUID loop and the Josephson junction barrier were in the same plane. With respect to the planar configuration, where the SQUID loop and the substrate lie in the same plane, this arrangement allows to reduce some spurious signals arising from the excitation magnetic field.

In the planar configuration, the excitation field for magnetic nano-object characterization is applied in the plane of the SQUID loop, resulting parallel to the plane of the junction barrier. Since, a JJ is sensitive to such magnetic field; a subsequent critical current reduction leads to SQUID performance degradation. Instead, in the case of a vertical configuration, if a fine alignment system is employed so that the excitation field is merely perpendicular to the SQUID substrate plane, only the stray fields due to the magnetization response of nano-objects are detected by the nanoSQUID. The nanoSQUID was based on a *Nb* thin film microstripline geometry, i.e. the same design employed for the low inductance superconductive interferometers [31]. The arms of the SQUID loop, separated by an insulating layer, lie directly on top of each other and are connected via two JJs (Fig. 34a). In this case, the thickness of bottom and top *Nb* were *200* and *160 nm* respectively while the *SiO$_2$* insulation layer was *225 nm* thick. This nanoSQUID exhibited a magnetic flux noise less than *250 n$\Phi_0$/Hz$^{1/2}$* up to a magnetic field of *50 mT* applied perpendicularly to the substrate plane while for a magnetic field of *500 mT*, the noise increased to *680 n$\Phi_0$/Hz$^{1/2}$*. Using the same technology, very recently, Bechstein et al. [189] have developed a family of nanoSQUID gradiometers in both series and parallel configuration (Fig. 34b and c). They employed JJs with a size of *(200×200) nm$^2$*, and *HfTi* barrier thickness of about *30 nm*. The gradiometer loops had an inner diameter of *840 nm* and a line-width of 300 nm. Gradiometer feedback and excitation coils were also integrated on the same chip of nanoSQUIDs, in order to enable sufficient coupling for feedback and working point adjusting.

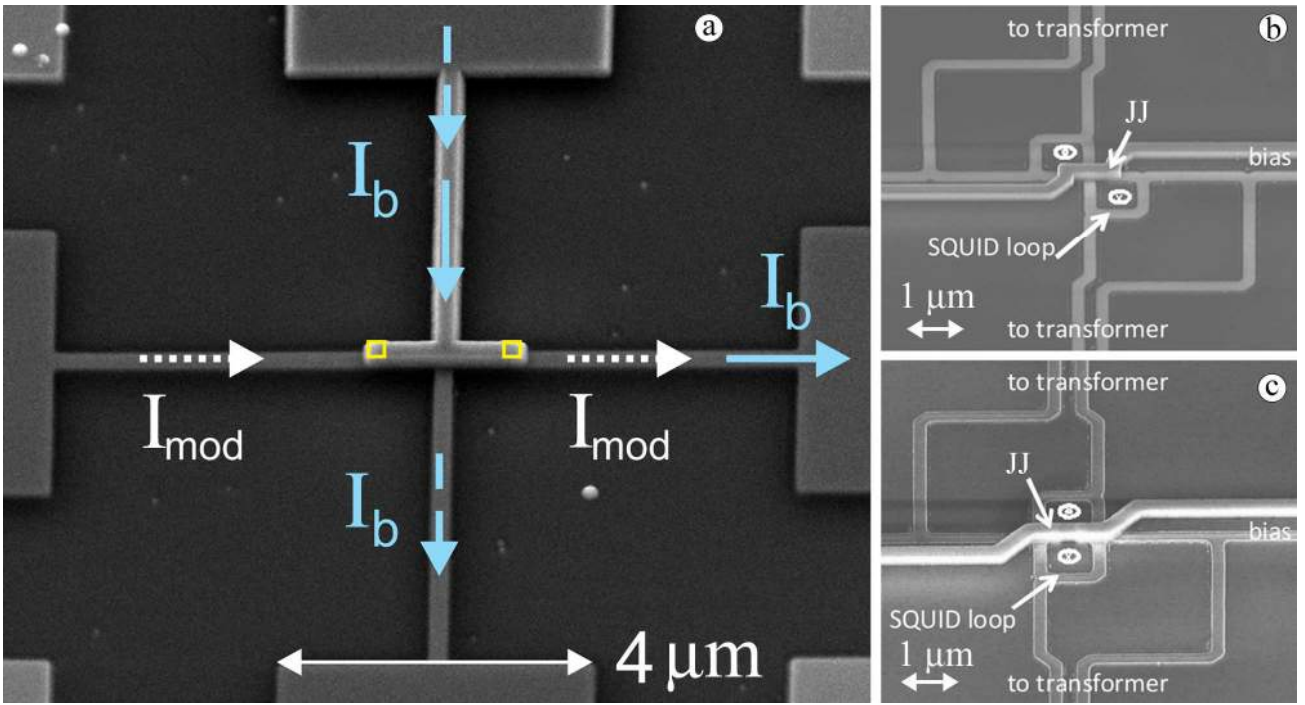

*Fig 34. a) SEM image of a nanoSQUID designed to operate in moderately high magnetic field. The Josephson junctions with HfTi barrier are indicated by the (yellow) box, while (cyan) arrows indicate symmetric and asymmetric current bias (adapted from [188]). b) Pictures of nanoSQUID series gradiometer and (c) parallel gradiometer (adapted from [189]).*

By using an array of 26-SQUID in series and a flux locked loop based readout at room temperature, a spectral density of white magnetic flux noise as low as *115 n$\Phi_0$/Hz$^{1/2}$* and *200 n$\Phi_0$/Hz$^{1/2}$* were measured at *T=4.2 K* for parallel and series configuration, respectively. A readout setup based on bias reversal scheme [44] in a two-stage configuration was successfully employed to reduce the low frequency noise. So that, a magnetic flux noise equal to *600 n$\Phi_0$/Hz$^{1/2}$* was achieved at *1 Hz*. Furthermore, the authors demonstrated that the gradiometer design, together with the special design of the coupling transformer coils, allows obtaining a very low level of nonlinearity.

NanoSQUIDs based on Superconductor/Normal metal-Insulator/Superconductor (SNIS) nano-junctions fabricated using an innovative three-dimensional FIB based nanomachining were developed by Granata et al. in both planar and vertical configurations [190-193]. The planar nanoSQUID consisted of a niobium superconducting loop (*0.2 μm$^2$*) interrupted by two square nanometric *Nb/Al-AlOx/Nb* Josephson junctions having an area of *(0.3×0.3) μm$^2$*. The thickness of both *Nb* electrodes was *350 nm* and the *Al* was *80 nm* thick and the AlOx grown in situ without breaking the vacuum during the oxidation phase, in pure oxygen. The nanosensors were designed in both planar and vertical configurations (Fig 35a and b).

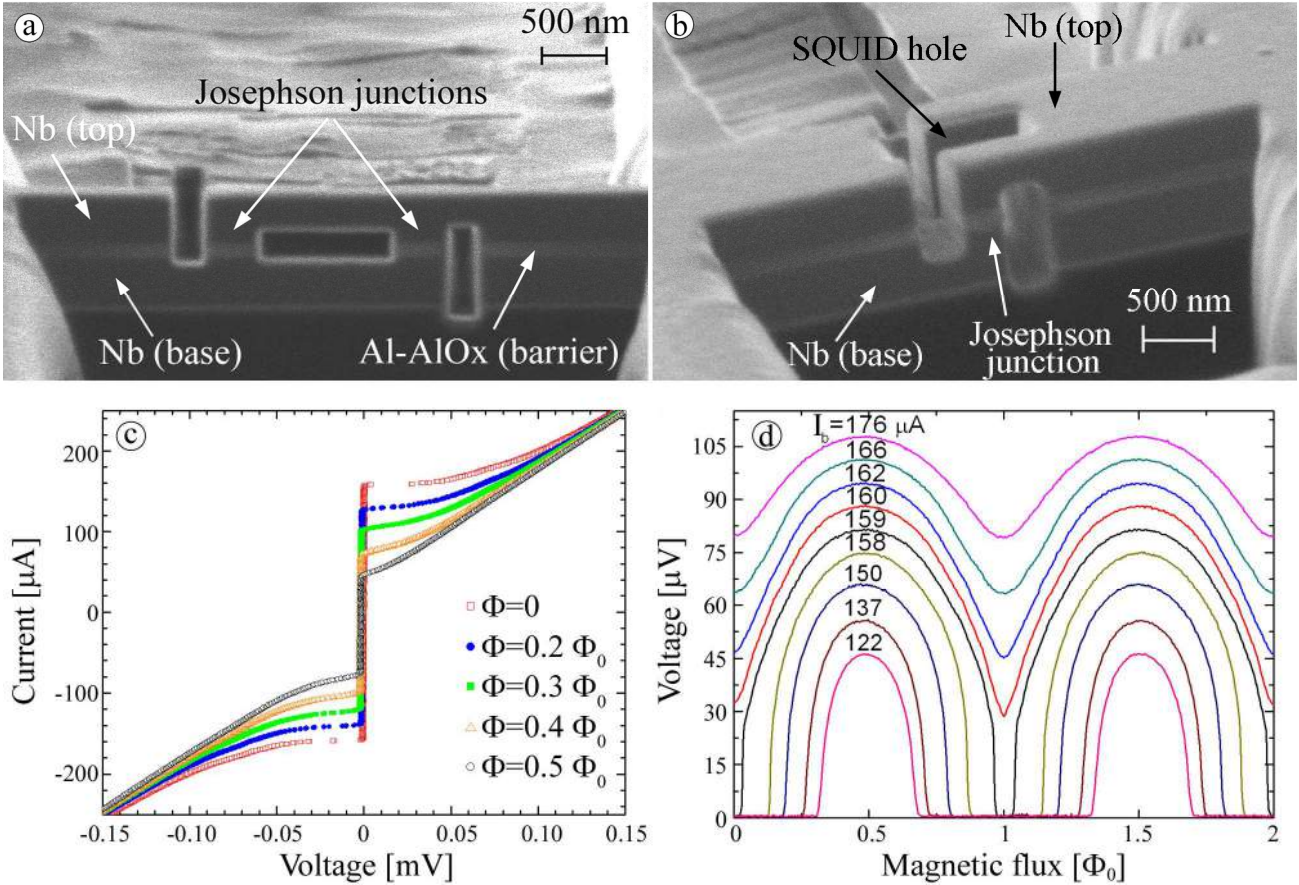

*Fig. 35 a) SEM images of a Nb-based nanoSQUIDs having both the junctions and hole in a vertical plane. b) NanoSQUIDs with hole in a planar plane with respect to the junctions. c) Current-voltage characteristics obtained for different magnetic flux linkages to the planar hole. d) Voltage as a function of external magnetic flux exhibited by planar nanoSQUID for different bias current values (adapted from [190] and [192]).*

The aspect ratio (ratio between the height and the width) referred to the single lamella is about *2*, resulting in a three-dimensional structure. Even in this case, the *I-V* characteristics did not exhibit hysteresis, allowing the device to operate as a magnetic flux to voltage transducer (Fig.35c). A modulation depth $\Delta I_c$ of *110 µA*, corresponding to about 70% of the maximum critical current was measured. The estimated value of $\beta_L$ was *0.55*, resulting in a SQUID inductance of *7 pH*, compatible with the inductance of the device including the kinetic inductance of the striplines. The *V-Φ* characteristics (Fig.35d) were resonance-free allowing the device to work in a wide region of bias points. This circumstance guarantees good stability during operation in both small signal mode and in flux-locked-loop configurations. A voltages swing of *80 mV* and responsivity value as high as *2.9 mV/Φ$_0$* for a bias current close to the critical current value was obtained, thanks to high values of both $R_d$ and modulation depth. The spectral density of the magnetic flux noise was measured in small signal mode by employing low-noise readout electronics in a direct coupling scheme. In the white noise region, the nanoSQUID exhibited an intrinsic magnetic flux noise level as low as *680 nΦ$_0$/Hz$^{1/2}$* corresponding to the spin noise in the loop centre of about *38 μ$_B$/Hz$^{1/2}$* and less than *10 μ$_B$/Hz$^{1/2}$* close to the edges and corners. The main performances of these nanoSQUID as a function of the temperature have been also investigated and compared with those of SQUIDs with nanobridges [194].

Ronzani et al [195] developed another kind of nanoSQUID based on SNS JJs. By using standard EBL, they fabricated a SQUID employing vanadium/copper/vanadium (*V/Cu/V*) JJs having width and length of *60* and *370 nm* respectively, while the SQUID loop area was *1.5 μm$^2$* (Fig.36a). The thickness of vanadium and copper were *150* and *20 nm* respectively. The *I-V* characteristics measured at *240 mK* were non-hysteretic, and the *I$_c$-Φ* were well fitted by standard SQUID theory indicating that the current phase relationship is sinusoidal, despite the structure resemble a long nanobridge. A voltage responsivity of *450 μV/Φ$_0$* was obtained by taking the derivative of the *V-Φ* characteristic (Fig.36a). The magnetic flux noise measured by using a direct-coupled amplifier was *2.8 μΦ$_0$/Hz$^{1/2}$* that enabling to estimate a spin sensitivity of *100 μ$_B$/Hz$^{1/2}$* in the optimal coupling regime. Repeated cool-down cycles without any appreciable performance degradation demonstrated the robustness and the reliability of this device. Based on the same *V/Cu/V* nanojunctions, the same authors developed also a double loop interferometer [196]. It consists of the parallel circuit of three Josephson junctions defining two superconducting loops (Fig.36b).

Within the experimental error (about $10^{-2} I_c$), such a nanodevice exhibited, at *240 mK,* a modulation depth equal to the whole critical current of the SQUID (Fig. 36b).

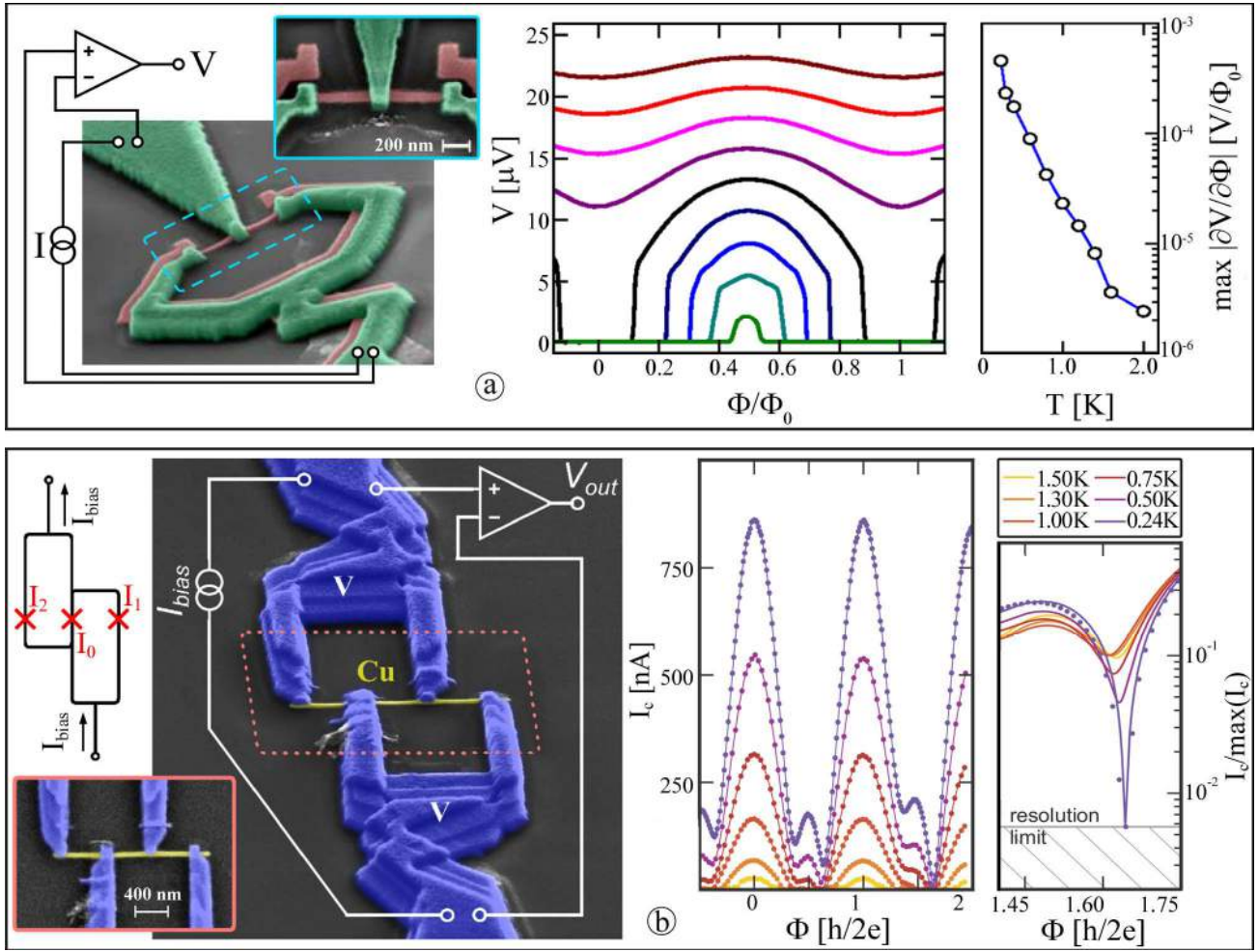

*Fig. 36 SEM images of superconducting nanodevices based on V/Cu/V Josephson junctions. a) NanoSQUID image with a superimposed measurement setup scheme. The inset displays a magnified view of the weak links. The voltage-magnetic flux characteristics, measured at 240 mK, for different values of bias current is shown in the middle figure while the influence of the temperature on the maximum responsivity ($V_\Phi$) is reported in figure on the right (adapted from [195]). b) Picture of double-loop superconducting interferometer. The figure reports also the measurement setup as overlapped draft and a scheme of working principle (top left). The lower inset depicts the top view of the metallic nanowire. At centre, the I-$\Phi$ curves measured at different temperatures are shown. Finally, the detailed view of minimum values of normalized critical current are shown (adapted from [196]).*

Very recently, M. Schmelz et al. [197] have developed an ultra low noise nanoSQUID based on standard superconductor/insulator/superconductor (SIS) Josephson junctions. The *SIS* junction based SQUIDs offer a higher voltage swing resulting in a better performance than comparable to *SNS* or *SNIS* junctions. They employed a technology based on sub-micron cross-type *Nb/AlOx/Nb* fabrication [198]. The nanoSQUID depicted in the Fig. 37a consists of a narrow *Nb/AlOx/Nb* trilayer strip and a *Nb* wiring layer having an u-shape. The square Josephson junctions, having a size of $(0.8 \times 0.8)$ μm$^2$, are formed by the overlap between the trilayer and the *Nb* wiring. The inner SQUID loop dimensions ranges from *0.5* to *10 μm*. At *T=4.2 K*, these nanoSQUIDs exhibited a voltage amplitude of about 300 μV and, consequently, very high voltage responsivity. The magnetic flux noise of the smallest nanoSQUID (loop area of *0.25 μm$^2$*), measured by using a SQUID series amplifier, was as low as *66 nΦ$_0$/Hz$^{1/2}$*. The estimated value of corresponding magnetic moment sensitivity was *7μ$_B$/Hz$^{1/2}$*, for a magnetic dipole lying at the centre of the square loop.

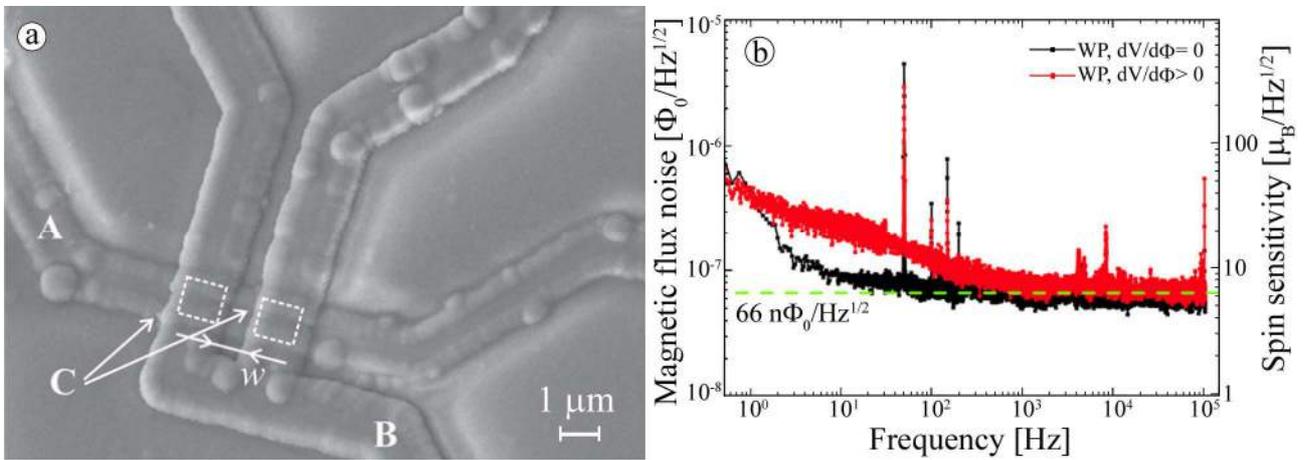

*Fig. 37 a) SEM image of all niobium based nanoSQUID with cross-type Josephson junctions pointed out by the dotted boxes. b) Flux noise spectral densities obtained with different responsivity. On the right axis, the corresponding spin sensitivity is reported (adapted from [197]).*

NanoSQUIDs based on superconducting aluminium tunnel junction were developed by Isciguro et al. [199]. The hole of nanoSQUID was *(0.5×0.5) μm$^2$* while the junction size was about *(0.1×0.1) μm$^2$*. At *T=75 mK*, the *I-V* characteristic showed an hysteretic behaviour with a critical current lower than *1 µA*. Despite the critical current modulation depth reached the 80% of $I_c$, the resulting flux to current transfer factor was small (less than *1 µA/Φ$_0$*) leading to a poor a magnetic flux sensitivity.

**3.4 NanoSQUIDs for scanning microscopy: nanoscale 3D pick-up loop, nanoSQUID on a tip.**

As will be shown in the next section, one of the key points is how to bring or integrate the nano-sample of interest in close proximity to the SQUID, in order to achieve an enough magnetic coupling to detect single spin or small spin populations. A different approach could be the development of SQUID for scanning magnetic microscopy (section 1.5) with an ultra high spatial resolution. Recent improvements of the spatial resolution includes the terraced scanning SQUID susceptometer with submicron pickup loops [200], the STM-SQUID which combine the scanning tunnel microscope with a SQUID [201], the thin-film microsusceptometer with integrated nanoloop [202].

Thereafter, we will describe two peculiar nanodevices, developed following the aforementioned intent. Romans et al. [203] fabricated a niobium SQUID with a freestanding nano pick-up loop made of tungsten. In such a way, it is possible to fabricate three-dimensional pick-up loop enabling the measurements of all spatial components of the magnetic field. Moreover, the nano pick-up loop can be specially engineered and deposited around the sample to investigate. The 3D pick-up loop fabrication was based on the FIB induced growth of metallic nanowire [204]. The nanodevice consists of a freestanding tungsten pick-up loop with an area of *1.6 µm$^2$*, connected to a single turn input coil via a planar striplines having a length and a width of *9 µm* and *0.5 µm* respectively (Fig.38). In order to reduce the parasitic flux capture area between the two nanostrips, these are coated by an insulating material and a superconducting shielding layer.

The input loop was magnetically coupled to a *Nb* microSQUID (area of *1.6 µm$^2$*) including two *70 nm*-wide Dayem nanobridges. Two parallel tungsten control lines, carrying equal and opposite currents, were integrated on the same chip and located close to the 3D pick-up loop (Fig.38).

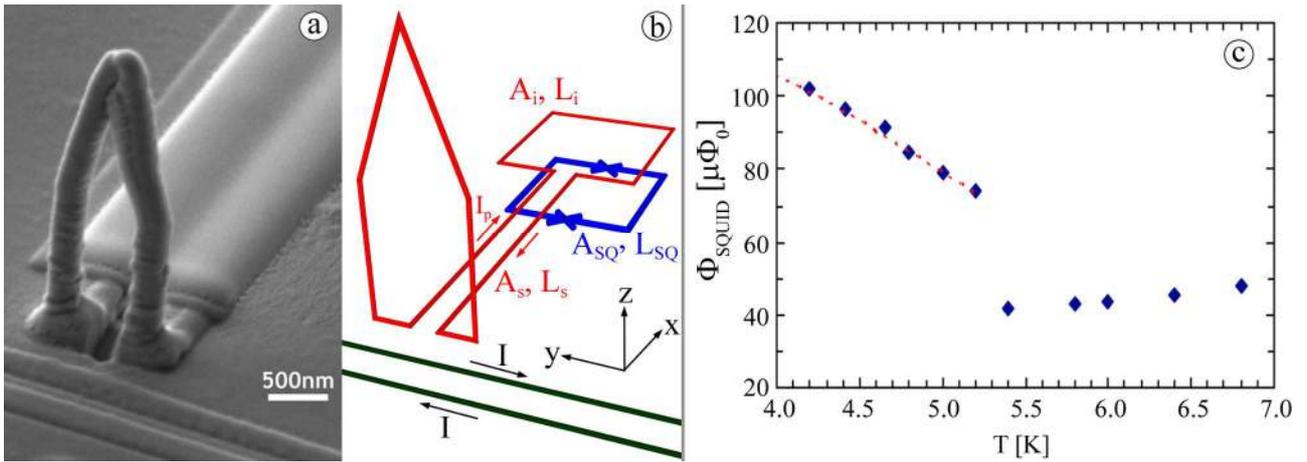

*Fig. 38 a) Scanning electron micrograph of a 3D tungsten pickup loop arranged perpendicularly to the SQUID plane and a sketch of full three-dimensional device. b) The total amount of magnetic flux induced into the SQUID by control current as a function of temperature. The (red) dashed line shows the theoretical prediction (adapted from [203]).*

Since the magnetic field generated by the control lines decays rapidly, these were used to demonstrate the effectiveness of the device to detect the localized source. In particular, the authors measured the magnetic flux due to control lines as function of the temperature noting an abrupt signal variation in correspondence of the tungsten critical temperature (about *5.4 K*) (Fig. 38c).

Finkler et al. and Vasyukov et al. [205-207] have developed an innovative technique for the realization of a nanoSQUID for ultra high sensitivity magnetic scanning microscopy. They fabricated a nanoSQUID on the apex of a sharp quartz tip obtaining magnetic moment sensitivity as low as *0.38 μB/Hz$^{1/2}$*. The nanoSQUIDs, made of three different superconducting materials (Al [205,206], *Nb* [207], and *Pb* [207]) were realized on pipettes with tip diameters in the range of *40-300 nm*. The tips were obtained by pulling a *1 mm* of quartz tube by a suitable puller instrument. The fig. 39 a-c depicts the *Al, Nb*, and *Pb* nanoSQUIDs with three different loop diameters. The fabrication procedure was based on three self-alignment steps. In the first one, the superconducting thin film was deposited onto the apex ring of the pipette. Such step forms the superconducting loop of the SQUID. After that, the pipette was tilted at an angle of about 100° with respect to the axis of the deposition source and a second superconducting film was deposited forming a first lead. Then the pipette was rotated at an angle of about *–100°*, followed by a third deposition on the other side to form the second lead. The short parts of the ring corresponding to the gaps between the leads act as two Dayem-bridges, while the two areas where the leads contact the ring form superconducting regions (Fig.39 a-c). The thickness of superconducting film depends on the materials. In particular, for the different nanoSQUIDs, the thicknesses of the leads and superconducting loop were respectively: *25 nm* and *17 nm* in the case of *Al*, *35 nm* and *23 nm* for *Nb* and, at last, *25 nm* and *15 nm* for the *Pb*.

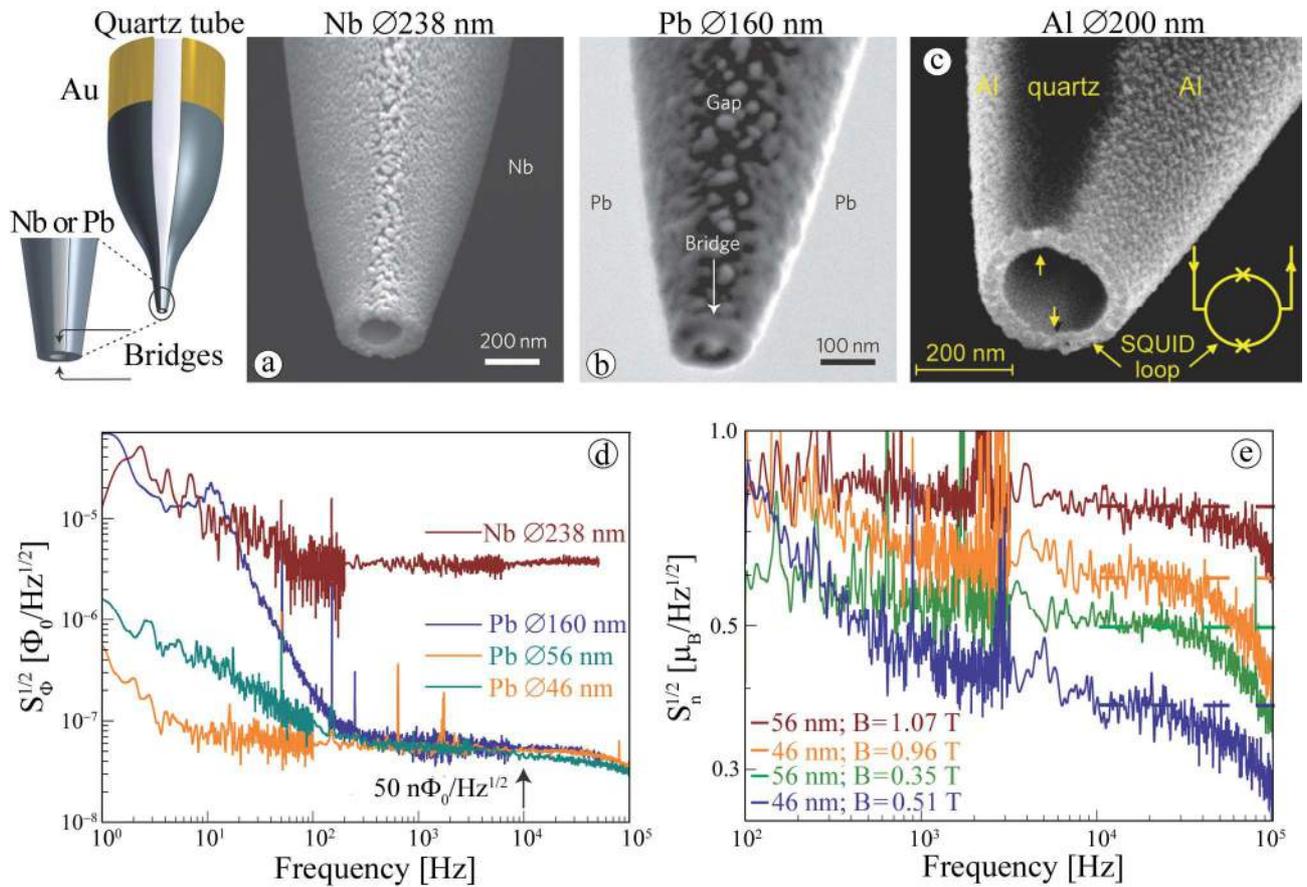

*Fig. 39 SEM images of nanoSQUIDs built on a quartz tip as schematized in the top left corner. a), b), and c) Pictures of Nb based, Pb based and Al based nanoSQUID respectively. d) Magnetic flux noise spectral densities exhibited by Nb and Pb nanoSQUIDs. e) Spectral densities of spin noise for Pb based nanoSQUIDs at different applied magnetic field (adapted from [205] and [207]).*

The *Nb* and *Pb* nanodevice were characterized at *T=4.2 K* whereas the *Al* device at *T=300 mK*. All devices showed a non-hysteretic *I-V* characteristics and a critical current modulation depth of about 25%, 50%, and 80% for *Nb*, *Al* and *Pb* based devices respectively. Such discrepancies of critical current modulation depth can be ascribed to different ratio between the bridge length and the superconducting coherence length. It was confirmed by the triangular shape of the *I-Φ* patterns in the case of *Nb* nanoSQUID that exhibited the lowest modulation depth. The magnetic flux noise spectra of *Nb* and *Pb* nanoSQUIDs for different loop diameters and different magnetic field expositions are reported in the Fig. 39d and 39e respectively. The *Pb* nanoSQUIDs exhibited a magnetic flux noise as low as *50 $n\Phi_0/Hz^{1/2}$*, which is one of the best values achieved. The magnetic moment sensitivity obtained by assuming a Bohr magneton in the centre of the loop was *0.38 $\mu_B/Hz^{1/2}$*. It is worth to point out that the spin sensitivity still remain below *1 $\mu_B/Hz^{1/2}$* for magnetic field as high as *1 T* as shown on Fig.39e, ensuring an ultra high magnetic moment sensitivity for all nanomagnetism measurements.

### 3.5 High critical temperature superconductor based nanoSQUIDs

Different high critical temperature (HTc) Josephson junctions have been extensively investigated, including bi-crystal, bi-epitaxial, step-edge, and superconductor–normal–superconductor (SNS) type [32]. High-Tc SQUID magnetometers are mostly fabricated on bicrystal, ramp edge and step edge substrates [33]. However, it is not easy to fabricate nanojunctions and nanoSQUIDs by using these methods. On the other hand, nanoSQUIDs made of HTc superconductors might extend the operational working temperature (from *mK* to above *77 K*) and the application of very strong magnetic fields (order of few T).

Wu et al. [208] fabricated the first HTs nanoSQUID based on nanobridge *YBa$_2$Cu$_3$O* (YBCO) junctions (Fig.40). The nanoSQUIDs, fabricated by using a FIB technique, had a minimum flux capture area of (*250×250) nm$^2$* and nanobridge sizes of *100 nm* (width) and *250 nm* (length). The nanoSQUIDs were fabricated on *SrTiO3* (STO) substrate and were patterned on an YBCO/*Au* bilayer having a thickness of *150/300 nm* respectively.

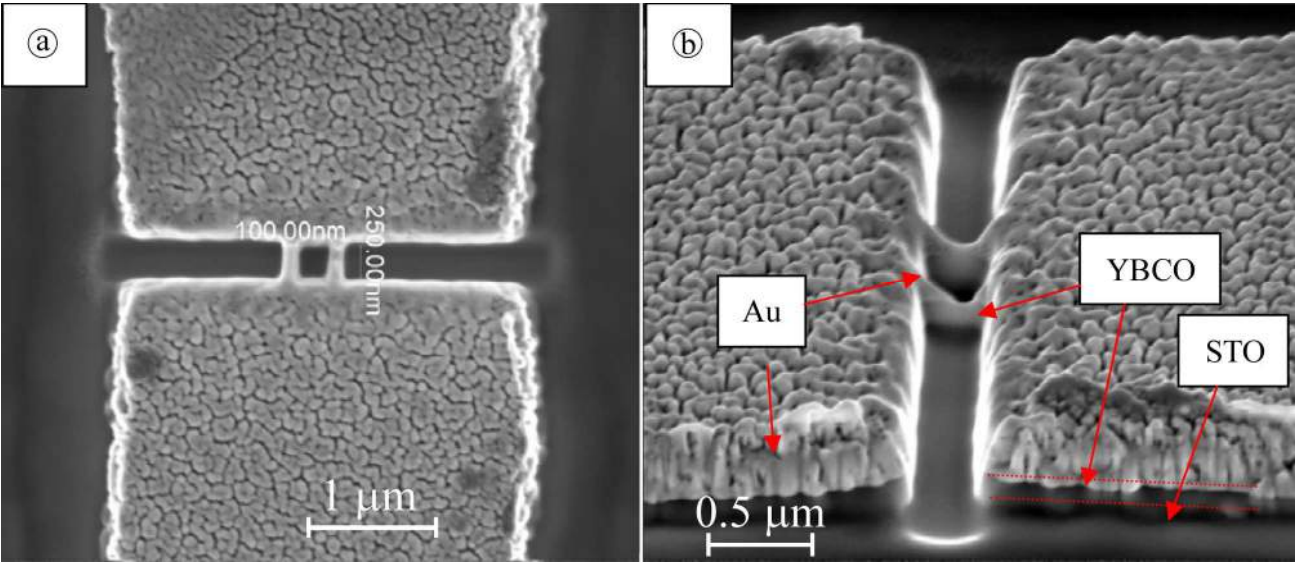

*Fig. 40 Picture of a high-$T_c$ NanoSQUID based on $YBa_2Cu_3O_{7-x}$ nanobridges. a, b) Images of the device by different view angle (adapted from [208]).*

As usual, the *Au* layer acts as both fabrication mask for FIB etching and protective layer. The *I-V* characteristic showed a non-hysteretic behaviour and a strong flux flow at temperature below *76 K*. These nanoSQUIDs exhibited low values of critical current and normal resistance ($I_cR_n$=5 $\mu V$) resulting in very low amplitudes of the *V-Φ* characteristic (few *μV*) and consequently in a poor magnetic flux sensitivity.

Schwartz et al [209] developed a HTc nanoSQUID based on grain boundary Josephson junction demonstrating low noise performance up to *B=1 T* applied parallel to the plane of the SQUID. The fabrication was based on the deposition of a bilayer YBCO/*Au* (*50/60 nm* thick) on a $SrTiO_3$ by-crystal substrate patterned by a FIB technique [140]. The Figs. 41a and 41b show the nanoSQUID with a hole size of (*300×400*) $nm^2$ and junctions having a width of *130 nm* and a length of *400 nm*. In order to effectively flux modulate the SQUID, a *90 nm* wide nano-constriction was included. The authors reported a fully characterization at *T=4.2 K* for applied magnetic field up to *B=3 T*. Due to a high normal resistance, a high $I_cR_n$ (*130 μV*) value was obtained. For *B=0 T*, a critical current modulation depth of more than 50% of $I_c$ and a voltage responsivity of *500 μV/Φ_0* were measured from the *I-Φ* and *V-Φ* curves, respectively. For *B=1 T*, the characteristics remained almost the same. However, due to a slight decrease of the critical current, the *V-Φ* became more smooth leading to a decrease of the voltage responsivity.

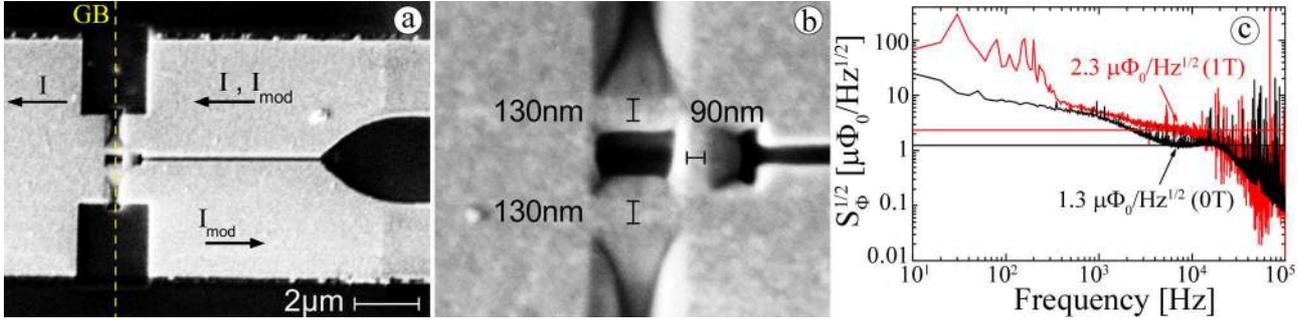

*Fig. 41 a) SEM image of the YBCO based nanoSQUID with a grain boundary Josephson junctions. b) Magnified view of the nanoSQUID to point out the hole and junctions. c) Flux noise spectral density of the device at zero (black) and non-zero (red) applied field (adapted from [209]).*

In the white region of the magnetic flux noise spectra (Fig. 41c) a noise level of *1.3* and *2.3* $\mu\Phi_0/Hz^{1/2}$ for *B= 0 T* and *B=1 T* respectively was measured. As expected, in both case, the flux noise increases at frequencies below 3 kHz. In fact, unlike LTc SQUID, the low frequency noise corner in the HTc SQUID is much higher (several kHz). However, at *B=1 T*, an additional low frequency noise due to the fluctuating Abrikosov vortices can be observed. The spin sensitivity was evaluated by following the method reported in [140] and summarized in the section 2.5. Assuming that the magnetic nano-object is placed on the top of nano-constriction at a distance of *10 nm*, they found a spin sensitivity of *62* and *110* $\mu_B/Hz^{1/2}$ for *B=0* and *B=1 T* respectively. Very recently, the same group has employed a similar YBCO nanoSQUID exibiting a very low white noise ($S_\Phi^{1/2}$=230 n$\Phi_0$/Hz$^{1/2}$ in FLL at T=4.2 K) to peform interesting measurements of nanomagnetism (see section 4.2).

A HTc nanoSQUID based on Dayem nanobridges and able to operate in the full temperature range below the transition temperature was developed by Arpaia et al. [210]. The fabrication process was based on the pulsed laser deposition of YBCO *50 nm* thick on a *MgO* substrate and patterned by EBL. The nanobridges had a width of *50 nm* and a length in the range *100-200 nm*, while the nanoSQUID loops range from *100* to *1000 nm*. The Fig.42 shows a nanoSQUID with a loop size of (*200×150*) *nm²*.

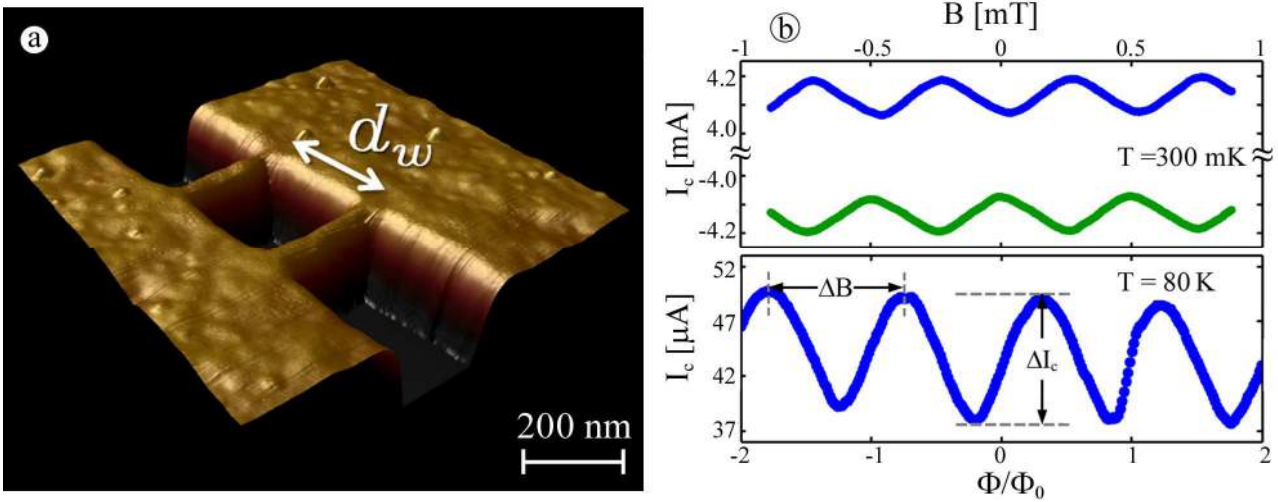

*Fig. 42 a) Atomic force micrograph of a Dayem bridge based high-Tc nanoSQUID coated with an Au layer. b) Critical current as a function of the applied magnetic field measured at T = 300 mK and close to the critical temperature (80 K) (adapted from [210]).*

The *I-Φ* curves for *T=300 mK* and *80 K* exhibited enough critical current modulation depth in both case confirming the capability of these nanosensor to work in a wide range of temperature. The authors obtained a convincing fit of current modulation depth by using an approximated expression of the CPR obtained from equation 31 [122]. In particular, they found that *ΔI$_c$/I$_c$=1/β$_L$*, which is standard relationship of a SQUID with high *β$_L$* (section 1.2). Also in this case the inductance of nanoSQUID (*L≈15 pH*) coincided with the kinetic inductance of the nanobridges. The intrinsic white magnetic flux noise measured above *10 kHz* was of *0.7 μΦ$_0$/Hz$^{1/2}$* and corresponds to a spin sensitivity of about *50 μ$_B$/Hz$^{1/2}$*. At frequencies below *10 kHz,* the noise was dominated by *1/f* noise probably due to critical current fluctuations.

**3.6 Other nanoSQUID types**

In many cases, the development of nanoSQUIDs provides new insights on both the Josephson coupling between the superconducting electrodes and the electronic transport properties of materials. In this section, we will shortly describe some nanoSQUIDs that, beyond technology and applications, are powerful probes of underlying physics of new material and tunnelling.

A very attractive nano-device is the nanoSQUID based on single-walled carbon nanotube (CNT) Josephson junctions developed by Cleuziou et al. [211,212]. The device, having an area of about 1 µm$^2$ consists of a superconducting loop formed by an *Al* structure closed by two CNT Josephson junctions [213] with a diameter of about *1 nm* and a length of *200 nm*. The fabrication procedure was based on the location of CNT by atomic force microscope technique followed by aligned electron beam lithography to pattern the SQUID loop and contacts. A palladium layer (*3 nm* thick) was used as a buffer layer to provide high transparency contacts to carbon nanotubes, while *50 nm* thick *Al* was employed as superconducting electrodes. The CNT nanoSQUIDs reported in Fig.43a shows fork geometry for the loop allowing to fabricate both junctions from the same nanotube. The two lateral gates were used to tune independently the electronic properties of each CNT junction while the back-gate adjusts the transparency of the CNT contact barrier. The critical currents of each junction can be tuned independently with gate voltage (Fig.43 b-e).

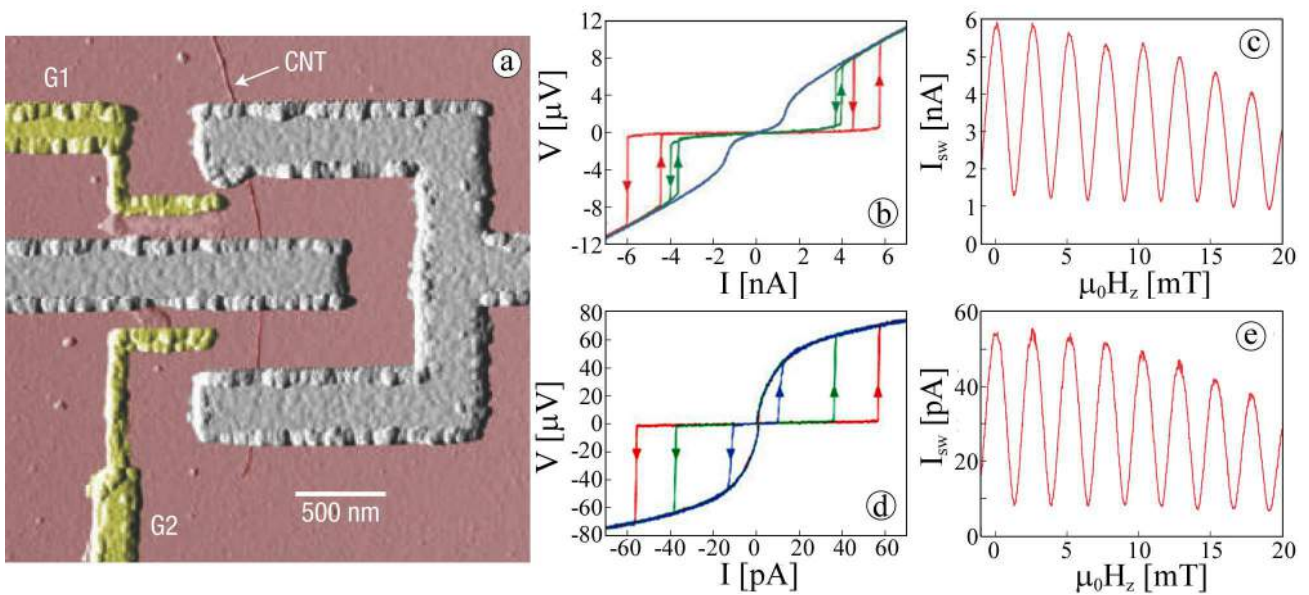

*Fig. 43 a) Atomic force microscopy image of the nanoSQUID built by Cleuziou and co-workers. The carbon nanotube (CNT) indicated in the figure by an arrow acts as Josephson junctions. b) and c) Current-voltage and critical current-magnetic flux characteristics measured at about 35 mK for*

*CNT tuned on resonance. d) and e) The same characteristics for both junctions tuned off resonance. Note that, there is a difference of almost two orders of magnitude between the two cases (adapted from [211]).*

In fact, since the electronic wave-functions in the nanotube are confined by the insulating contacts with the superconductors, the CNT junctions acts as a quantum dots with well-separated energy levels and can be tuned by voltage gate. When the energy level is adjusted to the superconductor Fermi energy, the CNTs are tuned on-resonance and a supercurrent flows troughs the junctions (Fig. 43b and 43c). On the other side, if the CNTs are tuned off-resonance, the critical current is much smaller (Fig. 43d and 43e). Moreover, the authors showed that it is possible to switch to a negative Josephson tunnelling by varying the back-gate voltage. In other words, the device allows to switch from 0-junction to π-junction by a gate control, making it very appealing also for quantum bit realization where π-junctions seem to exhibit a long decoherence time. In view of applications, a great advantage of this nanodevice is the possibility to achieve high magnetic coupling by placing the single molecule just on the nanotube. In fact, their comparable cross section of about $1nm^2$ allows the SQUID loop to collect up to the half amount of the magnetic field produced by the single nanoparticle. Furthermore, it is possible to tune the coupling between the sample and the sensor, by switching on and off the critical current via the lateral gate control.

Another example of superconducting quantum device acting as a "self-probe" to investigate the innovative material is the SQUID based on graphene junctions [214]. The possibility to realize a Josephson junction based on a graphene barrier was investigated in previous works [215,216]. Such a SQUID can be used as a tool to investigate the CPR of graphene Josephson junctions, transport mechanisms and the scattering length in graphene single layer. Due to "relativistic" band structure of the graphene [217], the investigations of electrical properties are very interesting and useful to understand the underlying physic of this material. In addition, if employed as a magnetic sensor, the carbon surface facilitates the chemical attachment of nano-magnets. The SQUID reported in [214] consists of *Al/Pd* rectangular loop interrupted by a single graphene sheet having a width of *4 μm* (Fig.43a). The distance between the *Al/Pd* electrodes (*50 nm*) sets the length of the junctions. The thickness of *Al* and *Pd* films were *50* and *3 nm* respectively. The *Pd* thin film ensures a low contact resistance to the graphene sheet. Characterized at *T=20 mK*, the device exhibited a critical current of about *10 μA* and a critical current modulation depth more than 90% of $I_c$, indicating a slight deviation of the CPR from the sinusoidal shape.

In the framework of development of micro and nanoSQUIDs operating at high magnetic fields, Mandal et al. developed a SQUID based on diamond [218]. This material when doped with boron above a critical level (0.25 atom %) shows superconducting properties with very high critical field [219,220]. The diamond SQUID, was formed by a (*2.5×2.5*) *μm²* loop including two nanobridges having a wide of *100 nm* and a length of *250 nm* (Fig. 44b). The fabrication procedure was based on the fabrication of a boron doped diamond film (*300 nm* thick) by microwave plasma-enhanced chemical vapour deposition. The films were patterned by EBL technique and oxygen plasma etching. Even if the critical temperature of the diamond superconductor film was about *3 K*, the device was tested at *400 mK* or at *40 mK* for high magnetic field measurements. The *I-V* characteristics were hysteretic and the critical current modulation depths were about 15% of $I_c$. The magnetic flux noise estimated by the current switching distribution (equation 18) was *50 $\mu\Phi_0/Hz^{1/2}$*. The characterization reported by the authors for perpendicular applied magnetic field in the range *0.2-4 T*, showed that the critical current decreases and the *I-V* characteristics became non-hysteretic (for *B>200 mT*) allowing to measure also the *V-Φ* characteristics. Voltage oscillations in the full range of applied magnetic field were observed, but their amplitudes were lower than *1 μV* giving a poor magnetic flux sensitivity if the device is employed as a voltage transducer.

Following the same fabrication procedure of Cleuzou et al. for the CNT SQUID, Spathis et al [221] developed another interesting nanodevice based on indium arsenide (*InAs*) nanowires and fabricated by EBL technique. The SQUID consisted of a vanadium-superconducting loop including two *InAs* weak links obtained by a single nanowire contacted by vanadium electrodes (Fig.44c).

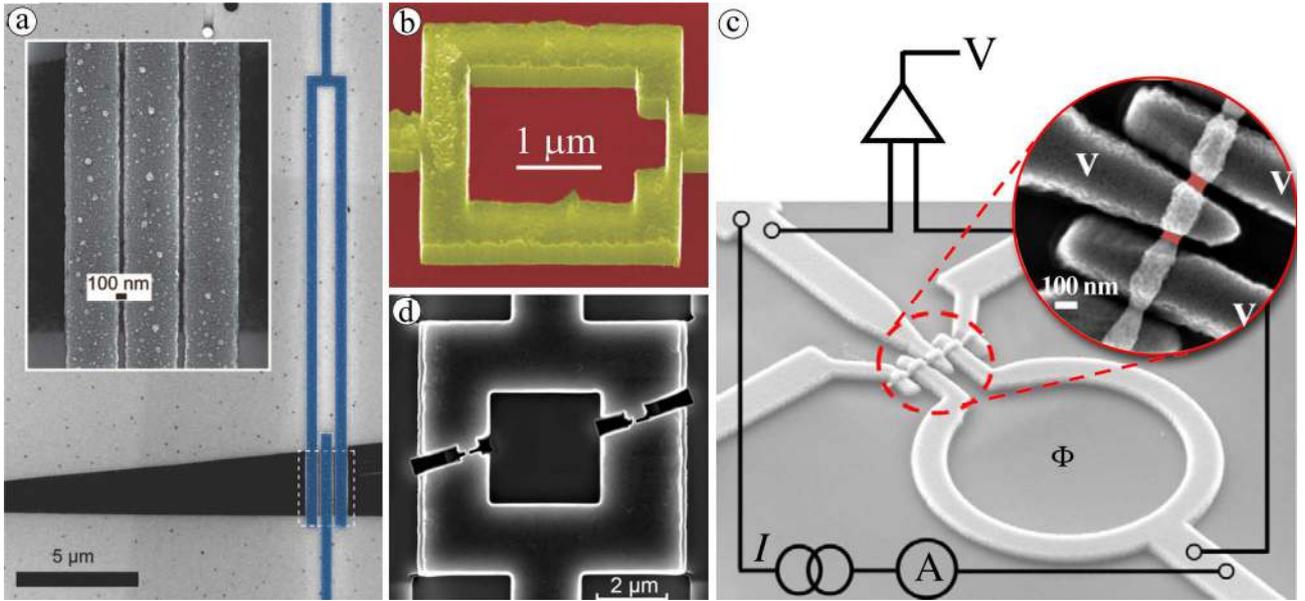

*Fig. 44 SEM images of microSQUIDs based on: a) Josephson nano-junctions with graphene barrier; b) boron-doped diamond; c) InAs nanowires; d) $MgB_2$ nanobridges (adapted from [214], [218], [221], and [226] respectively).*

The SQUID loop had a geometrical area of *12 μm²*, whereas the two nanowires had a diameter of *90 nm* and a length ranging between *20 nm-50 nm*. The devices were characterized at *244 mK*. The critical current ranged from *150 nA* to *350 nA* and the *I-V* characteristics showed no hysteresis up to a critical current of *250 nA*. It can be ascribed to a heating effect occurring when the device switch in the resistive branch. Despite the low critical current, due to high normal resistance (*200-300 Ω*), the product $I_cR_n$ was about *100 μV*. The critical current modulation depths were as high as 95% of $I_c$ indicating a sinusoidal CPR. However, both the voltage swing and voltage responsivity were lower than *10 μV* and *50 μV/$\Phi_0$* respectively, leading to a magnetic flux noise greater than *25 μ$\Phi_0$/Hz$^{1/2}$* if operating with a standard read-out electronics. Since the spin-orbit of *InAs* nanowire is strong, these device are very attractive for the investigation of the Majorana fermions [222].

At beginning of the new century (2001), a new type of superconductor was discovered, that is the *$MgB_2$* [223]. In view of applications, the main advantage is the relative long coherence length that weakens the grain boundary effect leading to a large critical current density. Moreover, as seen, the fabrication of SQUID based on nanobridges benefits by long coherence length. Therefore, since its discovery, several SQUID sensors based on *MgB₂* have been developed [224,225]. Micro-sized *MgB₂* SQUIDs based on nanobridges (*100 nm* in width) with a loop area of about *10 μm²* have been fabricated by using a FIB technique [226,227] (Fig.44d). The *I-V* characteristics were non-hysterestic in the full range of explored temperature [*4.2 K-33 K*]. The *V-Φ* characteristics showed a sinusoidal like behaviour and high amplitudes (*30 μV* at *T=33.5 K, 80 μV* at *T=15.2 K*, and *130 μV* at *T=5.9 K*). Unfortunately, magnetic flux noise measurement were not reported.

## 4. NanoSQUID applications

In this section, we will report the measurements techniques and the main applications of nanoSQUIDs. They include the nanomagnetism, the ultra high spatial resolution scanning microscopy, the nanoscale electromechanical system resonator, single photon detection, nanoelectronics and quantum computing.

### 4.1 Measurements methods

#### 4.1.1 Switching current measurements and cold mode

As seen in the previous sections, in many cases the nanoSQUIDs show a hysteretic *I-V* characteristic preventing the use of standard readout electronics based on the low noise voltage amplifier and the FLL circuitry to increase the linear dynamical range. However, as reported in the section 1.3, in the case of hysteretic *I-V,* the sensor can be used as magnetic flux to current transducer. In this mode, the switching currents are measured to obtain magnetic flux by dividing the critical current variation by a suitable transfer factor. This method has been extensively employed in many experiments [23, 106-107, 172-176]. Due to the thermal and quantum fluctuations, the critical current is a stochastic variable, and so it is affected by an intrinsic uncertainty. In order to measure effectively the switching current, a time of flight technique is typically used [228, 229]. For this aim, the SQUID is current biased with a triangular-shaped waveform at frequencies ranging from *100 Hz* up to *10 kHz*. The synchronism of the ramp generator is delayed and sent to start input of a time-to-amplitude converter (TAC). The device voltage is sent to a discriminator that provides the stop signal for the TAC at the time of the switching out of the zero voltage state. The TAC generates a voltage signal proportional to the time elapsed between the start and the stop inputs that is sent to an analog to digital converter. In this way, a count is assigned to the channel corresponding to the switching time. Several thousands events are to be recorded in order to get a suitable experimental histograms of the switching current. The $I_c$ measurement, for each exciting magnetic field value, is obtained by averaging the switching current values. In such a way it is possible to measure the switching current with an uncertain of about *1* part in *$10^4$*, resulting in an overall noise of the order of *$10^{-4}$ $\Phi_0$* [174-176] or a spectral density of magnetic noise of *10 $\mu\Phi_0/Hz^{1/2}$* [23]. A drawback of the averaging is the reduced bandwidth of the measurement set-up making this measurement technique suitables for slow or quasi-stationary signals.

It is well know that the Joule heating, occurring when the SQUID or a Josephson junction switch in the resistive state, can be a limiting factor for several experiments requiring an extremely low temperature (*mK* range) or a very low dissipation level such as macroscopic quantum tunnelling (MQT) investigations. In order to study the temperature dependence of the magnetization reversal and the MQT of magnetization, a suitable measurement technique called cold mode technique was developed [23, 106, 230]. The nanoSQUID is biased with a constant current close to the critical value while an orthogonal magnetic field provides a magnetic flux bias in a high responsivity point of the *I-Φ* characteristic. When the magnetization reversal of the particle occurs, the magnetic flux trough the SQUID causes the transition to the resistive state. Since the pulse voltage can be measured few nanoseconds after the magnetization reversal, a very precise switching field measurements can be performed. The nanoparticle is heated only after the reversal of magnetization (cold mode). As concern as the MQT of magnetization [231] the advantage of this method lies in the weak coupling between the nano-object under investigation and the nanoSQUID acting as a readout system. However, it worth to point out that the cold mode allows to measure only the switching field of magnetization reversal and not the magnetization before and during the magnetization reversal.

## 4.1.2 Feedback mode for increase the linear dynamic range

Due to non linearity of SQUID transfer function (*I-Φ*), the above mentioned method fails in the case of magnetic flux variation larger than $\Phi_0/2$. It occurs when a strong magnetic signal arises from nano-objects or when a high excitation magnetic field is required as in the magnetization curve measurements. In any case, data treatment is needed to extend the linear range of the SQUID output. To this aim, suitable feedback circuits have been implemented [23, 170, 174, 232]. The basic scheme, reported in the Fig. 45, features the standard flux locked loop mechanism. The variation of nanoparticles magnetization induces a change of the magnetic flux into the SQUID loop, which results in a variation of the SQUID mean critical current $\Delta I_c$. This value is used to calculate the feedback current $\Delta I_F$ to send into an integrated coil producing a counter magnetic flux into the nanoSQUID in order to keep the switching current constant at a working point. So, as in the standard FLL, the flux bias point is locked and the variation of the magnetic flux is given by *ΔΦ=ΔI_F//I_{Φ,F}*; where $I_{\Phi,F}$ is the feedback coil sensitivity $I_{\Phi,F}=\partial I_F/\partial \Phi$, that is the feedback current required to induce one flux quantum in the SQUID. It can be easily evaluated by sending a current in the feedback coil and measuring the corresponding flux quanta coupled in the SQUID loop. This procedure is reiterated for each excitation field value and a customized data analysis program sums the measured magnetic flux variations. A solenoid coil can be used to provide a uniform exciting magnetic field $B_{ex}$ in the plane of the nano-sensor. However, a small misalignment between the excitation magnetic field and chip surface is unavoidable and it is responsible for a parasitic magnetic flux threading the SQUID loop, proportional to the excitation magnetic field. Since, the exciting magnetic field can be very high (up to few T), the residual normal component may be greater than the signal arising from the magnetic nanoparticles and, therefore, its contribution to the magnetic flux has to be suitably subtracted. To this aim, a nanoSQUID without MNP acting as a reference can be integrated on the same chip to measure the parasitic magnetic flux $\Phi_p=B_p A_S$ to be subtract ($A_S$ is the effective nanoSQUID area and $B_p$ the normal component of the exciting magnetic field) [132]. Since $B_p=B_{ex} \sin(\alpha)$, the misalignment angle $\alpha$ can be evaluated by $\alpha=\arcsin(\Phi_P/(A_S \cdot B_{ex}))$.

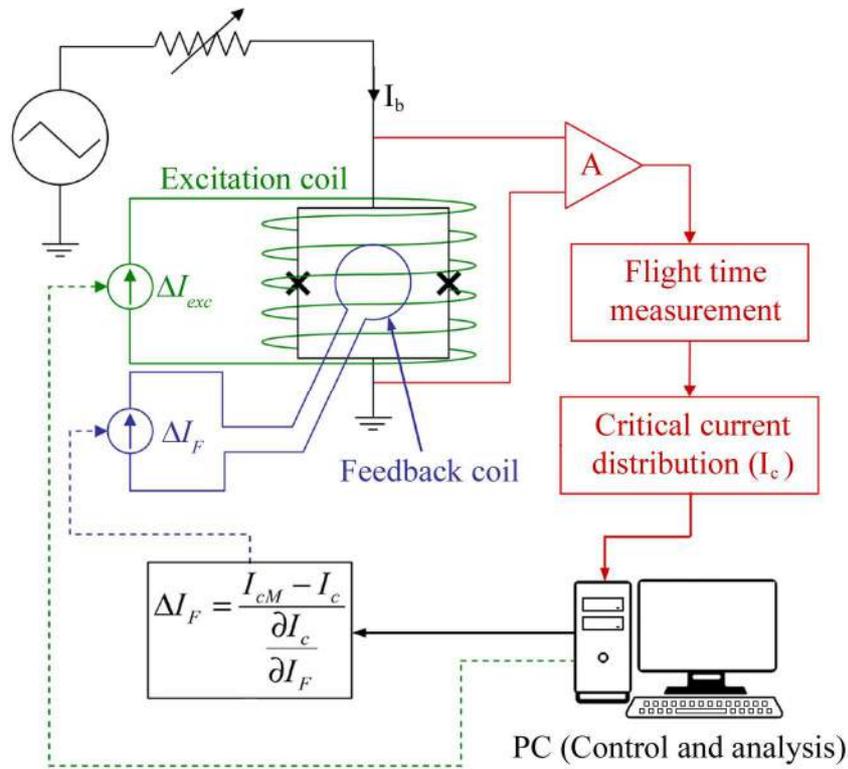

*Fig. 45 A Scheme of the experimental set-up to increase the dynamical linear range of a hysteretic nanoSQUID during the nanoparticle magnetization measurements. In order to reduce the stochastic noise, the variation of the critical current can be obtained by averaging the switching current events measured by using a time of flight technique [174-176].*

As alternative, a fine alignment system based on a 3D vector magnet can be used [170]. Such a system allows a precise alignment of the magnetic field in the SQUID plane by combining field generated along the three axis of magnet ($\mathbf{B}=B_x\mathbf{i}+B_y\mathbf{j}+B_z\mathbf{k}$).

### 4.1.3 Attachment and manipulation of magnetic nano-objects on a nanoSQUID

The positioning of the magnetic nano-object on a SQUID is an important issue. In fact, as seen in the previous sections the SQUID response depends on the position of the sample within the loop. In particular, the highest magnetic coupling is obtained for sample positioned directly on the nanobridge or the loop edges.

Several techniques are used to place the nano-objects on the SQUIDs. The simplest technique consists in depositing onto nanoSQUID chip a single drop of a solution containing the magnetic nanoparticles (MNPs) dispersed in a solvent. The drop is left to dry leaving the nanoparticles stuck on the chip due to Van der Waals force [23, 174-176, 232]. If a suitable pipette is used, it is possible deposit a drop of about *1 μl* corresponding to about *1 pg*. The position of the nanoparticles respect to the SQUID loop can be verified by scanning electron microscopy. As expected, in this case, the distribution of the MNPs is random and it is not possible to put one or some MNPs in a desiderate position. Therefore, after the magnetization measurements, the position and size of the nanoparticles are determined by scanning electron microscopy.

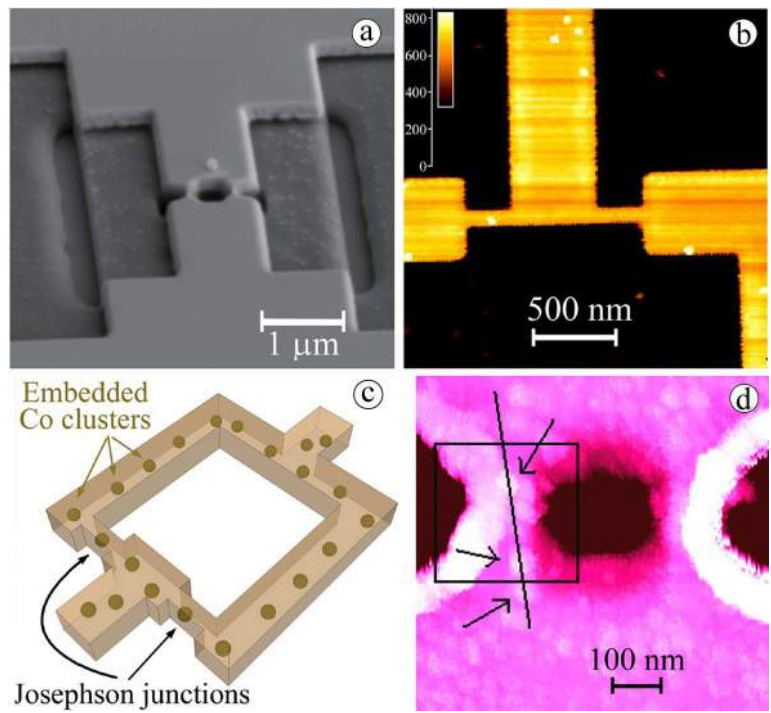

*Fig. 46 a) Image of a nanoSQUID with a FePt nanobead positioned by SEM technique (adapted from [235]). b) Metallic aerosol particles positioned onto nanobridges by AFM technique (adapted from [23]). c) Sketch of a nanoSQUID with nanoparticles embedded within his body [107]. d) AFM image of a nanoSQUID showing ferritin nanoparticles (indicated by the arrows) located onto nanobridges by using a self-assembly monolayer and EBL (adapted from [242]).*

Another simple method to deposit MNPs on the sensitive area of the nanoSQUID is the electrospray or spray technique [23], which has already successfully used to produce quantum dots and nanoparticles and to deposit thin film of biological, organic and inorganic materials. This method could be particularly suitable for nanoSQUID based on carbon nanotubes. In addition, the chemical functionalization of the carbon nanotube and the controlled attaching of molecules with adapted ligand [234] could be another effective possibility for this hybrid nanodevice.

Accurate and reversible MNPs positioning method are obtained by using SEM [235,236] or AFM [237-239]. In the SEM case, a micromanipulator brings a sharpened carbon fibre probe close to single nano-object (nanobead or nanoparticle). In order to weld the tip to the nano-object, a burst from the electron beam deposits amorphous carbon acting as ligand. After that, the probe reaches the desiderata position (loop edge), the nanosample is attached to the nanoSUID surface by using another brief electron beam exposure and the tip is withdrawn. It is worth noting that this method is reversible, that is, after the measurement the nano-object can be removed or located in another place by using the same procedure. In addition, the use of thin carbon fibre tips is better than more traditional electrochemically sharpened metal tips, and provides an excellent way to protect the SQUID loop and particle, as the tips are very flexible. The Fig. 46a shows a nanoSQUID with a single *FePt* particle bead positioned close to the edge..

Nanoparticles can be manipulated and arranged in a very effective way also by using the AFM, in which the tip is used as tool to push particles along smooth substrates. Unlike the SEM, the AFM can be also used in the non-contact mode both for locating and moving nanoparticles. The advantage of this method is the possibility of seeing the nano-object moving in real time. In addition, the apex wear and the tip degradation due to particle sticking are avoided [238, 239]. In the Fig.46b, a picture of a micro-SQUID with metallic aereosol nanoparticles manipulated by using an AFM is reported.

Another effective method consists in embedding the MNPs directly in the nanobridge [107,240,241]. In particular, the cluster of nanoparticles are simultaneously deposited together with niobium film in ultra high vacuum condition, so that the MNPs are buried in the superconducting film (Fig.46c). This method improves the magnetic flux coupling and protect the MNPs against the oxidation. It has been successfully employed to perform magnetic measurements on cobalt and iron nano-cluster containing up to 1000 atoms [240, 241]. In these experiments the cobalt and iron cluster were produced by a laser vaporization and inert gas condensation while a *15-20 nm* thick *Nb* film was deposited by electron gun evaporator. Afterwards, an EBL provided the pattern of both loop and nanobridge junctions. Since the SQUID loops had micrometric size, only the clusters embedded in the nanobridge produced an appreciable magnetic signal.

An original method to attach MNPs on a nanoSQUID was developed by Lam et al [242]. They developed a technique for attaching ferritin-based nanoparticles on the SQUID using self-assembly monolayers and electron beam lithography. The self-assembly monolayers are monomolecular layers which are spontaneously formed after a suitable chemical procedure. They act as linker molecules allowing the binding between the nanoSQUID surface and the ferritin nanoparticles. In particular, to attach ferritin molecules to the *Au* thin film coating the niobium SQUID, a self-assembled monolayer of sulphur-terminated linker molecules was used. Before the nanoparticle attachment, the nanoSQUID was covered with a PMMA layer and an EBL pattern was performed in order to open nanometric windows ($200 \times 200$ *nm²*) on the top of the nanobridges. In such a way, a local attachment of the linker molecules and ferritins in the uncovered area was obtained (Fig.46d).

**4.2 Nanomagnetism applications**

The main application of nanoSQUID is in nanomagnetism, which is the study of small magnetic systems including: nanoparticles, nanobeads, nanotube, nanocluster etc. The interest for this branch of condensed physics goes back to the end of 1940s, when the pioneering works of Neel [243], hypothesized the measurement of the magnetic properties of the individual nanoparticle. Recently, there is a growing interest for magnetic nanoparticle applications in biology and nanomedicine, as well as for the study of underlying physics. In particular, the measurements of magnetic relaxation process is very useful for both basic physics investigations like the measurements of the anisotropy constant [244], quantum tunnelling of magnetization [107] and for drug delivery applications or immunoassay techniques [245]. Among the several techniques tools employed to investigate the magnetic nano-objects [14-21], those based on nanoSQUIDs allow the most detailed and precise study of magnetic objects at nanometric scale. In fact, as above mentioned, nanoSQUIDs exhibit an ultra high magnetic moment sensitivity, approaching to the single Bohr magnetons per unit of bandwidth.

In this section, we will address the main nanomagnetism measurements, performed by using micro and nanoSQUIDs. We will start with some of the pioneering works carried out by the group of Institut Nèel in Grenoble which successfully employed *Nb* micro-sized SQUIDs (*0.5-2.0 μm* in diameter) based on nanobridges to perform very interesting and useful nanomagnetism investigations. Many of their experiments were focused on the investigation of the magnetization reversal of magnetic nanoparticles (that is the reversal of its magnetic moment), which is of great relevance for applications in spintronic and basic physics. In the case of a single-domain nanoparticle, the model of uniform rotation of the magnetization (Stoner–Wohlfarth model) can describe the magnetization reversal [246]. The basic assumption of this model can be summarized as follow. In a particle of an ideal magnetic material, the exchange energy holds all spins tightly parallel to each other. In this case, the exchange energy is constant and it plays no role in the energy minimization. Hence, there is competition only between the anisotropy energy of the particle and the effect of the applied field. In the framework of this model, the potential energy of a particle is given by:

$$E = KV\, sin^2\, \phi - \mu_0 M_s V H\, cos(\phi - \theta) \qquad (42)$$

Where *KV* is the uniaxial anisotropy energy which depends on the shape of the particle, *V* is its volume, $M_S$ is the spontaneous magnetization and *H* the magnitude of the applied field; finally, ϕ and *θ* are the angles of the magnetization and the applied field respectively, with respect to the easy axis of magnetization. The energy potential has two wells corresponding to the two stable orientations of the magnetization (Fig. 47a). For given values of *θ* and *H*, the magnetization is oriented along an angle ϕ that locally minimizes the energy. When a magnetic field is applied, the height of the energy barrier decreases and one of the two wells can become metastable. The magnetization reversal is defined by the minimal field value at which the energy barrier between the metastable minimum and the stable one vanishes. This process is analogue to the supercurrent decay in a Josephson junction or the decay from the metastable flux statesin a rf SQUID; also in these cases, there is multi or double well potential and the escape process occurs when the bias current or the bias magnetic flux reaches the critical values [228, 247, 248]. Therefore, the magnetization reversal in a nanoparticle is a stochastic process characterized by thermal and quantum fluctuations. If the thermal energy is low enough (low temperature), the quantum tunnel through the potential barrier can occur as in the Josephson devices.

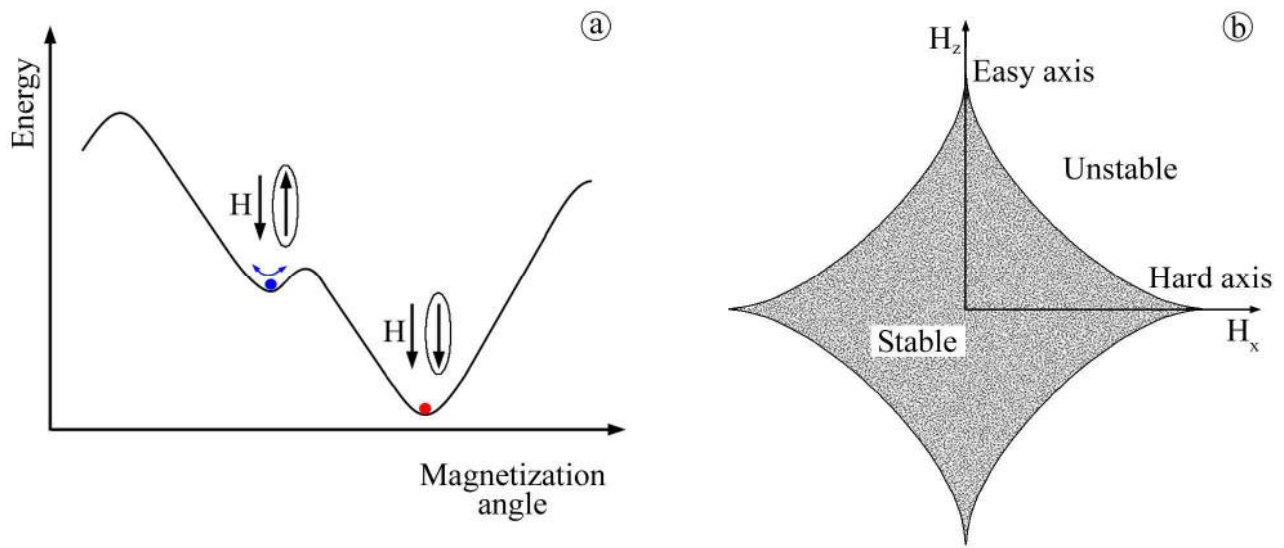

*Fig. 47 a) Schematic draw of the potential energy of a magnetic nanoparticle as a function of the magnetization angle. The potential energy has two wells corresponding to two stable orientations of the magnetization. b) Field dependence of the switching field (Stoner–Wohlfarth astroid). The magnetization switches from the metastable well to the stable one when the applied field crosses the Stoner–Wohlfarth astroid.*

The curve relative to switching field as a function of the direction of the applied field (Fig. 47b) is called the Stoner–Wohlfarth astroid and provides information about the stability of the nanoparticle magnetization. This topic is exhaustively described in a Werndorfer's review article [107].

The first magnetization measurements of individual single-domain nanoparticles, at very low temperatures, by using a SQUID were presented by Wernsdorfer et al. [249]. They used a *Nb* micro-bridge dc-SQUID fabricated by EBL while the nanoparticles (*Co, Ni, TbFe$_3$,* and *Co$_{81}$Zr$_9$Mo$_8$Ni$_2$*) were fabricated by sputtered thin films and defined by either ion beam etching or lift-off technique. The nanoparticle clusters were deposited on the SQUID loop using a PMMA mask. The external field was applied in the plane of the SQUID, thus only the flux induced by the stray field of the sample magnetization was detected. They measured, at *T=0.2 K*, both hysteresis loop and switching magnetic field by using the SQUID as magnetic flux to current converter and as a switching detector respectively. In the Fig. 48a, typical hysteresis loops for *Co, Ni* and *Co$_{81}$Zr$_9$Mo$_8$Ni$_2$* particles with dimension of either (*200×100*) nm$^2$ or (*100×50*) nm$^2$ and thickness between *8* and *30 nm* are reported.

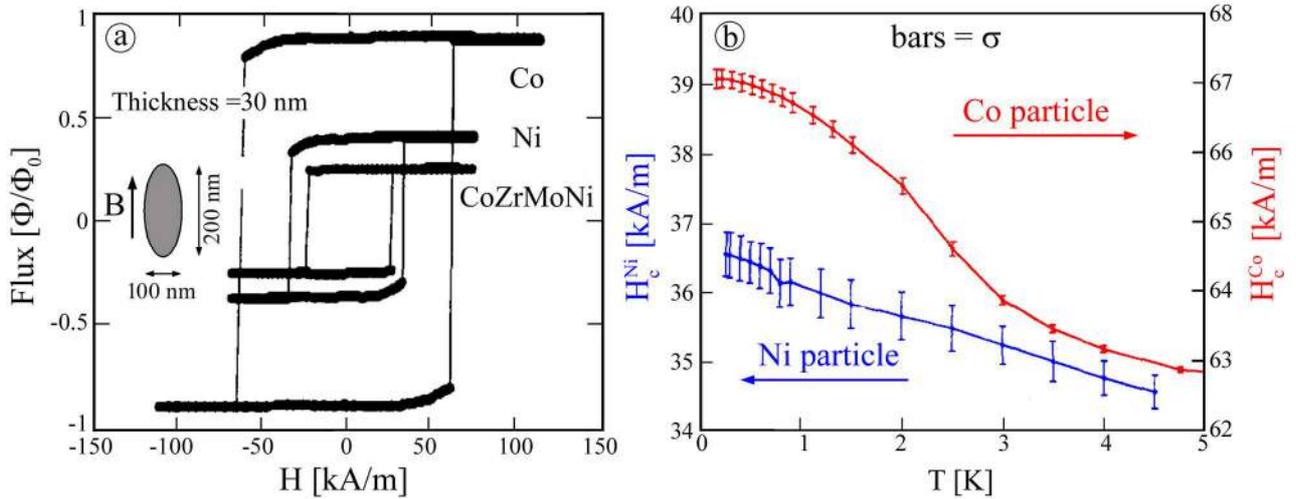

*Fig. 48 a) Hysteresis loops relative to different nanoparticles, measured by a microSQUID at T=0.2 K. b) Variation of the mean switching field versus temperature for both Ni and Co nanoparticles (adapted from [249]).*

The magnetization reversal takes less than *100 µs* and its behaviour is close to the magnetization of a single domain particle described by the Stoner–Wohlfarth model. The variation of the mean switching field $H_c$, of *Ni* and *Co* nanoparticles as a function of temperature is shown in Fig.48b. The vertical bars indicate the width of the switching field distribution.

Soon after, magnetization measurements with a sensitivity of about 10$^4$ µ$_B$ were performed, allowing the first studies of magnetization reversal of a small number of crystalline single-domain *Co* clusters of *2-5 nm* in diameter [250].

Later, the magnetization measurements of isolated *Ni* nanowires, with a diameter ranging from 40 and 100 nm, and of individual ferromagnetic nanoparticles (ellipsoidal *Co* particles with a diameter of *25 nm*) were successfully performed [251,252] at temperature between *0.2 and 6.0 K*.

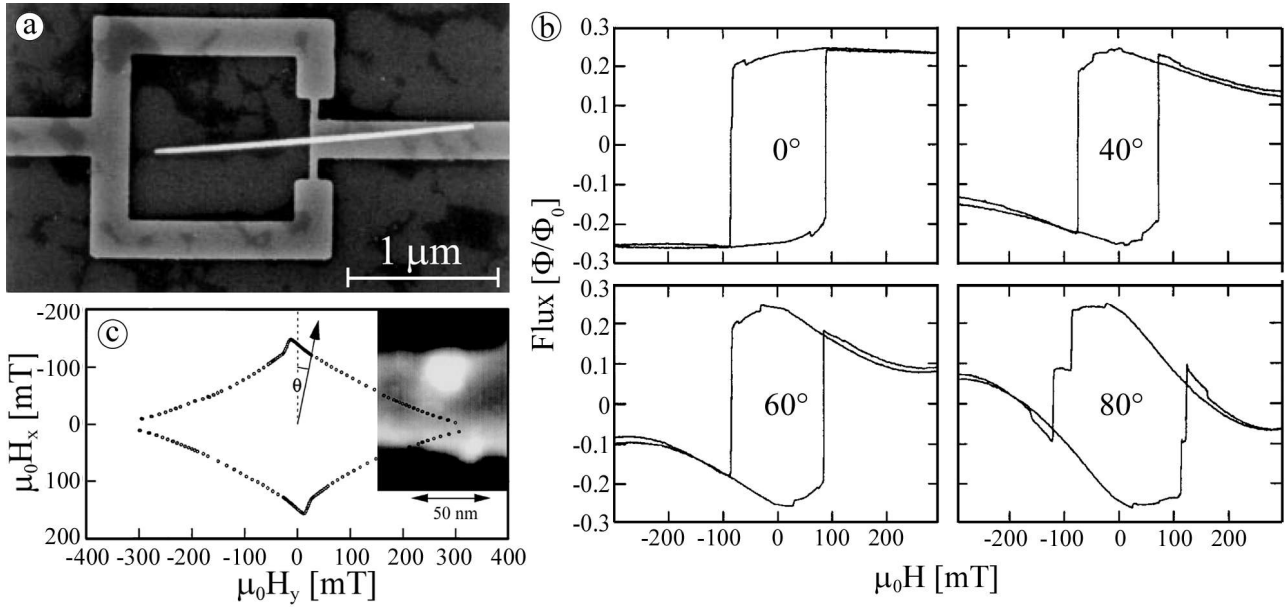

*Fig. 49 a) SEM images of microSQUID with a Ni nanowire having a diameter of 65 nm (adapted from [251]). b) Hysteresis loops of the Ni wire at several values of the angle between the applied field and the wire axis (adapted from [251]). c) Angular dependence of the switching field of an ellipsoidal Co nanoparticle (25 nm in diameter) deposited on a microbridge; the inset shows a cobalt nanoparticle (white spot) on the nanobridge (adapted from [252]).*

The nanowires and the nanoparticles were dispersed in a chloroform and ethanol solvent respectively and a single drop was placed on a chip containing about one hundred SQUIDs (see section 4.2). In order to study the stochastic nature of the switching field, its distribution and switching time were measured for both nanowire and *Co* nanoparticles. Fig. 49a shows a SEM image of a microSQUID with a single *Ni* nanowire. The hysteresis loops, measured by a microsQUID for different angles (*θ)* between the magnetic field and the wire axis, are reported in Fig. 49b. The Fig.49c shows the angular dependence of the switching field for an ellipsoidal cobalt nanoparticle. These measurements provided very useful information about the process of magnetization reversal in both nanowire and nanoparticles. In particular, the data analysis showed that, in the explored temperature range (*0.2-6.0 K*), the magnetization reversal of a well-prepared ferromagnetic nanoparticles can be described by thermal activation over a single-energy barrier as stated by Neel-Brown model [253]. Also for smallest nanowires, the measurements revealed a thermal activated switching following the Arrhenius law according with the aforementioned model. In a successive experiment, Wernsdorfer et al. [254] carried out switching field measurements of individual ferromagnetic *BaFeCoTiO* (*10-20 nm*) nanoparticles in the temperature range of *0.1-6 K*. A cold mode technique was used in order to minimize the dissipation effects of the SQUID readout (see section 4.2). For *T<0.4 K* a strong deviation from Neel-Brown theory was observed, which was quantitatively explained by the theory of macroscopic quantum tunnelling of magnetization without dissipation. In particular, they observed a thermal saturation of the switching field distribution width indicating a saturation of the escape rate.

Later on, Bonet et al. [255] performed three-dimensional measurements of switching fields of *BaFeO* nanoparticles in the size range of *10-20 nm*, by using an improved SQUID technique, in which, at first, the SQUID was switched off and a field was applied in an arbitrary direction, which may or not cause a magnetization switching. Then, the SQUID was turned on and a second field was applied in the plane of the SQUID to probe the resulting magnetization state. In such a way, the entire field space can be scanned. Thanks to this improved technique, Jamet et al. investigated the magnetic anisotropy in a single *1000*-atom cobalt cluster [240,241]. They measuring the three-dimensional diagram of the magnetization switching fields of a 3 nm cobalt cluster directly buried within the superconducting film of a micro-SQUID. The data analysis demonstrated that magnetic anisotropies of such individual nanoparticles was dominated by surface anisotropy.

Therion et al. [256] proposed a new method to overcome the restrictions due to high magnetic fields required to reverse the magnetization of a nanoparticle. They applied a constant magnetic field well below the switching field combined with a radio frequency (RF) field pulse. When the frequency of RF field matches the precession frequency of the magnetization, energy can be pumped into the system leading to magnetization reversal from the metastable to the stable well. The effectiveness of this method was demonstrated on a *Co* nanoparticle having a diameter of *20 nm*. The Fig. 50b shows a SEM image of a SQUID based on nanobridge upon which two *Co* nanoparticle were placed. The two dimensional switching field maps for static (black line) and for different frequencies (colour lines) are reported in the Fig. 50c. The reversal magnetization occurs only for field outside these curves (Stoner–Wohlfarth astroids). In specific field regions, the switching field is strongly reduced by the RF pulse. It is possible to achieve a switching field reduction of about *100 mT* with an RF pulse amplitude of few *mT* (at 4.4 GHz).

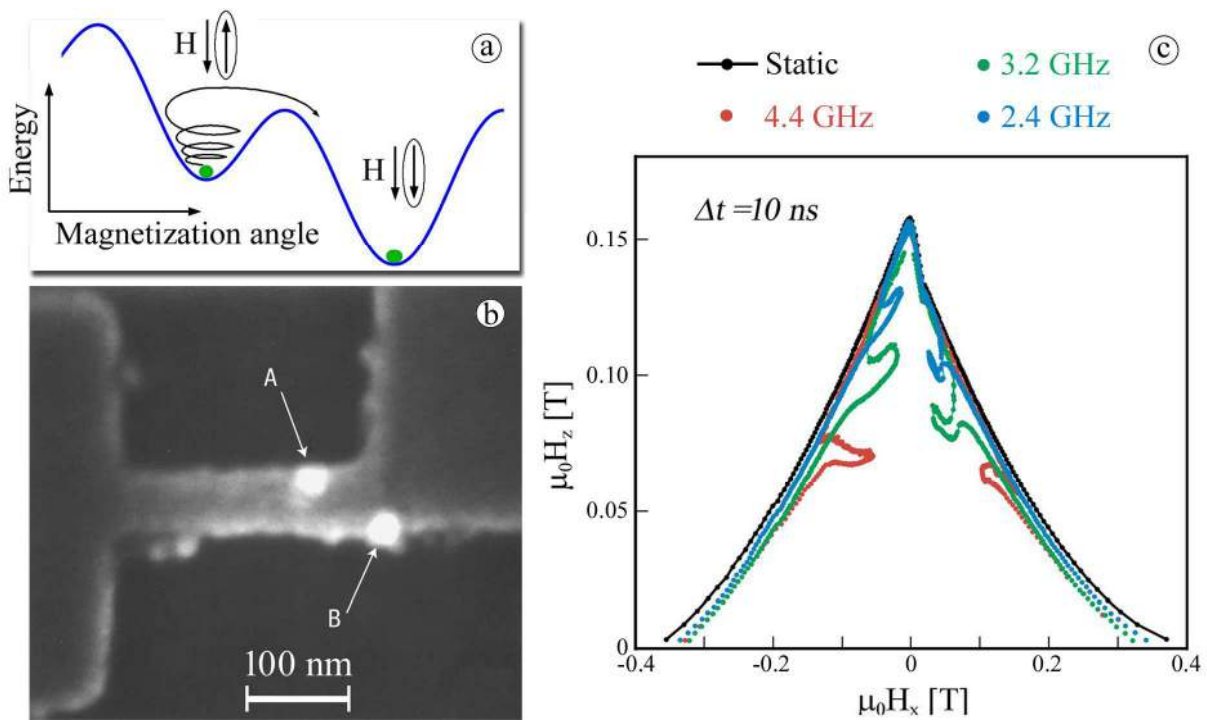

*Fig. 50 a) Potential energy of a magnetic nanoparticle versus magnetization angle. An additional alternative field induces oscillations of the magnetization in the energy wells. b) SEM image of Co nanoparticles (white spot) on the SQUID nanobridge. c) Stoner–Wohlfarth astroids for different RF magnetic pulses. This technique allows to reduce drastically the switching fields (adapted from [256]).*

It is worth to mention the use of arrays of microSQUIDs to investigate the magnetic properties of macroscopic samples by measuring the magnetic field arising from its magnetization.

When a SQUID of the array is very close to the sample, it locally measures the magnetization reversal whereas when the SQUID is far away, it integrates over a bigger sample volume. Such device type has been successfully employed to investigate array of magnetic dots, crystal of magnetic molecular cluster and nucleation and depinning of magnetic domain walls in thin films [257,258].

Another interesting nanomagnetism application of microsized SQUIDs was proposed by Martinez-Perez at al. [259], that used a microSQUID based on Josephson tunnel junctions to measure the susceptibility of a single layer of nanomagnet dots deposited on the most sensitive SQUID area by dip pen nanolithography [260]. The nanomagnets consisted of cobalt oxide nanoparticles (*12 nm* in diameter) synthesized inside the protein nanocavity of horse spleen apoferritin. The estimated magnetic moment of a single nanoparticle was about *12 $\mu_B$*. They measured the susceptibility of a nanoparticle monolayer corresponding to about $10^7$ molecules down to *13 mK* and found that each molecule preserved its magnetic properties.

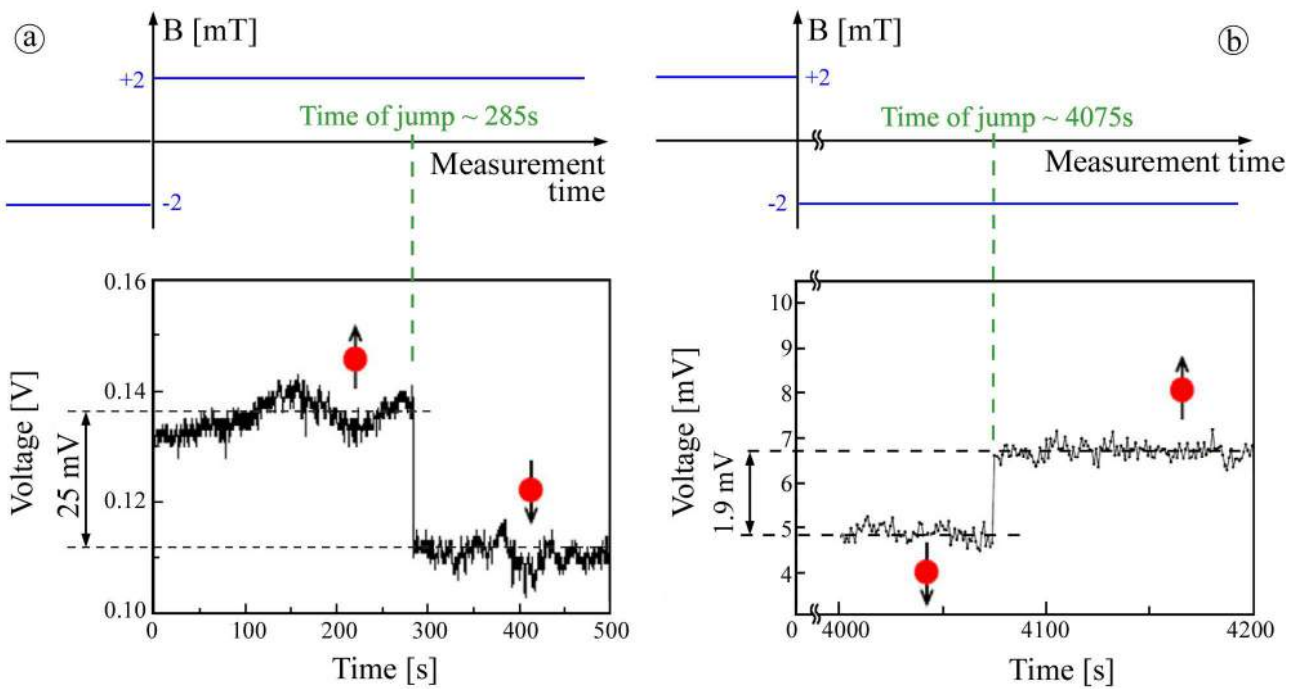

*Fig. 51 Measurements of magnetization reversal of ferritin nanoparticles attached to the gold coating of a nanoSQUID having a flux capture area of about 0.04 µm². The variation of applied magnetic field was ± 4 mT. The magnetization reversal occurs after 285s and 4075 s for a) and b) respectively (adapted from [242]).*

The first use of a SQUID with a nano-sized loop to measure the magnetic response of magnetic nanoparticles was done by Volarick and Lam [145]. They employed a gold-shunted nanoSQUID (similar to that reported in Fig.21), to detect the magnetization reversal from ferritin nanoparticles, obtained by horse-spleen. The single nanoparticle had a diameter of *8 nm* and a magnetic moment of *300 $\mu_B$*. A protein shell having a diameter of *12 nm*, surrounded the nanoparticles core keeping them separated, so that, the nanoparticles behave as isolated nanomagnets.

The nanoparticles were attached to the gold overlay above one of the nanobridge by using the technique of chemical self-assembly described in the section 4.1.2 [242]. They applied a magnetic field of approximately *–2 mT* or *2 mT* perpendicular to the nanoSQUID plane and recorded the SQUID output voltage for a total time ranging from *0.5 to 96 h*. When the magnetization reversal occurs a voltage jump was observed (Fig. 51). The amplitudes of corresponding magnetic flux coupled to the SQUID, obtained by the ratio of the voltage jumps and the voltage responsivity, were appreciably greater than the intrinsic noise of nanosensors. The authors repeated the same measurements using nanoSQUIDs with no nanoparticles or with the linker molecules without ferritin attached. The measurements did not show any flux jumps confirming that their origin was due to the magnetization reversal of the ferritin nanoparticles. However, the simulation performed by the authors provided an upper limit for of the magnetic flux change into the SQUID due to the flipping ferritin moments. So that only about half of the observed jumps can be ascribed to the magnetization reversal of ferritin nanoparticles.

By using the nanoSQUID shown in the Fig. 28, Russo et al. [174] and Granata et al. [190, 232] investigate the magnetic properties of iron oxide nanoparticle having a diameter ranging from *4 to 8 nm*. The nanoparticles were synthesized by thermal decomposition of metalorganic precursors in the presence of oleic acid and oleylamine as surfactants and organic solvent with high boiling point [174]. A solenoid surrounding the chip supplied the excitation field co-planar to the SQUID. They measured the field dependence of magnetization at *T=4.2 K* (Fig.52 a and b) for two different sizes of nanoparticles. The magnetic hysteresis loop indicate that the blocking temperature is well above 4.2 K for both nanoparticle types. The sigmoid shape of the virgin curve (Fig. 52b) suggests the presence of dipole–dipole interparticle interactions, which tend to resist the magnetization process. The applied field corresponding to the change in curvature of the virgin curve are about *100* and *350 Gauss* for smallest and biggest nanoparticle respectively. They correspond to the magnetic fields amplitude needed to overcome the magneto-crystalline anisotropy as well as the interparticle interactions. The hysteretic loops reported in the Fig.51a, show a coercive field $H_c \cong 290\ Gauss$ for *8 nm* MNPs diameter and a $H_c \cong 100\ Gauss$ for *4 nm* MNPs diameter indicating an increase of anisotropy as the particle size increases.

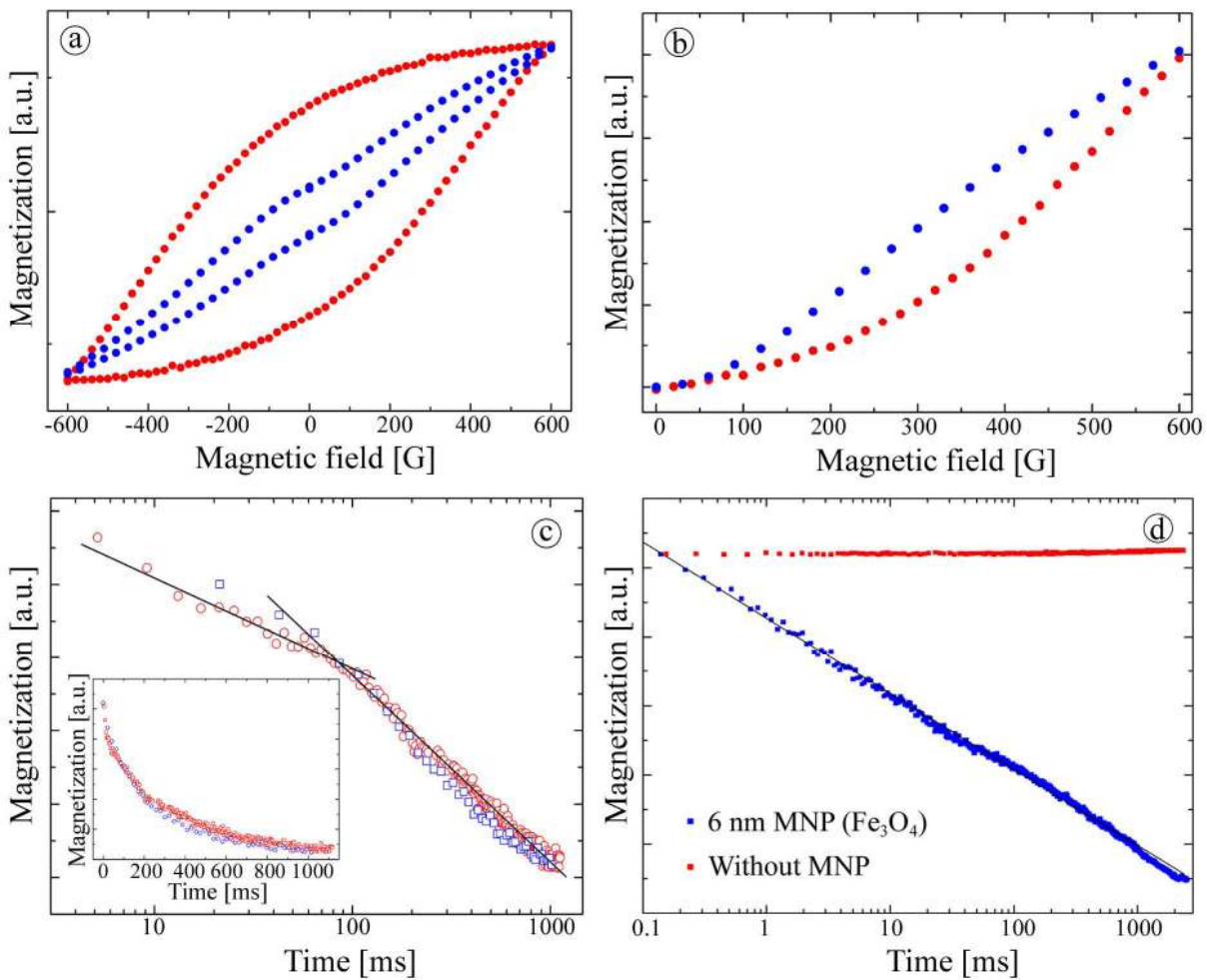

*Fig. 52 a) Magnetic field dependence of magnetization for iron oxide nanoparticle having a diameter of 4 nm (blue dots) and 8 nm (red dots) measured at T = 4.2 K by using a nanoSQUID shown in Fig. 28 b) first magnetization curve (adapted from [232]). c) Magnetic relaxation measurement at T=4.2 K for 8 nm iron oxide nanoparticles diameter (red circle) compared with a*

*measurement performed by a commercial SQUID instrumentation (blue squares). The inset shows the same measurements in a linear scale (adapted from [174]). d) Magnetic relaxation measurements for a 6 nm iron oxide nanoparticle performed by the 3D nanoSQUID shown in Fig. 35. The blue square dots refer to the same measurement without nanoparticles (adapted from [190]).*

Usually, magnetic anisotropy increases as particle size decreases due to the increase of surface component of anisotropy. The authors provided an explanation of this unexpected behaviour ascribing it to the presence of oleic acid, covalently bounded to the particle surface. Relaxation measurements at *T=4.2 K* for MNP having a diameter of 8 and 6 nm are reported in the Fig. 52 b and c. The nanoparticles were cooled in a magnetic field of *10 mT* from *300 K* to *4.2 K*, then the magnetic field was switched off and the remnant magnetic moment was measured for approximately *1000* seconds. The data were compared with those obtained by using a commercial system Quantum Design SQUID Magnetometer (Fig. 52c, blue squares), confirming the effectiveness of the nanoSQUID measurements. In addition, the nanoSQUID was capable to analyse the magnetic relaxation behaviour in short time regime. This feature, unlike the commercial set-up, allows to point out the slower magnetic relaxation for short times with respect to that for longer times. The relaxation measurements reported in the Fig.52 d was performed using the 3D nanoSQUID based on S/N-I/S tunnel junctions shown in Fig.35 and concern iron oxide nanoparticles having a diameter of *6 nm*. An exponential decay typical of relaxation phenomena was observed. The red squares in the Fig.52 d refers to the same measurements without nanoparticles. After inverting the magnetic field, the nanoSQUID goes rapidly to another constant value, having a time dependence negligible with respect to the signal arising from the nanoparticle relaxation. This check measurement guarantees that the eventual vortex penetration during the field cooling does not affect the relaxation measurement.

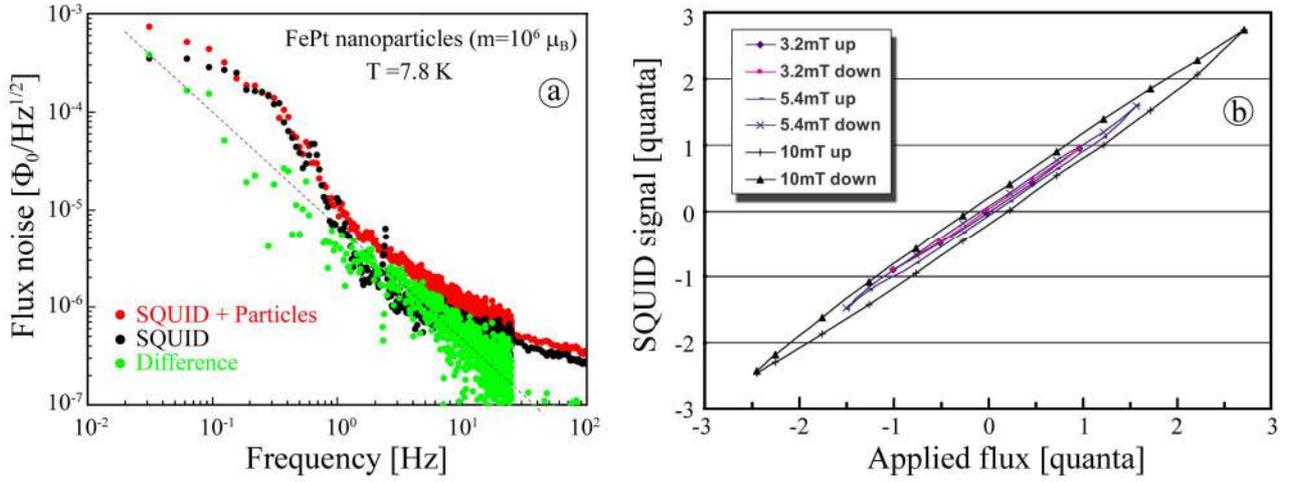

*Fig. 53 a) Magnetic flux noise spectra of a nanoSQUID (shown in Fig. 46a) with (red dots) and without (black dots) FePt nanobead (adapted from [236]). b) Magnetization loop of a single FePt nanobead obtained by the measurement of the magnetic hysteresis of the output voltage due to the magnetic nano-object (adapted from [235]).*

Hao et al. [235, 236], by using the nanoSQUID shown in Fig. 46a, measured the magnetic response of a single iron-platinum (*FePt*) nanobead with 40:60 (Fe:Pt) composition and fabricated from a water-based solution of *FePt* nanoparticles with an average diameter of 3.8 nm. The single nanobead, having a diameter ranging from 70 to 200 nm, was positioned onto SQUID by using the SEM technique in the previous section. The magnetic flux noise spectral densities with and without nanobeads at *T=7.8* is reported in the Fig. 53a. At high frequencies, the noise does not increase whereas at low frequencies an additional noise appears as highlighted by the difference between the two curve with and without nanobeads. This behaviour may arise from the individual low-frequency moment fluctuations below the blocking temperature of the particles [236]. The magnetization loops of a single *150 nm* sized nanobead, measured at *T=7.8 K* and for increasing magnetic field up to *10 mT* are shown in Fig. 53b. The curves were obtained by measuring the magnetic hysteresis of the periodic response of SQUID due to the presence of *FePt* nanobead. The presence of a remnant, moment also at the lowest field sweep, indicates that the temperature of *7.8 K* is already below the blocking temperature. The author estimated a nanobead dipole moment of $10^6$ $\mu_B$.

Before moving on the other nanoSQUID applications, it is worth to mention a very recent measurements carried out by employing a HTc nanoSQUID. Schwarz et al. have successfully used a low noise YBCO nanoSQUID based on grain boundary junctions to measure at *T=4.2 K* the magnetization reversal of iron nanowire with a diameter of *39 nm* (Fig. 254). The latter was encapsulated in a carbon nanotube and positioned close to the SQUID loop by manipulator inside a FIB-SEM system. Switching of the magnetization was detected at a magnetic field of *±100 mT*, which was in very good agreement with estimated value [261].

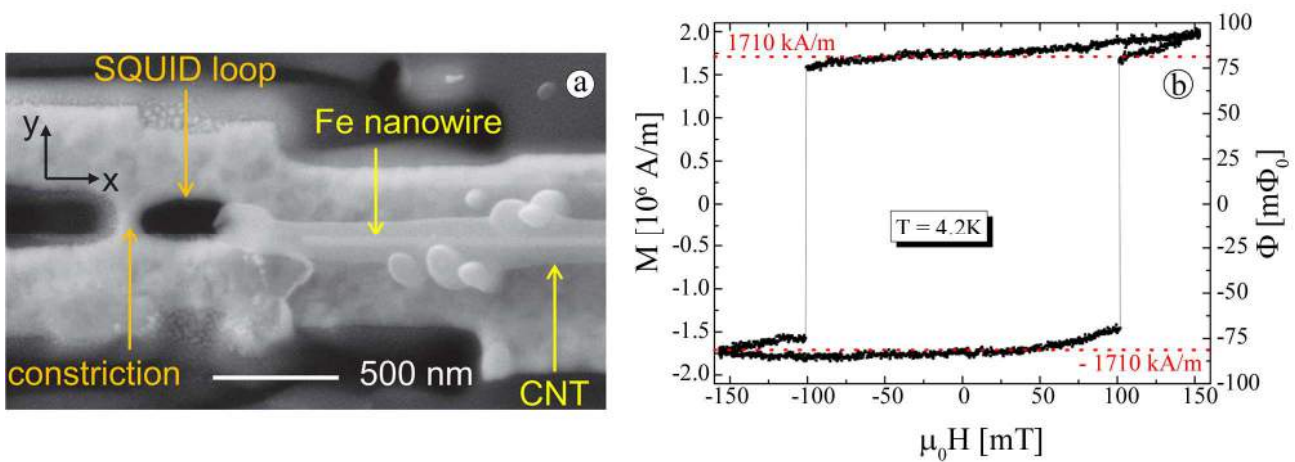

*Fig. 54 a) SEM Image of YBCO nanoSQUID with Fe nanowire located close to the SQUID loop. b) Histereis loop of Fe nanowire. The switching of the magnetization occurs at a filed of about 100 mT. The horizontal red dotted lines indicate the literature value $M_s=±1710$ kA/m. (Adapted from [235]).*

**4.3 Other applications**

**4.3.1 Scanning magnetic microscopy with an ultra high spatial resolution**

As mentioned in the section 1.5, micro sized SQUIDs have been widely employed for scanning magnetic microscopy, allowing very interesting application in view of fundamental studies of superconductor [102], magnetic materials and micro-sized magnetic beads [262-264]. In a SQUID, scanning magnetic microscope (SSM) the spatial resolution is limited by the size of the micro pick-coil and the distance from the sample under investigation, typically, of the order of few μm. The development of nanoSQUIDs fabricated on a quartz tip (section 3.4) could appreciably increase the spatial resolution of SSMs. To demonstrate the effectiveness of nanoSQUID for magnetic microscopy with an ultra-high spatial resolution, Finkler et al. developed [206] a SSM based on an *Al* nanoSQUID fabricated on a quartz sharp pipette with a tip diameter ranging from *100* to *400 nm*. The nanosensor was glued to a quartz tuning fork allowing scanning within few *nm* from the surface of the sample to investigate. Such innovative instrumentation was employed to image the vortex lattice and local magnetic response of different materials (*Al, Nb* and *NbSe$_2$*) with spatial resolution of about *200 nm*.

Later, Vasukov et al. [207] employed the *Pb* nanoSQUID shown in the Fig. 39 to image vortices in a *Nb* thin film (Fig. 55) achieving an even better spatial resolution. During the measurement the nanoSQUID was placed at a constant height of *50 nm* above the sample and the scanning was performed by an attocube-integrated scanner. The scan areas were $(1 \times 1)$ μm$^2$ (Fig. 54a) and $(300 \times 300)$ nm$^2$ (Fig. 55b); the applied fields were *B = 28 mT* and *0.2 T* for measurements reported in Fig. 54a and 54b respectively. The pronounced peaks displayed in the figure correspond to the vortex centres. The minimum distance between the vortices was *330 nm* (Fig. 55a) and *120 nm* (Fig. 55b)

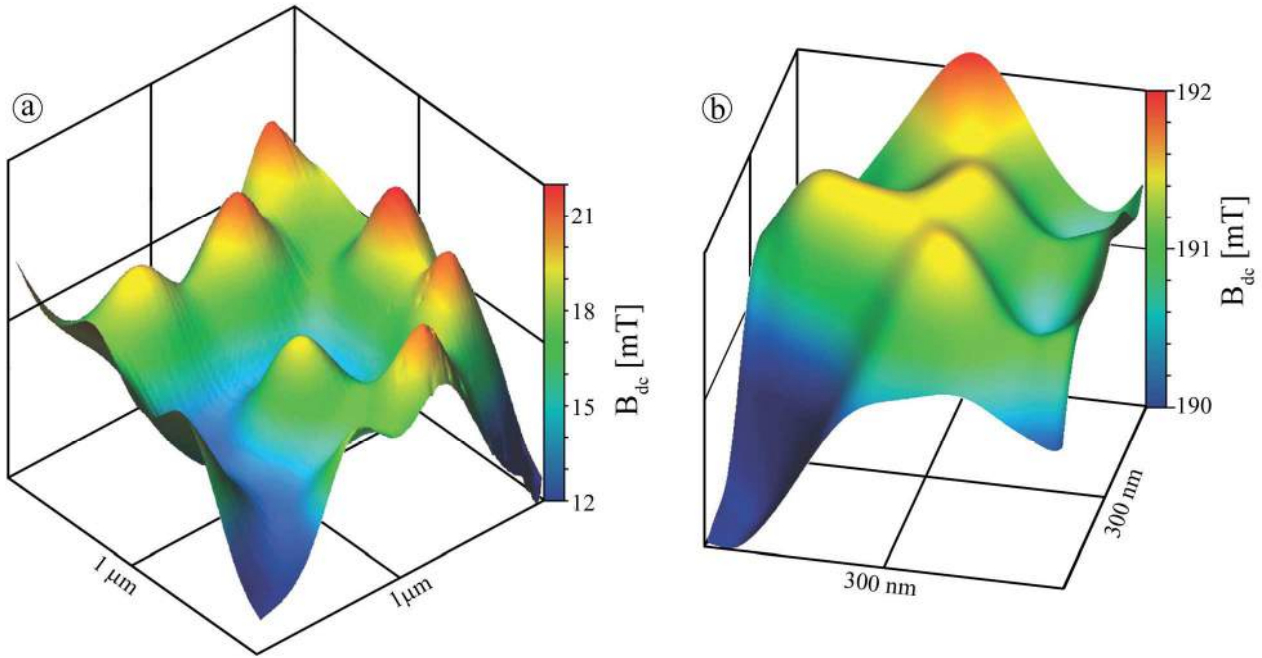

*Fig. 55 Magnetic field distribution generated by an a.c. transport current flowing in a 3 μm wide Nb strip. The measurements were performed by two different Pb nanoSQUIDs fabricated on a*

*quartz tip having a diameter of 117 nm (right scanning) and 88 nm (left scanning) (adapted from [207]).*

Nagel et al. [265] developed a multifunctional sensor consisting of a low temperature magnetic force microscope obtained by using a *Ni* nanotube, protruded from the cantilever end by few *μm*, as a ferromagnetic tip for a *Nb* nanoSQUID. A nichel nanotube, protruded from the cantilever end by few *μm*. The cantilever was perpendicular to the nanoSQUID, which was moved by a three dimensional piezoelectric positioning stage (Attocube, System AG). The nanoSQUID, in vertical configuration, was based on sub-micrometric *Nb/HfTi/Nb* Josephson junctions (see Fig. 34). They determined the spatial dependence of magnetic coupling between the *Ni* nanotube and the nanoSQUID, obtaining useful information about magnetic coupling factor, which is directly related to the spin sensitivity (formulas 27, 28). In this way, it is possible to experimentally identify the position of highest sensitivity of a nanoSQUID. By using the magnetic force microscope, they detected the Abrikosov vortices, which are trapped within the nanoSQUID structure when high magnetic field is applied. The authors noted that the *Ni* nanotube can be also considered as a nano-sample to investigate and, in this case, the nanoSQUID is used either as a detector of *Ni* nanotube displacements or as a local probe of the stray field produced by the *Ni* nanotube. As displacement detector, a sensitivity as low as *110 fm/Hz$^{1/2}$* at an operation temperature of *4.3 K* was estimated.

**4.3.2 Nanoscale electromechanical system (NEMS) resonator readout**

As stated in the introduction section, a SQUID is able to detect, with an unequalled sensitivity, a magnetic flux or any physical quantities that can be converted into magnetic flux such as magnetic fields, currents, voltage, displacements and temperature. As concern as the displacement detection, satisfactory results have been achieved by using micro-sized SQUID [266-267]. Ultra high sensitive detection of displacements at nanoscale level and in the *GHz* range plays a key role for nanoelectromechanical system (NEMS). In fact, the small displacements of these miniaturized devices induce very low signals which are overwhelmed by parasitic background. Therefore, efficient actuation and sensitive detection at the nanoscale remains a challenge. Due to nanoscale effects, NEMS present interesting and unique characteristics, which deviate greatly from their predecessor microelectromechanical systems (MEMS). The main potential applications of NEMS based devices are single molecules biosensing, information storage, nanoscale refrigerator and high sensitivity sensors of mass, force, heat capacities [268].

In the recent years, the quantum detection group of NPL laboratories (UK) proposed nanoSQUIDs as an effective readout of a NEMS resonator [157, 269-271]. The basic principle of the measurement lies on the coupling between the electrically conducting resonator and the SQUID loop. This produces a SQUID inductance variation and a consequent change of the output signal.

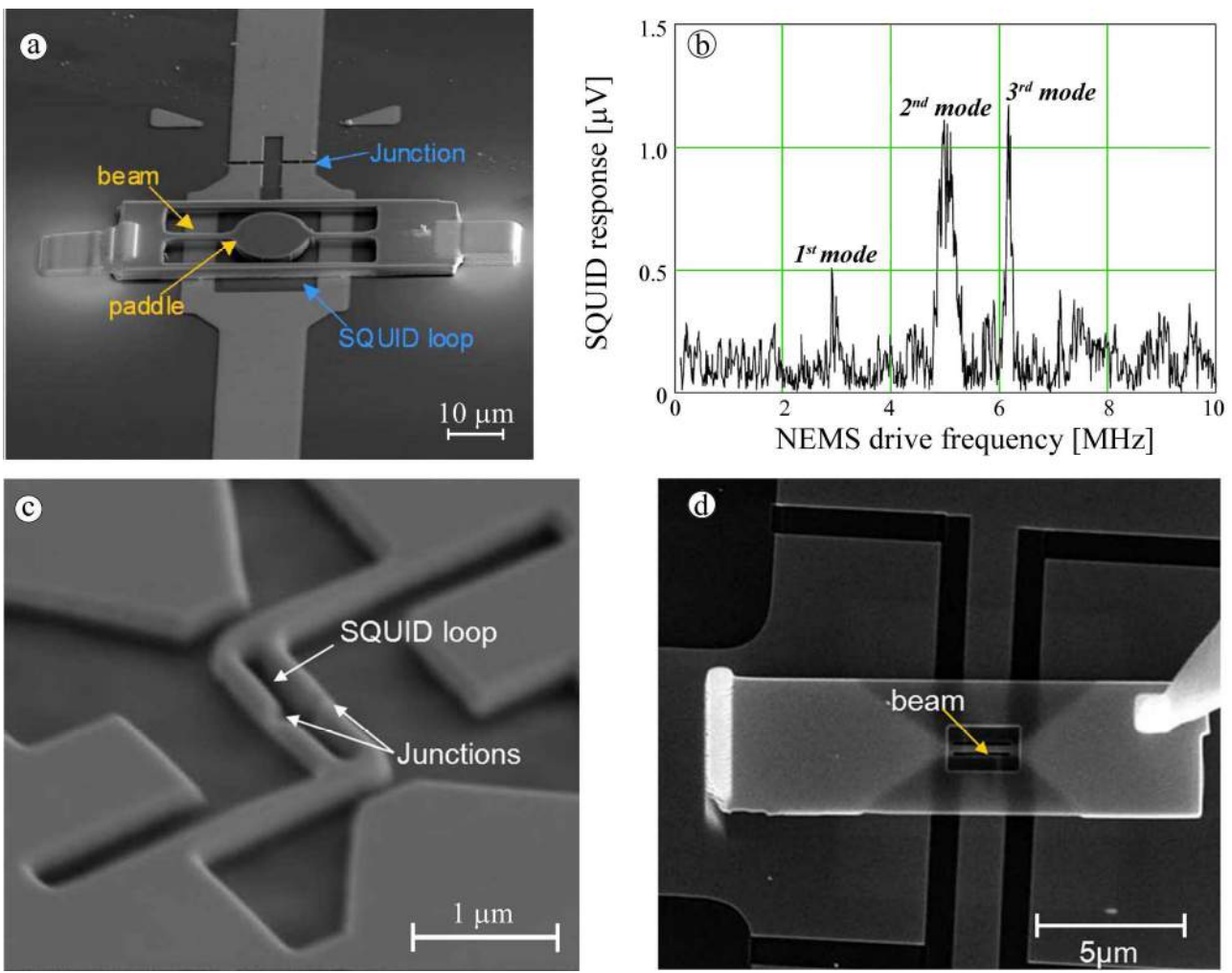

*Fig. 56 a) SEM images of a microSQUID including a silicon resonator for nanoelectromechanical applications (adapted from [157]). The response to an excitation by Lorentz drive force method is shown in b) (adapted from [157]). c) and d) SEM images of a nanoSQUID in a slot shape with and without the carbon nano-resonator (adapted from [271]).*

The use of a SQUID with a nano-loop optimizes the magnetic coupling between the nanometric moving parts of NEMS resonator and the SQUID. The Fig. 56 reports the nanodevices fabricated by NPL group [157, 271].

The SQUID of Fig.56a was fabricated by FIB, starting from a bilayer of *Nb/W* and included two Dayem nanobridges having both length and width of *60-80 nm*. This SQUID is similar to that depicted in Fig.23a but with a flux capture area of about *(20x20) μm$^2$*. The silicon resonator, positioned over the nanoSQUID loop by a carbon fibre microprobe, has been also fabricated by FIB milling and consist of a central beam (*1 μm* in width, *40 μm* in length and *800 nm* tick) including at its centre a circular paddle (*15 μm* in diameter). The resonator was excited by a Lorentz force drive method. A radio frequency current is sent into the central wire of the resonator while a magnetic field is applied in the plane of the SQUID. So, an oscillating Lorentz force (**F**=-*i l* x **B**) acts on the wire of length *l*. The resonant frequencies were calculated by using a suitable software exploiting the finite elements method [157]. The frequency values relative to z-displacements was about *2.2 kHz* where those relative to *x* and *y* torsion ranges from *2.2* to *7.8 MHz*. A preliminary measurement at *T=7.7 K* showing the SQUID output is reported in the Fig. 56b. The applied magnetic field and the RF current were *0.22 T* and *33 μA* respectively. The peaks in the figure, probably, correspond to lowest frequencies of resonant mode of the paddle. In order to increase the magnetic coupling and to extend the usable excitation field, an improved nanodevice has been developed by the same group [270]. The large SQUID loop area was replaced by a slotted nanoSQUID having an area of *(0.9×0.1) μm$^2$* (Fig. 56c) and the resonator, consisting of a double clamped *Si* beam (*0.1 μm* wide and *1 μm* length), was attached on top of the SQUID slot (Fig. 56d).

### 4.3.3 Single photon and macromolecule detection

The development of ultra high sensitive single photon detectors plays a crucial role for several applications in science and technology. In medical imaging, for instance, single photons are detected in PET (Positron Emission Tomography) and CT (Computer Tomography) scanners and, more recently, in laser optical imaging. Lifetime fluorescence measurements using single photon counting is also used in the diagnosis of some medical conditions. It is also widely used in analytical chemistry for determining the chemical recipe of samples. Moreover, single photon detection is extensively used in scientific research in the fields of particle physics, astrophysics, quantum cryptography and materials science. Superconducting transition-edge bolometer are among the most sensitive detectors of electromagnetic radiation extending in wavelength from x-rays to the far infrared [271, 272]. In this framework, a nanoSQUID based device has been proposed as a promising superconducting single photon and macromolecule detector [273-276], in which, a radiation absorber (superconductive thin film patch) is placed within the loop of nanoSQUID and is maintained just below its critical temperature (Fig. 57a). The SQUID has higher transition temperature than the absorber. When a photon hits the absorber, its temperature increases slightly, causing a variation of the inductance ($L$) of the SQUID due to the change of the London penetration depth ($\lambda$) of the absorber. The temperature variation caused by the incident photons increases by decreasing the heat capacitance of the absorber. Therefore, in order to obtain a suitable sensitivity, the absorber could be as small as possible, and the SQUID loop could have having the same size of the absorber. A quantitative estimation of the detector sensitivity can be obtained by calculating the variation of the SQUID voltage output with the temperature:

$$\frac{dV}{dT} = \frac{dV}{dL}\frac{dL}{d\lambda}\frac{d\lambda}{dT} \tag{43}$$

By using the detailed calculation of the second member reported in [272], the above formula becomes:

$$\frac{dV}{dT} \approx \frac{6\pi R_{dyn} I_c \mu_0^2 a}{L^2} \frac{\lambda(0) T^3}{\left(1-\left(\frac{T}{T_c}\right)^4\right)^{3/2} T_c^4} \tag{44}$$

Where $R_{dyn}$ is the dynamic resistance of the nanoSQUID and $a$ its loop radius. Using realistic values for all parameters included in the (44), a sensitivity of *10 mV/K* can be estimated. Considering a low noise amplifier with $S_V^{1/2}= 1nV/Hz^{1/2}$, a temperature change of *0.1 μK* per bandwidth unit should be detected. It corresponds to an energy sensitivity as low as $10^{-25}$ J/Hz for a Nb absorber having a diameter of 0.8 μm, a thickness of 50 nm and operating at T=0.1 K.

In the device shown in the Fig. 57b [276], the Nb nanoSQUID had a loop area of (200×200) $nm^2$. The film thicknesses of the nanoSQUID and the Nb absorber were 20 nm and 14 nm respectively resulting in a critical temperature of 7.9 K for the SQUID and 6.3 K for the absorber.

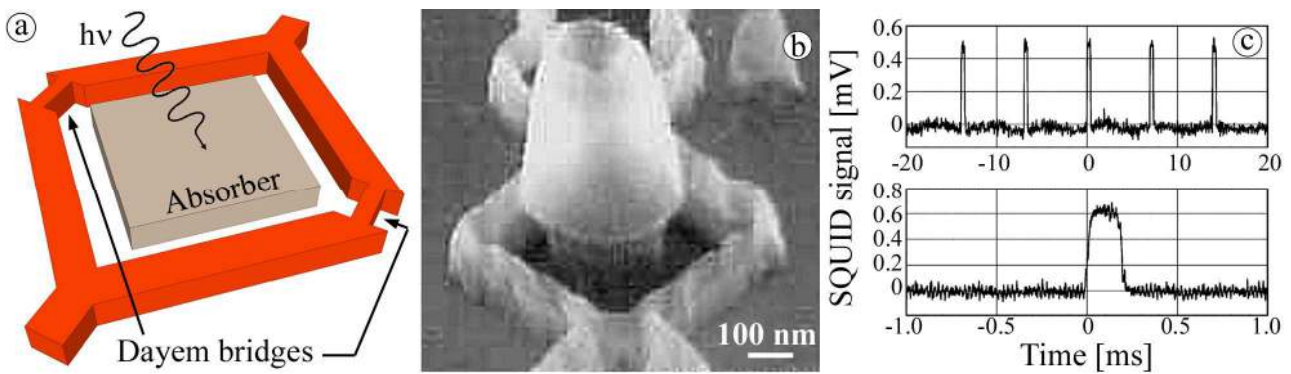

*Fig. 57 a) Scheme of a nanoSQUID based device for single photon detection. b) SEM image of a Nb nanoSQUID with a Nb absorber located inside the loop. c) NanoSQUID response to chopper laser light at T=6.3 K (adapted from [276]).*

Preliminary test were performed at temperature of about to 6.3 K using a chopped He-Ne laser to heat the absorber. Voltage pulses corresponding to the laser light exposure were recorded at the SQUID output with a SNR=50, demonstrating the feasibility of the detector (Fig.57c). Another possible application of this nano-detector could be in energy-resolving detector of massive (>200 amu) molecular or polymeric species. In the conventional mass spectrometer, the sensitivity is related to the velocity of the molecule, which decrease inversely with its mass making it inefficient for big molecules. Instead, in this detector, the signal is directly related to the molecule kinetic energy, which is independent on the mass for a given ion beam [273].

**4.3.4 Nanoelectronics and Quantum computing**

Tejada et al. [277] proposed a new magnetic approach for quantum bits, based on the employment of both ferromagnetic and antiferromagnetic nanoparticle clusters. Such magnetic systems could be suitable candidates for quantum hardware for two main reasons. Very small nanoparticles have large total spin (of the order of few hundreds $\hbar$), so they are easier to measure with respect to single spin. The second reason is their high magnetic anisotropy barrier leading to an energy separation between the two lowest levels inside the potential well of a few Kelvins. Moreover, nanoparticle clusters can be considered as a mesoscopic system. In fact, their magnetic state are defined by the collective motions of all their constituent particles and, at the same time, they consist of thousand nucleons and electrons.

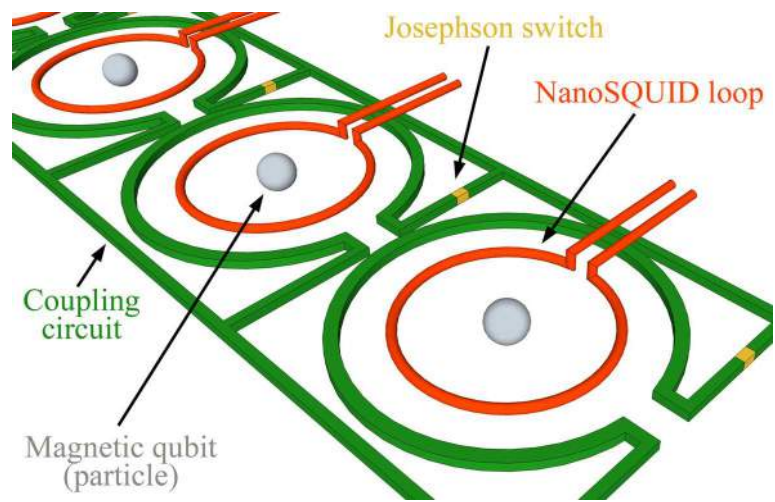

*Fig. 58 Schematic draw of the realization of coupled controlled quantum bits. The nanoSQUID loops (orange colour) are magnetically coupled to nanoparticles quantum bit (grey colour) [277].*

The Fig.58 shows a schematic example of the realization of coupled controlled quantum bit. The magnetic qubits (grey colour) are arranged in one or two-dimensional lattice and coupled to both superconducting circuit (green colour) containing Josephson switch (yellow colour) and loops of nano-SQUIDs (red colour). The quantum state of each qubit can be manipulated by sending electromagnetic signals to superconducting coupling circuit whereas the state measurement can be performed by the individual nanoSQUID. The required sensitivity is about (*10-100*) $\mu_B$ that is comparable with the sensitivities exhibited by recent nanoSQUIDs.

Another prospective application of NanoSQUIDs is in the nanoelectronics. In this framework, the nanoSQUID have been proposed as a flux flow transistor [278] and trapped-vortex memory device [278]. As regard as the transistor, Lam [279] demonstrated that a Josephson nanodevice, including a nanoSQUID (similar to that shown in Fig. 21) and a single *Nb* strip, can acts as a gate and employed as three-terminal device (Fig. 59). The width of the strip is *1 μm* and the separation from the nanoSQUID is *0.15 μm*. By sending a suitable current in the *Nb* strip, it is possible to couple a magnetic flux into the SQUID loop and consequently modulate the current-voltage characteristics. If the SQUID is biased with a current a little bit lower than the critical value, the current flowing in the gate (control line current) can modulate the output voltage. For the device reported by Lam, a gate current $\Delta I_g = 130\ \mu A$ was required to obtain a maximum critical modulation $\Delta I_c = 40\ \mu A$ corresponding to a voltage output of *180 mV* resulting in a current gain $g_c = \Delta I_c/\Delta I_g = 0.3$ and a transresistance $r_m = \Delta V/\Delta I_g = 1.3\ \Omega$. The main advantages of the superconducting three-terminal based on nanoSQUIDs is the large-scale integration.

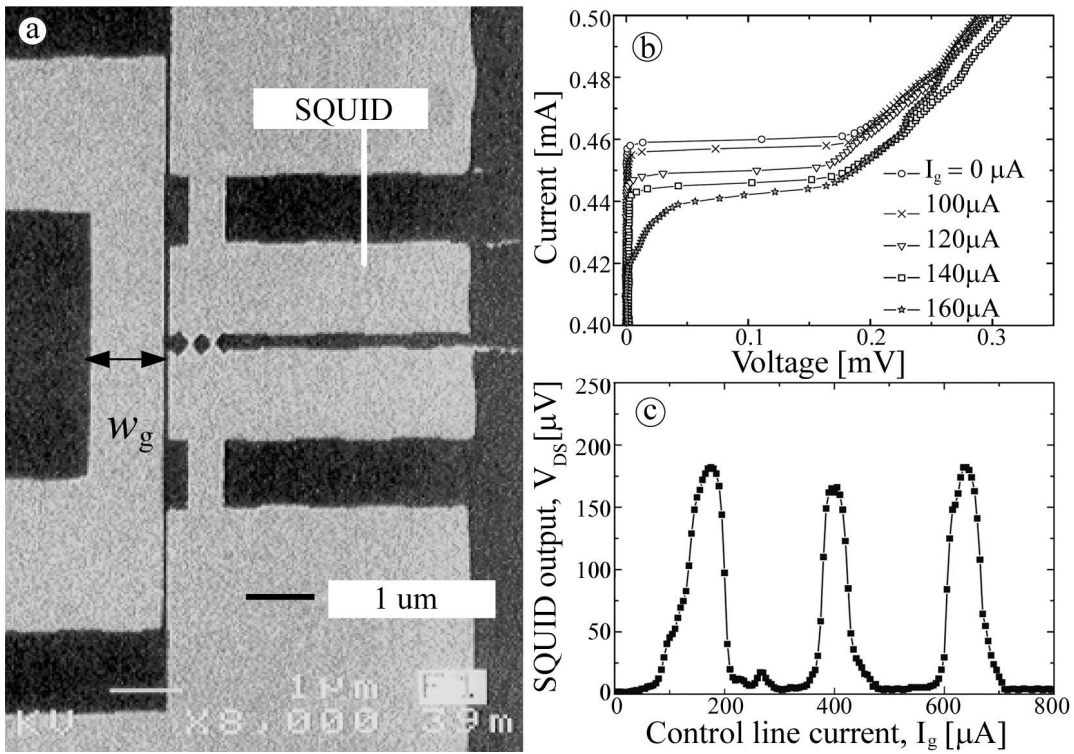

*Fig. 59 a) SEM Image of a three-terminal nanodevice based on a nanoSQUID coupled to a control gate line located very close to the SQUID loop.  b) I-V characteristics for different control line current values. c) NanoSQUID output as a function of the control line current (adapted from [278]).*

As seen in the section 1.4, magnetic vortices can penetrate in the superconducting structures and cause low frequency noise. In fact, to avoid the entry of vortices or their activated thermally motion, careful designs of the device has to be take into consideration. However, single vortex at localized trapping site can be used as a memory bit as proposed in the late '70s [280]. In the early '80s, Uehara and Nagata [281] investigated a memory cell based on multiple trapped vortices in superconducting thin film. The main obstacle to develop suitable memory device was the poor scale integration of the single cell memory, which had a size of some tens of *μm*. Exploiting the submicron size of nanoSQUIDs, Lam and Gnanarajan [278] proposed them as trapped-vortex memory device with the potentiality for high-density structures. In this case, the two logic states ('0' and '1') of the memory cell are related to the presence or not of a trapped vortex in a pinning centre. A pinned vortex causes a magnetic hysteresis in the $I_c$-$\Phi$ characteristics of the nanoSQUIDs. The occurring remnant magnetic field can be measured by the $I_c$ value of nanoSQUID. They investigated the vortex penetration in the nanoSQUID body by measuring the $I_c$-B and the current-voltage characteristics for different applied magnetic fields. An appreciable magnetic hysteresis due to vortex penetration was observed in the magnetic pattern of the nanoSQUID, suggesting the possibility to apply the nanoSQUID as memory cells.

## 5. Summary and conclusions

In summary, we have presented a review about nanoSQUID including the foundamental, the fabrication technique, the different type of nanodevice and the applications.

Since the realization of the first nanoSQUID (2003), the fabrication, the design and the characterization of these nanosensors has been appreciably improved reaching spin sensitivity less than one electron spin per unit bandwidth. It is due to the progress, during the last ten years, in nanofabrication techniques, low noise readout circuits and performance simulations.

Different nanoSQUIDs have been recognized. Dayem or $HT_c$ nanoSQUIDs are superior when high magnetic fields ( > 0.5 T) are involved, since nanoSQUIDs based on trilayer Josephson junctions (SNS or SNIS) show a critical current modulation due to the residual magnetic field parallel to the SQUID loop, preventing the correct operation of the SQUID. Moreover, $HT_c$ nanoSQUIDs allow to operate in a wider range of the temperature (from mK to above 77 K). On the other hand, the SNS or SNIS nanoSQUIDs exhibit a critical current modulation depth of more than 70%, compared to the 20%–30% values for most of Dayem nanoSQUIDs resulting in an improved responsivity and magnetic flux resolution.

It is worth to emphasize the development of nanoSQUIDs based on innovative technique such as those based on CNT or nanoSQUID on a tip. In addition to promising applications in nanomagnetism, CNT nanoSQUIDs also offer the possibility to investigate interesting physical phenomena such as Kondo physics [283], and the generation of entangled electrons pairs in a carbon nanotube [284]. The nanoSQUID fabricated on the apex of a quartz tip combine an ultra high magnetic moment sensitivity (0.38 $\mu_B/Hz^{1/2}$) with a spatial resolution lower than 1 $\mu$m, providing a new tool for quantitative nanoscale scanning magnetic microscopy with single spin sensitivity.

As regard as the applications, while many of the SQUID applications involve the detection of tiny magnetic signals from relatively distant objects, the devices presented in this review are aimed to the detection of nearby nanoscale objects. One of the most interesting application is the investigation of single magnetic nanoparticle. In fact, although techniques for measuring the magnetic properties of large particles or collections of nanoparticles are well established, single nanoparticle measurements are less straightforward and become increasingly challenging as the particle dimensions and magnetic moments are reduced. The investigation of nanoparticles at nanoscale level could be of significance not only for the information technology and the quantum computing but also for future medical and biological applications. In this context, the numerous experiments on single nanoparticle, nanowire and nanobead, performed by using nanoSQUIDs, have demonstrated the effectiveness of these quantum nanodevices to investigate the magnetic properties of the matter at nanoscale level. Other stimulating applications could be the single photon detection, the readout of nano-electromechanical system resonator, nanoelectroning and quantum computing.

In conclusion, we would like to stress that, even if the nanoSQUID remains an object of dedicated research, the obtained results are very encouraging in view of a wide employment of this nanodevice for severals nanoscale applications.

## References

[1] The SQUID handbook vol I : Fundamentals and technology of SQUIDs and SQUID systems" J. Clarke and A. I. Braginski eds. Wiley-VCH Verlag GmbH & Co. KgaA, Weinheim, (2004).

[2] The SQUID handbook vol II: Fundamentals and technology of SQUIDs and SQUID systems" J. Clarke and A. I. Braginski eds. Wiley-VCH Verlag GmbH & Co. KgaA, Weinheim, (2006).

[3] R. K. Fagaly, Superconducting quantum interference device instruments and applications, Rev. Sci. Instrum. 77 (2006 ) 101101.

[4] Applied Superconductivity: Handbook on Devices and Applications, P. Seidel ed., Wiley, Weinheim (2015).

[5] P. D. Nation, M. P. Blencowe, A. J. Rimberg, E. Buks, Analogue Hawking radiation in a dc-SQUID array transmission line, Phys. Rev.Lett, 103 (2009) 087004.

[6] C. M. Wilson, G. Johansson, A. Pourkabirian, M. Simoen, J. R. Johansson, T. Duty, F. Nori, and P. Delsing, Observation of the dynamical Casimir effect in a superconducting circuit, Nature, 479 (2011) 376.

[7] M. Veldhorst, C. G. Molenaar, X. L. Wang, H. Hilgenkamp, and A.Brinkman, Experimental realization of superconducting quantum interference devices with topological insulator junctions, Appl. Phys.Lett., 100 (2012) 072602.

[8] F. Giazzotto, M.G. Martinez-Perez, The Josephson heat interferometer, Nature, 492 (2012) 401.

[9] J. E. Carlstrom, G. P. Holder, E. D. Reese, Cosmology with the Sunyaev–Zel'dovich Effect, Ann. Rev.Astron. Astrophys., 40 (2002) 643–680.

[10] F. Marin et al., Gravitational bar detectors set limits to Planck-scale physics on macroscopic variables, Nature Phys., 9 (2012) 71–73.

[11] P. A. R. Ade et al., Detection of B-Mode Polarization at Degree Angular Scales by BICEP2, Phys. Rev.Lett, 112 (2014) 241101.

[12] D. Gatteschi, R. Sessoli, Quantum Tunneling of Magnetization and Related Phenomena in Molecular Materials, Angew. Chem., Int. Ed., 42 (2003) 268-287.

[13] S. D. Bader, Colloquium: Opportunities in nanomagnetism, Rev. Mod. Phys. 78 (2006) 1-15.

[14] A. M. Chang, H. D. Hallen, L. Harriott, H. F. Hess, H. L. Kao, J. Kwo, R. E. Miller, R. Wolfe, J. van der Ziel, T. Y. Chang, Scanning Hall probe microscopy., Appl. Phys. Lett., 61 (1992)1974-1976.

[15] M. Bode, Spin-polarized scanning tunnelling microscopy, Rep. Prog. Phys., 66 (2003) 523

[16] Y. Martin, H. K. Wickramasinghe, Magnetic imaging by 'force microscopy' with 1000 Å resolution. Appl. Phys. Lett., 50 (1987) 1455–1457.

[17] R. E. Dunin-Borkowski, M. R. McCartney, R. B. Frankel, D. A. Bazylinski, M. Posfai, P.R. Buseck, Magnetic microstructure of magnetotactic bacteria by electron holography. Science 282 (1998)1868–1870

[18] J. R. Mazeet et al., Nanoscale magnetic sensing with an individual electronic spin in diamond, Nature, 455 (2008) 644–647.

[19] J. M. Taylor et al., High-sensitivity diamond magnetometer with nanoscale resolution, Nature Phys. 4, (2008) 810–816.

[20] M. S. Grinolds, S. Hong, P. Maletinsky, L. Luan, M. D. Lukin, R. L. Walsworth, A. Yacoby, Nanoscale magnetic imaging of a single electron spin under ambient conditions, Nature Phys., 9 (2013) 215–219.

[21] D. Le Sage, K. Arai, D. R. Glenn, S. J. DeVience, L. M. Pham, L. Rahn-Lee, M. D. Lukin, A. Yacoby, A. Komeili, R. L. Walsworth, Optical magnetic imaging of living cells, Nature, 496 (2013) 486–491.

[22] C. P. Foley and H. Hilgenkamp, hy NanoSQUIDs are important: An introduction to the focus issue, upercond. Sci. Technol., 22 (2009) 064001.

[23] W. Wernsdorfer, From micro to nano-SQUIDs: Applications to nanomagnetism, Supercond. Sci. Technol., 22 (2009) 064013.

[24] L Hao, D Cox, P See, J Gallop. O Kazakova, Magnetic nanoparticle detection using nano-SQUID sensors, J. Phys. D: Appl. Phys, 43 (2010) 474004.

[25] R. C. Jaklevic, J. Lambe, A. H. Silver, J. E. Mercereau, Quantum interference effects in Josephson tunneling, Phys. Rev.Lett., 12 (1964) 159-160.

[26] A.H. Silver, J.E. Zimmerman, Quantum states and transitions in weakly connected superconducting rings, Phys. Rev., 157 (1967) 317–341.

[27] R. Cantor, in SQUID Sensors: Fundamentals, Fabrication and Applications, edited by H. Weinstock, Dordrecht: Kluwer Academic Publisher, Series E: Applied Sciences-Vol. 329 (1996), 179.

[28] B. D. Josephson, Possible new effects in superconductive tunneling, Phys. Lett., 1 (1962) 251–253.

[29] R. Doll, M. Nabauer, Experimental proof of magnetic flux quantization in a superconducting ring, Phys. Rev. Lett., 7 (1961) 51–52.

[30] D. D. Coon, M. D. Fiske, Josephson ac and step structure in the supercurrent tunneling characteristics, Phys. Rev., 138 (1965) A744–A746.

[31] A. Barone, G. Paterno, Physics and Applications of the Josephson Effect. New York: John Wiley & Sons. (1982).

[32] D. Koelle, R. Kleiner, F. Ludwig, E. Dantsker, J. Clarke, High transition temperature superconducting quantum interference devices, Rev. Mod. Phys., 71(1999) 631–686.

[33] F. Tafuri, J. Kirtley, Weak links in high critical temperature superconductors, Rep. Prog. Phys., 68 (2005) 2573.

[34] W. Meissner, R. Oschsenfeld, Eh neuer Effekt bei Eintritt der Supraleitfahigkeit (A new effect concerning the onset of superconductivity), Die Naturwissenschaften, 21 (1933) 787

[35] W.C. Stewart, Current-voltage characteristics of Josephson junctions, Appl. Phys. Lett., 12 (1968) 277–280.

[36] D.E. McCumber, Effect of ac impedance on dc voltage-current characteristics of Josephson junctions, J. Appl. Phys., 39 (1968) 3113–3118.

[37] R. F. Voss, R. B. Laibowitz, A. N. Broers, S. I. Raider, C. M. Knoedler and J M Viggiano, Ultra low noise Nb DC SQUIDs, IEEE Trans. Magn.,17 (1981) 395-399.

[38] C. Tesche, J.Clarke, DC SQUID: noise and optimization, J. Low. Temp. Phys, 29 (1977) 301–331.

[39] T.Ryhanen, H. Seppa, R. Illimoniemi, J. Knuutila, SQUID magnetometers for low-frequency applications, J. Low Temp. Phys., 76 (1989), 287–386.

[40] V. J. de Waal, P. Schrijner, R. Llurba, Simulation and optimization of a DC squid with finite capacitance, J. Low Temp. Phys., 54 (1984) 215–232.

[41] J. Clarke "SQUID fundamentals", in SQUID sensors: fundamentals, fabrication and application, H. Weinstock Ed., Dordrecht: Kluwer Academic Publisher, Series E: Applied Sciences-Vol. 329 (1996), 1-62.

[42] J. Clarke, W. M.Goubau, M.B. Ketchen, Tunnel junction dc SQUID: fabrication, operation and performance, J. Low Temp. Phys., 25 (1976), 99–144.

[43] F. Wellstood, C. Heiden, J. Clarke, Integrated dc SQUID magnetometer with a high slew rate, Rev. Sci. Instrum., 55 (1984) 952–957.

[44] D. Drung, "Advanced SQUID read-out electronics" in SQUID sensors: fundamentals, fabrication and application, H. Weinstock Ed., Dordrecht: Kluwer Academic Publisher, Series E: Applied Sciences-Vol. 329 (1996) 63-116.

[45] D. Drung, R. Cantor, M. Peters, A. J. Scheer, and H. Koch, "Low-noise high-speed dc superconducting quantum interference device magnetometer with simplified feedback electronics" Appl. Phys. Lett., 57 (1990) 406-408.

[46] R. F. Voss, Noise characteristics of an ideal shunted Josephson junction, J. Low. Temp. Phys., 42 (1981) 151–163.

[47] O. Astafiev, Yu. A. Pashkin, Y. Nakamura, T. Yamamoto, J. S. Tsai, Quantum Noise in the Josephson Charge Qubit, Phys. Rev. Lett., 93 (2004), 267007.

[48] R. F. Voss, R. B. Laibowitz and A. N. Broers, Niobium nanobrldge dc SQUID, Appl. Phys. Lett. 37 (1980) 656

[49] D. D. Awschalom, J. B. Rozen, M. B. Ketchen, W. J. Gallagher, A. W. Kleinsasser, R. L. Sandstrom, B. Bumble, Low-noise modular microsusceptometer using nearly quantum limited de SQUIDs, Appl. Phys. Lett., 53 (1988), 2108

[50] B. Chesca, Analytical theory of dc SQUIDS operating in the presence of thermal fluctuations, J. Low.Temp.Phys., 112 (1998) 165–196.

[51] C. Granata, A. Vettoliere, R. Russo, M. Russo, and B. Ruggiero, Critical current noise investigations in underdamped Josephson devices, Phys. Rev. B, 83 (2011) 092504.

[52] J. Clarke, G. Hawkins, Flicker (1f) noise in Josephson tunnel junctions, Phys. Rev. B, 14 (1976) 2826

[53] Quantum Computing and Quantum Bits in Mesoscopic Systems, edited by A. J. Leggett, B. Ruggiero, P. Silvestrini (Plenum, New York, 2004).

[54] D. J. Van Harlingen, T. L. Robertson, B. L. T. Plourde, P. A. Reichardt, T. A. Crane, John Clarke, Decoherence in Josephson-junction qubits due to critical-current fluctuations, Phys. Rev. B, 70 (2004) 064517

[55] R. de Sousa, K. B. Whaley, T. Hecht, J. von Delft, and F. K. Wilhelm, Microscopic model of critical current noise in Josephson-junction qubits: Subgap resonances and Andreev bound states, Phys. Rev. B, 80 (2009) 094515.

[56] L. Faoro and L. B. Ioffe, Microscopic origin of low-frequency flux noise in josephson circuits, Phys. Rev. Lett., 100 (2008) 227005.

[57] M. Constantin and C. C. Yu, Microscopic model of critical current noise in Josephson junctions, Phys. Rev. Lett., 99 (2007) 207001.

[58] S. M. Anton et al., Low-frequency critical current noise in Josephson junctions induced by temperature fluctuations, Appl. Phys. Lett., 101 (2012) 092601.

[59] S. Sendelbach, D. Hover, A. Kittel, M. Muck, J. M. Martinis, and R. McDermott, Magnetism in SQUIDs at Millikelvin Temperatures, Phys. Rev. Lett., 100 (2008) 227006.

[60] R. H. Koch, D. P. Di Vincenzo, J. Clarke, Model for 1/f flux noise in SQUIDs and qubits, Phys. Rev. Lett., 98 (2007) 267003.

[61] J. Eroms, L. C. van Schaarenburg, E. F. C. Driessen, J. H. Plantenberg, C. M. Huizinga, R. N. Schouten, A. H. Verbruggen, C. J. P. M. Harmans, and J. E. Mooij, Low-frequency noise in Josephson junctions for superconducting qubits, Appl. Phys. Lett., 89 (2006) 122516

[62] F. C. Wellstood, C. Urbina, and John Clarke, Flicker (1/f) noise in the critical current of Josephson junctions at 0.09–4.2 K, Appl. Phys. Lett., 85 (2004) 5296-5298.

[63] R.H. Koch, J.Clarke, W.M. Goubau, J.M. Martinis, C.M. Pegrum, and D.J. Van Harlingen, Flicker (1/f) noise in tunnel junction dc-SQUIDs, J. Low Temp. Phys., 51 (1983) 207–224.

[65] E. Dantsker, S. Tanaka, P. A. Nilsson, R. Kleiner, J. Clarke, Reduction of 1/f noise in high-T c dc superconducting quantum interference devices cooled in an ambient magnetic field, Appl. Phys. Lett., 69 (1996) 4099-4101.

[66] E. Dantsker, S. Tanaka, J. Clarke, High-T c SQUIDs with slots or holes: low 1/f noise in ambient magnetic fields, Appl. Phys. Lett., 70 (1997) 2037–2039.

[67] A. B. M. Jansman, M. Izquierdo, J. Flokstra, H. Rogalla, Slotted high-Tc dc SQUID magnetometers, IEEE Trans. Appl. Supercond., 9 (1999) 3290–3293.

[68] R.H. Koch, J.Z. Sun, V. Foglietti, W.J. Gallagher, Flux dam: a method to reduce extra low frequency noise when a superconducting magnetometer is exposed to a magnetic field, Appl. Phys. Lett., 67 (1995) 709–711.

[69] S. M. Anton et al., Magnetic Flux Noise in dc SQUIDs: Temperature and Geometry Dependence, Phys. Rev. Lett., 110 (2013) 147002.

[70] T. Lanting, M. H. Amin, A. J. Berkley, C. Rich, S.-F. Chen, S. LaForest, R. de Sousa, Evidence for temperature-dependent spin diffusion as a mechanism of intrinsic flux noise in SQUIDs, Phys. Rev. B, 89 (2014) 014503.

[71] C. Granata, A. Vettoliere and M. Russo, Miniaturized superconducting quantum interference magnetometers for high sensitivity applications, Appl. Phys. Lett., 91 (2007) 122509.

[72] D. Drung, S. Knappe, H. Koch, Theory for the multiloop dc superconducting quantum interference device magnetometer and experimental verification, J. Appl. Phys., 77 (1995) 4088-4098.

[73] M. Schmelz, R. Stolz, V. Zakosarenko, T. Schonau, S. Anders, L. Fritzsch, M. Muck, H-G Meyer, Field-stable SQUID magnetometer with sub-fT $Hz^{-1/2}$ resolution based on sub-micrometer cross-type Josephson tunnel junctions, Supercond. Sci. Technol., 24 (2011) 065009.

[74] R. Monaco, C. Granata, R. Russo, A. Vettoliere, Ultra-low-noise magnetic sensing with long Josephson tunnel junctions, Supercond. Sci. Technol., 26 (2013) 125005.

[75] C. Granata, A. Vettoliere, R. Monaco, Noise performance of superconductive magnetometers based on long Josephson tunnel junctions, Supercond. Sci. Technol., 27 (2014) 095003.

[76] A. Vettoliere, C. Granata, R. Monaco, Long Josephson Junction in Ultralow-Noise Magnetometer Configuration, IEEE Trans. Magn., 51 (2015) 9000104.

[77] J. Luomahaara, V. Vesterinen, L. Gronberg, J.Hassel, Kinetic inductance magnetometer, Nature Commun., (2014) DOI: 10.1038/ncomms5872

[78] M.B. Ketcken, Design of improved integrated thin-film planar dc SQUID gradiometers, J. Appl. Phys., 58 (1985) 4323.

[79] B. Muhlfelder, W. Johnson, M. W. Cromar, Double transformer coupling to a very low noise SQUID, IEEE Trans. Magn., 19 (1983) 303-307

[80] V. Polushkin, E. Gu, D. Glowacka, D. Goldie, J. Lumley, A tightly coupled dc SQUID with an intermediary transformer, Physica C, 367 (2002) 280-284.

[81] D. Drung, J. Beyer, M. Peters, J.-H. Storm, T. Schurig, Novel SQUID current sensors with high linearity at high frequencies, IEEE Trans. Appl. Supercond., 19, (2009) 772-777.

[82] C. Granata, A. Vettoliere, M. Russo, An ultralow noise current amplifier based on superconducting quantum interference device for high sensitivity applications, Rev. Sci. Instrum, 82 (2011) 013901.

[83] V Zakosarenko, M Schmelz, R Stolz, T Sch¨onau, L Fritzsch, S Anders, H-G Meyer, Femtoammeter on the base of SQUID with thin-film flux transformer, Supercond. Sci. Technol., 25 (2012) 095014.

[84] M. B. Ketchen, J. M. Jaycox, Ultra-low-noise tunnel junction dc SQUID with a tightly coupled planar input coil, Appl. Phys. Lett., 40 (1982) 736-738

[85] M.B. Ketchen "Integrated thin-film dc SQUID sensors" IEEE Trans. on Magn, 23 (1987)1650-1657.

[86] M.B. Ketchen, W.J. Gallagher, A. W. Kleinsasser, S. Murphy and J. R. Clem "dc SQUID flux focuser", in SQUID '85: Superconducting Quantum Interference Devices and their Applications, H. D. Hadlbohm and H. Lubbig Ed., Walter de Gruyter, Berlin, New York, 1985, pp.865-871.

[87] Vu L.N., M. S. Wistrom, D. J. Van Harlingen, Imaging of magnetic vortices in superconducting networks and clusters by scanning SQUID microscopy, Appl. Phys. Lett., 63 (1993)1693-1695.

[88] J.R. Kirtley and J.P. Wikswo, "Scanning SQUID Microscopy", Annual Review of Materials Science, 29 (1999) 117-148.

[89] S. M. Frolov, M. J.A. Stoutimore, T. A. Crane, D. J. Van Harlingen, V. A. Oboznov, V. V. Ryazanov, A. Ruosi, C. Granata, M.Russo, Imaging spontaneous currents in superconducting arrays of π-junctions, Nature Phys., 4 (2008) 32.

[90] C.Granata, A.Vettoliere, R. Vaccarone, M. Russo, Low Critical Temperature dc-SQUIDs for High Spatial Resolution Applications", IEEE Trans. on Appl. Supercond., 17 (2007) 796-799.

[91] D. Cohen and E. Halgren, Magnetoencephalography, Encyclopedia of Neuroscience, 5 (2009) 615-622.

[92] K.Sternickel and A. I. Braginski "Biomagnetism using SQUIDs: status and perspectives" Supercond. Sci. Technol., 19 (2006) 160-171.

[93] M. Hämäläinen, R. Hari, R. Ilmoniemi, J. Knuutila, and O. Lounasmaa, "Magnetoencephalography - theory, instrumentation, and applications to noninvasive studies of the working human brain", Rev. Mod. Phys., 65 (1993) 413-497.

[94] C. Del Gratta, V. Pizzella, F. Tecchio, and G. L. Romani., "Magnetoencephalography - a noninvasive brain imaging method with 1 ms time resolution", Rep. Prog. Phys., 64 (2001) 1759-1814.

[95] D. N. Paulson, R. M. Toussaint, R. L Fagaly, Superconducting magnetometer system for detecting lung contaminants, IEEE Trans. Magn., 23 (1987) 1315–1318.

[96] B..M. Mackert, Magnetoneurography: theory and application to peripheral nerve disorders., Clin Neurophysiol., 115 (2004) 2667-2676

[97] G. M. Brittenham, D. E. Farrell, J. W. Harris, E. S. Feldman, E. H. Danish, W. A. Muir, J. H. Tripp, E. M. Bellon., Magnetic-susceptibility of human iron stores, New Eng. J. Med., 307 (1982) 1671–1675,

[98] D. N. Paulson, R. L. Fagaly, R. M. Toussaint, and R. Fischer, Biomagnetic susceptometer with SQUID instrumentation, IEEE Trans. Magn., 27 (1991) 3249–3252.

[99] R. Kotitz, H. Matz, L.Trahms, H. Koch, W. Weitschies, T. Rheinlander, W. Semmler, T. Bunte, SQUID based remanence measurements for immunoassays, IEEE Trans. on Appl. Supercond., 7 (1997) 3678-3681.

[100] W.G. Jenks, S. S. H. Sadeghi, J. P. Wikswo, SQUIDs for NDE, J. Phys. D, 30 (1997) 293–323.

[101] G.B. Donaldson, The use of SQUIDs for NDE, in SQUID sensors: fundamentals, fabrication and application, edited by H. Weinstock, Dordrecht: Kluwer Academic Publisher, Series E: Applied Sciences-Vol. 329 (1996), 599–628.

[102] J.R. Kirtley, Fundamental studies of superconductors using scanning magnetic imaging, Rep. Prog. Phys., 73 (2010)126501.

[103] R. F. Voss, R. B. Laibowitz, M. B. Ketchen, A. N. Broers, 1980 *Superconducting Quantum Interference Devices and Their Application*; Eds Hahlbohm H D and Luebbig H (de Gruyter, New York) 365.

[104] M. B. Ketchen, D. D. Awschalom, W. J. Gallagher, A. W. Kleinsasser, R. L. Sandstrom, J. R. Bozen, B. Bumble, Design, fabrication and performance of integrated miniature SQUID susceptometer, IEEE Trans. Magn., 25 (1989)1212.

[105] J. Gallop, W. J. Radcliffe, An absolute magnetometer, IEEE Trans. Magn., 21 (1985) 602.

[106] W. Wernsdorfer Magnetometrie micro-SQUID pour l'etude de particules ferromagnetiques isolees aux echelles sub-microniques, PhD Thesis Joseph Fourier University, Grenoble (1996).

[107] W. Wernsdorfer, Classical and quantum magnetization reversal studied in nanomete sized particles and clusters, Adv. Chem. Phys., 118 (2001) 99.

[108] S. K. H. Lam, D. L. Tilbrook, Development of a niobium nanosuperconducting quantum interference device for the detection of small spin populations, Appl. Phys. Lett., 82 (2003) 1078–1080.

[109] W. Wernsdorfer, D. Mailly, A. Benoit, Single nanoparticle measurement techniques, J. Appl. Phys., 87 (2000) 5094.

[110] C. Granata, A. Vettoliere, P. Walke, C. Nappi, and M. Russo, Performance of nano superconducting quantum interference devices for small spin cluster detection, J. Appl. Phys., 106 (2009) 023925.

[111] M. Schmelz, R. Stolz, V. Zakosarenko, S. Anders, L. Fritzsch, H. Roth, H.G. Meyer, Highly sensitive miniature SQUID magnetometer fabricated with cross-type Josephson tunnel junctions, Physica C, 478 (2012) 77.

[112] P E Lindelof, Superconducting microbridges exhibiting Josephson properties, Rep. Prog. Phys. 44, (1981) 60.

[113] P.W. Anderson and A. H. Dayem, Radio-Frequency Effects in Superconducting Thin Film Bridges, Phys. Rev. Lett., 13 (1964) 195.

[114] T. Van Duzer and C. W. Turner, Principles of Superconductive Devices and Circuits, 2. ed. Upper Saddle River, NJ: Prentice Hall Inc., (1999).

[115] G. N. Gol'tsman, O. Okunev, G. Chulkova, A. Lipatov, A. Semenov, K. Smirnov, B. Voronov, A. Dzardanov, C. Williams, and R. Sobolewski, Picosecond superconducting single-photon optical detector, Appl. Phys. Lett., 79 (2001) 705.

[116] M. Ejrnaes, R .Cristiano, O. Quaranta, S. Pagano, A. Gaggero, F. Mattioli, R. Leoni, B. Voronov, G. Gol'tsman, A cascade switching superconducting single photon detector, Appl. Phys. Lett., 91 (2007) 262509.

[117] D. E. Prober, Superconducting terahertz mixer using a transition-edge microbolometer, Appl. Phys. Lett., 62 (1993) 2119.

[118] J. E. Mooij and Y. V. Nazarov, Superconducting nanowires as quantum phase-slip junctions, Nature Phys., 2 (2006) 169-172.

[119] K.K. Likharev, Superconducting weak links, Rev. Mod. Phys., 51 (1979) 101-158

[120] A. A. Golubov, M. Yu. Kupriyanov, E. Il'ichev, The current-phase relation in Josephson junctions, Rev. Mod. Phys., 76 (2004) 411-469

[121] L.G. Aslamazov and A.I. Larkin, Josephson Effect in Superconducting Point Contacts, JETP Lett., 9 (1969) 87-91

[122] K.K. Likharev and L.A. Yakobson, Steady state properties of superconducting bridges, Zh. Tekhn. Fiz. 45, 1503 (1975) [Sov. Phys.Tech. Phys. 20, 950 (1975)]

[123] S.S. Pei, J.E. Lukens and R.D. Lukens, Measurement of the size dependence of the current-phase relation in microbridge Josephson junctions, Appl. Phys. Lett., 36 (1980) 88-90.

[124] K. Hasselbach, D. Mailly, J. R. Kirtley, Micro-superconducting quantum interference device characteristics, J. Appl. Phys., 91 (2002) 4432.

[125] G. J. Podd, G. D. Hutchinson, D. A. Williams, and D. G. Hasko, Micro-SQUIDs with controllable asymmetry via hot-phonon controlled junctions, Phys. Rev. B, 75 (2007) 134501.

[126] C. Granata, A. Vettoliere, M. Russo, B. Ruggiero, Noise theory of dc nano-SQUIDs based on Dayem nanobridges, Phys. Rev. B, 84 (2011) 224516.

[127] W. J. Skocpol, M. R. Beasely, M. Tinkam, Self-heating hot spot in superconducting thin-film microbridges, J. Appl. Phys., 45 (1974),4054-4066.

[128] I. O. Kulik, A. N. Omelyanchuk, Contribution to the microscopic theory of the Josephson effect in superconducting bridges, Pis'ma Zh. Eksp. Teor. Fiz., 21 (1975) 216 [JETP Lett., 21 (1975) 96].

[129] K. D. Usadel, Generalized Diffusion Equation for Superconducting Alloys, Phys. Rev. Lett. 25, (1970) 507-509.

[130] L.P. Gor'kov, On the energy spectrum of superconductors, Zh. Eksp. Teor., 34 (1958) 735 [Sov. Phys. JETP, 7 505 (1958)].

[131] L.P. Gor'kov, Theory of superconducting alloys in strong magnetic field near critical temperature, Zh. Eksp. Teor., 37 (1959) 1407 [Sov. Phys. JETP, 10 998 (1960)].

[132] A. G. P. Troeman, S. H. W. van der Ploeg, E. Il'Ichev, H. G. Meyer, A. A. Golubov, M. Yu. Kupriyanov, H. Hilgenkamp, Temperature dependence measurements of the supercurrent-phase relationship in niobium nanobridges, Phys. Rev. B, 77 (2008) 024509.

[133] R. Vijay, J. D. Sau, Marvin L. Cohen, I. Siddiqi, Optimizing Anharmonicity in Nanoscale Weak Link Josephson Junction Oscillators, Phys. Rev. Lett., 103 (2009) 087003.

[134] K. Hasselbach, C. Veauvy, D. Mailly, MicroSQUID magnetometry and magnetic imaging, Physica C, 332 (2000) 140–147.

[135] J. Gallop, P.W. Josephs-Franks, J. Davies, L. Hao, J. Macfarlane, Miniature dc SQUID devices for the detection of single atomic spin-flips, Physica C, 368 (2002) 109–113.

[136] J. C. Gallop, SQUIDs: Some limits to measurement, Supercond. Sci. Technol., (2003) 1575–1582.

[137] P. Josephs-Franks, L. Hao, A. Tzalenchuk, J. Davies, O. Kazakova, J. C. Gallop, L. Brown, J. C. Macfarlane, Measurement of the spatial sensitivity of miniature SQUIDs using magnetic-tipped STM, Supercond. Sci. Technol., 16 (2003) 1570-1574.

[138] D. L. Tilbrook, NanoSQUID sensitivity for isolated dipoles and small spin populations, Supercond. Sci. Technol., 22 (2009) 064003.

[139] V. Bouchiat, Detection of magnetic moments using a nano-SQUID: Limits of resolution and sensitivity in near-field SQUID magnetometry, Supercond. Sci. Technol., 22 (2009) 064002.

[140] J. Nagel, K. B. Konovalenko, M. Kemmler, M. Turad, R. Werner, E. Kleisz, S. Menzel, R. Klingeler, B. Buchner, R. Kleiner, D. Koelle, Resistively shunted $YBa_2Cu_3O_7$ grain boundary junctions and low-noise SQUIDs patterned by a focused ion beam down to 80 nm linewidth, Supercond. Sci. Technol., 24 (2011) 015015.

[141] R. Wölbing, T. Schwarz, B. Müller, J. Nagel, M. Kemmler, R. Kleiner, D. Koelle, Optimizing the spin sensitivity of grain boundary junction nanoSQUIDs -towards detection of small spin systems with single-spin resolution, Supercond. Sci. Technol., 27 (2014) 125007.

[142] J. R. Kirtley, M. B. Ketchen, K. G. Stawiasz, J. Z. Sun, W. J. Gallagher, S. H. Blanton and S. J. Wind, High-resolution scanning SQUID microscope, Appl. Phys. Lett., 66 (1995) 1138-1140.

[143] S. Chatraphorn, E. F. Fleet and F. C. Wellstood, Relationship between spatial resolution and noise in scanning superconducting quantum interference device microscopy, J. Appl. Phys., 92 (2002) 4731-4740.

[144] J. R. Bradley, G. S. Nestor and J. P. Wikswo Jr, Using a magnetometer to image a two-dimensional current distribution, J. Appl. Phys., 65 (1989) 361–72.

[145] P. F. Vohralik and S. K. H. Lam, NanoSQUID detection of magnetization from ferritin nanoparticles, 22 (2009) 064007.

[146] L. Hao, J. C. Macfarlane, J. C. Gallop, E. Romans, D. Cox, D. Hutson, and J. Chen, Spatial Resolution Assessment of Nano-SQUIDs Made by Focused Ion Beam, IEEE Trans. Appl. Supercond., 17 (2007) 742-745.

[147] E. H. Brandt, J.R. Clem, Superconducting thin rings with finite penetration depth, Phys. Rev. B, 69 (2004) 184509.

[148] P. Zeeman and A. D. Fokker (eds), H. A. Lorentz Collected Papers (The Hague, Holland: Martinus Nijhoff) vol III (1936) pp 1–11.

[149] M. A McCord, M. J. Rooks, Electron Beam Litography in SPIE Handbook of Microlithography, Micromachining and Microfabrication: Vol. 1, Chapter 2, P. Rai-Choudhury Editor, (2000) ISBN: 0-8194-2378-5, pp. 139-250

[150] H. D. Wanzenboeck, S. Waid, Focused Ion Beam Lithography in Recent Advances in Nanofabrication Techniques and Applications, Chapter 2, Bo Cui Editor, (2011) ISBN: 978-953-307-602-7, InTech, pp.27-50

[151] V. Bouchiat, M. Faucher C. Thirion and W. Wernsdorfer, T. Fournier and B. Pannetier, Josephson junctions and superconducting quantum interference devices made by local oxidation of niobium ultrathin films, Appl. Phys. Lett., 79 (2011) 123-125.

[152] M. Faucher, T. Fournier, B. Pannetier,, C. Thirion, W. Wernsdorfer, J.C. Villegier, V. Bouchiat, Niobium and niobium nitride SQUIDs based on anodised nanobridges made with an atomic force microscope, Physica C, 368 (2002) 211–217.

[153] S. K. H Lam, Noise properties of SQUIDs made from nanobridges, Supercond. Sci. Technol., 19 (2006) 963–967.

[154] A. G. P. Troeman, H. Derking, B. Boerger, J. Pleikies, D. Veldhuis, and H. Hilgenkamp, NanoSQUIDs based on niobium constrictions, Nano Lett., 7 (2007) 2152–2156.

[155] L. Hao, J. C. Macfarlane, J. C. Gallop, D. Cox, J. Beyer, D. Drung, and T. Schuring, Measurement and noise performance of nano-superconducting-quantum interference devices fabricated by focused ion beam, Appl. Phys. Lett., vol. 92 (2008) 192507–102509.

[156] L. Hao , D. C. Cox, J. C. Gallop, Characteristics of focused ion beam nanoscale Josephson devices, Supercond. Sci. Technol., 22 (2009) 064011.

[157] L. Hao, D. C. Cox, J. C. Gallop, E. J. Romans, J. C. Macfarlane, and J. Chen, Focused Ion Beam NanoSQUIDs as Novel NEMS Resonator Readouts, IEEE Trans. Appl. Supercond., 19 (2009) 693–696.

[158] E. J. Romans, S. Rozhko, L. Young, A. Blois, Ling Hao, D. Cox, John C. Gallop, Noise Performance of Niobium Nano-SQUIDs in Applied Magnetic Fields, IEEE Trans. Appl. Supercond., 21 (2011) 404-407.

[159] S. Rozhko, T. Hino, A. Blois, L. Hao, J. C. Gallop, D. C. Cox, and E. J. Romans, Study of Low-Frequency Noise Performance of Nanobridge-Based SQUIDs in External Magnetic Fields, IEEE Trans. Appl. Supercond., 23 (2013) 1601004.

[160] T. Patel, B. Li, J. Gallop, D. Cox, K. Kirkby, E. Romans, J. Chen, A. Nisbet, L. Hao, Investigating the Intrinsic Noise Limit of Dayem Bridge NanoSQUIDs, IEEE Trans. Appl. Supercond., 25 (2015) 1602105.

[161] D. Drung, C. Aßmann, J. Beyer, A. Kirste, M. Peters, F. Ruede, and Th. Schurig, Higly Sensitive and Easy-to-Use SQUID Sensors, IEEE Trans. Appl. Supercond., 17 (2007) 699-704.

[162] F. Ruede, S. Bechstein, L. Hao, C. Aßmann, T. Schurig, J. Gallop, O. Kazakova, J.Beyer, D. Drung, Readout System for NanoSQUID Sensors Using a SQUID Amplifier, IEEE Trans. Appl. Supercond., 21 (2011) 408-411.

[163] C. Granata, E. Esposito, A. Vettoliere, L. Petti, M. Russo, An integrated superconductive magnetic nanosensor for high-sensitivity nanoscale applications," Nanotechnology, 19 (2008) 275501–275506

[164] A.Vettoliere, C. Granata, E. Esposito, R. Russo, L. Petti, B. Ruggiero, M. Russo, Performance of High-Sensitivity Nano-SQUIDs Based on Niobium Dayem Bridges, IEEE Trans. Appl. Supercond., 19 (2009) 702-705.

[165] R. Vijay, E. M. Levenson-Falk, D. H. Slichter, I. Siddiqi, Approaching ideal weak link behavior with three dimensional aluminum nanobridges, 96 (2010) 223112.

[166] N. Antler, E. M. Levenson-Falk, R. Naik,Y.-D. Sun, A. Narla, R. Vijay, I. Siddiqi, In-plane magnetic field tolerance of a dispersive aluminum nanobridge SQUID magnetometer, Appl. Phys. Lett., 102 (2013) 232602.

[167] E. M. Levenson-Falk, R. Vijay, N. Antler, I Siddiqi, A dispersive nanoSQUID magnetometer for ultra-low noise, high bandwidth flux detection, Supercond. Sci. Technol., 26 (2013) 055015.

[168] M. Hatridge, R. Vijay, D. H. Slichter, J. Clarke, and I. Siddiqi, Phys. Rev. B, 83 (2011) 134501.

[169] S. K. H. Lam, J. R. Clem, and W. Yang, A nanoscale SQUID operating at high magnetic fields, Nanotechnology, 22 (2011) 455501.

[170] L. Chen, W.Wernsdorfer, C. Lampropoulos, G. Christou, I. Chiorescu, On-chip SQUID measurements in the presence of high magnetic fields, Nanotechnology, 21 (2010) 405504.

[171] C. Granata, A. Vettoliere, R. Russo, E. Esposito, M. Russo, and B. Ruggiero, Supercurrent decay in nano-superconducting quantum interference devices for intrinsic magnetic flux resolution, Appl. Phys. Lett., 94 (2009) 0625030.

[172] R. Russo, C. Granata, P. Walke, A. Vettoliere, E. Esposito, M. Russo, NanoSQUID as magnetic sensor for magnetic nanoparticles Characterization, J. Nanopart. Res., 13, (2011) 5661–5668.

[173] E. Esposito, C. Granata, A. Vettoliere, R. Russo, D. Peddis, M. Russo, Nano Superconducting QUantum Interference Device Sensors for Magnetic Nanoparticle Detection, J. Nanosci. Nanotechnol.,12 (2012) 7468-7472.

[174] R. Russo, C. Granata, E. Esposito, D. Peddis, C. Cannas, A. Vettoliere, "Nanoparticle Magnetization Measurements by a High Sensitive Nano-Superconducting Quantum Interference Device," Appl. Phys. Lett., 101 (2012) 122601.

[175] E. Esposito, C. Granata, M. Russo, R. Russo, A. Vettoliere, High Sensitive Magnetic Nanosensors Based on Superconducting Quantum Interference Device, IEEE Trans. Magn., 49 (2013) 140

[176] C. Granata, R. Russo, E. Esposito, A. Vettoliere, D. Peddis, A. Musinu, B. Ruggiero, Dino Fiorani, M. Russo, Hysteretic NanoSQUID Sensors for Investigation of Iron Oxide Nanoparticles, Trans. Appl. Supercond., 23 (2013) 1602305.

[177] J. Clarke, J. L. Paterson, Josephson-Junction Amplifier, Appl. Phys. Lett., 19 (1971) 469-471.

[178] D. Hazra, J. R. Kirtley, K. Hasselbach, Nano-superconducting quantum interference devices with continuous read out at milliKelvin temperatures, Appl. Phys. Lett., 103 (2013) 093109.

[179] A. Blois, S. Rozhko, L. Hao, J. C. Gallop, and E. J. Romans, Proximity effect bilayer nano superconducting quantum interference devices for millikelvin magnetometry, J. Appl. Phys., 114 (2013) 233907

[180] D. Hazra, J. R. Kirtley, K. Hasselbach, Nano-superconducting quantum interference devices with suspended junctions, Appl. Phys. Lett., 104 (2014) 152603.

[181] S.-B. Lee, D. G. Hasko, H. Ahmed, Thermally switched superconducting weak-link transistor with current gain, Appl. Phys. Lett., 76 (2000) 2295-2297.

[182] G. D. Hutchinson, H. Qin, D. G. Hasko, D. J. Kang, D. A. Williams, Controlled-junction superconducting quantum interference device via phonon injection, Appl. Phys. Lett., 84 (2004) 136-138.

[183] G. J. Podd, G. D. Hutchinson, D. G. Hasko, D. A. Williams, Hot-Phonon Controlled MicroSQUIDs With Independently Controlled Junctions, IEEE Trans. Appl. Supercond., 17 (2007) 710-713.

[184] G. C. Tettamanzi, C. I. Pakes, S. K. H. Lam, S. Prawer, Flux noise in ion-implanted nanoSQUIDs, Supercond. Sci. Technol., 22 (2009) 064006.

[185] J. Nagel, O. F. Kieler, T. Weimann, R. Wolbing, J. Kohlmann, A. B. Zorin, R. Kleiner, D. Koelle, M. Kemmler, Superconducting quantum interference devices with submicron Nb/HfTi/Nb junctions for investigation of small magnetic particles, Appl. Phys. Lett., 99 (2011) 032506.

[186] D. Hagedorn, R. Dolata, F.-I. Buchholz, and J. Niemeyer, Properties of SNS Josephson junctions with Hf Ti interlayers, Physica C, 372–376 (2002) 7-10

[187] D. Hagedorn, O. Kieler, R. Dolata, R. Behr, F. Muller, J. Kohlmann, J. Niemeyer, Modified fabrication of planar sub-µm superconductor–normal metal–superconductor Josephson junctions for use in a Josephson arbitrary waveform synthesizer, Supercond. Sci. Technol., 19 (2006) 294-298.

[188] R. Wölbing, J. Nagel, T. Schwarz, O. Kieler, T.J. Weimann, J. Kohlmann, A. B. Zorin, M. Kemmler, R. Kleiner, D. Koelle, Nb nano superconducting quantum interference devices with high spin sensitivity for operation in magnetic fields up to 0.5 T, Appl. Phys. Lett., 102 (2013) 192601.

[189] S. Bechstein, F. Ruede, D. Drung, J.-H. Storm, O. F. Kieler, J. Kohlmann, T. Weimann, T. Schurig, HfTi-nanoSQUID gradiometers with high linearity, Appl. Phys. Lett., 106 (2015) 072601.

[190] C. Granata, A. Vettoliere, R. Russo, M. Fretto, N. De Leo, V. Lacquaniti, Threedimensional spin nanosensor based on reliable tunnel Josephson nano-junctions for nanomagnetism investigations, Appl. Phys. Lett., 103 (2013) 102602.

[191] C. Granata, A. Vettoliere, R. Russo, M. Fretto, N. De Leo, E. Enrico,V. Lacquaniti, Ultra High Sensitive Niobium NanoSQUID by Focused Ion Beam Sculpting, J. Supercond. Nov. Magn., 28 (2015) 585–589.

[192] C. Granata,, A. Vettoliere, M. Fretto, N. De Leo, V. Lacquaniti, Vertical nano superconducting quantum interference device based on Josepshon tunnel nanojunctions for small spin cluster detection, J. Magn. Magn. Mat., 384 (2015) 117–121.

[193] C. Granata, A. Vettoliere, B. Ruggiero, M.o Russo, M. Fretto, V. Lacquaniti, L. Boarino, N. De Leo, Low Noise NanoSQUIDs Based on Deep Submicron Josephson Tunnel Junctions, IEEE Trans. Appl. Supercond., 25 (2015) 1600905.

[194] R. Russo, C. Granata, A. Vettoliere, E. Esposito, M. Fretto, N. De Leo, E. Enrico, V. Lacquaniti, Performances of niobium planar nanointerferometers as a function of the temperature: a comparative study, Supercond. Sci. Technol., 27 (2014) 044028.

[195] A. Ronzani, M. Baillergeau, C. Altimiras, and F. Giazotto, Micro-superconducting quantum interference devices based on V/Cu/V Josephson nanojunctions, Appl. Phys. Lett., 103 (2013) 052603.

[196] A.Ronzani, C. Altimiras, F. Giazotto, Balanced double-loop mesoscopic interferometer based on Josephson proximity nanojunctions, Appl. Phys. Lett., 104 (2014) 032601.

[197] M. Schmelz, Y. Matsui, R. Stolz, V. Zakosarenko, T. Schönau, S. Anders, S. Linzen, H. Itozaki, H-G Meyer, Investigation of all niobium nano-SQUIDs based on sub-micrometer cross-type Josephson junctions, Supercond. Sci. Technol., 28 (2015) 015004

[198] S. Anders, M. Schmelz, L. Fritzsch, R. Stolz, V. Zakosarenko, T. Schönau, H. G. Meyer, Supercond. Sci. Technol., 22 (2009) 064012.

[199] R. Ishiguro et al., Development of nano and micro SQUIDs based on Al tunnel junctions, Journal of Physics: Conference Series, 568 (2014) 022019.

[200] N. C. Koshnick, M. E. Huber, J. A. Bert, C. W. Hicks, J. Large, H. Edwards, K. A. Moler, A terraced scanning superconducting quantum interference device susceptometer with submicron pickup loops, Appl. Phys. Lett., 93 (2008) 243101.

[201] N. Watanabe, Y.Miyato, S. Matsusawa, M. Tachiki, T. Hayashi, H. Itozaki, Fine probe for an STM-SQUID probe microscope, IEEE Trans. Appl. Supercond., 23 (2013) 1601804.

[202] D. Drung, J.H. Storm, F. Ruede, A. Kirste, M. Regin, T. Schurig, A. M. Repollés, J. Sesé, F. Luis, Thin-Film microsusceptometer with integrated nanoloop, IEEE Trans. Appl. Supercond., 24 (2014) 1600206.

[203] E. J. Romans, E. J. Osley, L. Young, P. A. Warburton, W. Li, Three-dimensional nanoscale superconducting quantum interference device pickup loops, Appl. Phys. Lett., 97 (2010) 222506.

[204] W. Li and P. A. Warburton, Low-current focused-ion-beam induced deposition of three-dimensional tungsten nanoscale conductors, Nanotechnology 18, (2007) 485305.

[205] A. Finkler, Y. Segev, Y. Myasoedov, M. L. Rappaport, L. Ne'eman, D. Vasyukov, E. Zeldov, M. E. Huber, J. Martin, A. Yacoby, Self-aligned nanoscale SQUID on a tip, Nano Lett., 10 (2010) 1046–1049.

[206] A. Finkler, D. Vasyukov, Y. Segev, L. Ne'eman, E. O. Lachman, M. L. Rappaport, Y. Myasoedov, E. Zeldov, M. E. Huber, Scanning superconducting quantum interference device on a tip for magnetic imaging of nanoscale phenomena, Rev. Sci. Instrum., 83 (2012) 073702

[207] D. Vasyukov, et al., A scanning superconducting quantum interference device with single electron spin sensitivity, Nat. Nanotech. 8 (2013) 639–644.

[208] C. H. Wu, Y. T. Chou, W. C. Kuo, J. H. Chen, L. M. Wang, J. C. Chen, K. L. Chen, U. C. Sou, H. C. Yang, J. T. Jeng, Fabrication and Characterization of High-Tc $YBa_2Cu_3O_{7-x}$ NanoSQUIDs Made by Focused Ion Beam Milling, Nanotechnology, 19 (2008) 315304.

[209] T. Schwarz, J. Nagel, R. Wolbing, M. Kemmler, R. Kleiner, and D. Koelle, Low-Noise Nano Superconducting Quantum Interference Device Operating in Tesla Magnetic Fields, ACS nano, 7 (2013) 844–850.

[210] R. Arpaia, M. Arzeo, S. Nawaz, S. Charpentier, F. Lombardi, T. Bauch, Ultra low noise $YBa_2Cu_3O_{7-\delta}$ nano superconducting quantum interference devices implementing nanowires, App. Phys. Lett., 104 (2014) 072603.

[211] M. Aprili, The nanoSQUID makes its debut, Nature Nanotech., 1 (2006) 15-16.

[212] J.-P. Cleuziou, W. Wernsdorfer, V. Bouchiat, T. Ondarcuhu, M. Monthioux, Carbon nanotube superconducting quantum interference device, Nature Nanotech., 1 (2006) 53–59.

[213] A. Y. Kasumov et al., Supercurrents through single-walled carbon nanotube,. Science 397, (1999) 598–601.

[214] C. Girit, V. Bouchiat, O. Naaman, Y. Zhang, M. F. Crommie, A. Zettl, I. Siddiqi, Tunable Graphene dc Superconducting Quantum Interference Device, ACS nano, 9 (2009) 198-199.

[215] H. B. Heersche, P. Jarillo-Herrero, J. B. Oostinga, L. M. K. Vandersypen, A. F. Morpurgo, Bipolar supercurrent in graphene, Nature, 486 (2007) 56-59

[216] X. Du, I. Skachko, E. Y. Andrei, Josephson current and multiple Andreev reflections in graphene SNS junctions, Phys. Rev. B, 77 (2008) 184507.

[217] A. K. Geim, K. S. Novoselov, The rise of graphene, Nat. Mater., 6 (2007) 183.

[218] S. Mandal, T. Bautze, O. A. Williams, C . Naud, E. Bustarret, F. Omnes, P. Rodiere, T. Meunier, C. Bauerle, L. Saminadayar, The Diamond Superconducting Quantum Interference Device, ACS nano, 5 (2011) 7144–7148.

[219] E. A. Ekimov, A. A. Sidorov, E. D. Bauer, N. N. Melnik, N. J. Curro, J. D. Thompson, S. M. Stishov, Superconductivity in Diamond, Nature, 428 (2004) 542 – 545.

[220] X. Blase, E. Bustarret, C. Chapelier, T. Klein, C. Marcenat, Superconducting Group-IV Semiconductors., Nat. Mater., 8 (2009) 375 – 382.

[221] P. Spathis, S. Biswas, S. Roddaro, L. Sorba, F. Giazotto, F. Beltram, Hybrid InAs nanowire–vanadium proximity SQUID, Nanotechnology 22 (2011) 105201.

[222] R. M. Lutchyn, J. D. Sau, S. Das Sarma, Majorana fermions and a topological phase transition in semiconductor–superconductor heterostructures, Phys. Rev. Lett., 105 (2010) 077001.

[223] J. Nakamatsu, N. Nakagawa, T. Muranaka, Y. Zenitani, J. Akimitsu, Superconductivity at 39 K in magnesium diboride, Nature, 410 (2001) 63-64

[224] A. Brinkman J.M. Rowell, MgB 2 tunnel junctions and SQUIDs, Physica C, 456 (2007) 188–195

[225] Y. Harada, K. Kobayashi, M. Yoshizawa, $MgB_2$ SQUID for Magnetocardiography, in Superconductors - Properties, Technology, and Applications, edited by Y. Grigorashvili, (Publisher InTech, 2012), 389-404.

[226] S.-G. Lee, S.-H. Hong, W. K. Seong, W. N. Kang, All focused ion beam fabricated MgB2

inter-grain nanobridge dc SQUIDs, Supercond. Sci. Technol., 22 (2009) 064009

[227] S.-H. Hong, S.-G. Lee, W. K. Seong, W. N. Kang, Fabrication of $MgB_2$ nanobridge dc SQUIDs by focused ion beam, Physica C, 470 (2010) S1036–S1037.

[228] B. Ruggiero, C. Granata, V.G. Palmieri, A. Esposito, M. Russo, P Silvestrini, Supercurrent decay in extremely underdamped Josephson junctions, Phys. Rev. B, 57 (1998) 134-137

[229] P. Silvestrini, B. Ruggiero, C. Granata, E. Esposito, Supercurrent decay of Josephson junctions in non-stationary conditions: experimental evidence of macroscopic quantum effects, Phys. Lett. A, 267 (2000) 45-51

[230] W. Wernsdorfer, D. Mailly, A. Benoit, Single nanoparticle measurement techniques J. Appl. Phys., 87 (2000) 5094.

[231] W .Wernsdorfer, E. Bonet Orozco, K. Hasselbach, A. Benoit, D. Mailly, O. Kubo, H. Nakano, B. Barbara, Macroscopic quantum tunneling of magnetization of single ferrimagnetic nanoparticles of barium ferrite, Phys. Rev. Lett., 79 (1997) 4014

[232] C. Granata, R. Russo, E. Esposito, A. Vettoliere, M. Russo, A. Musinu, D. Peddis, D. Fiorani, Magnetic properties of iron oxide nanoparticles investigated by nanoSQUIDs, Eur. Phys. J. B, 86 (2013) 272.

[234] L. Bogani, C. Danieli, E. Biavardi, N. Bendiab, A. L. Barra, E. Dalcanale, W. Wernsdorfer, A. Cornia. Single-molecule-magnet carbon-nanotube hybrids, Angew. Chem. Int. Edn Engl., 48 (2009) 746.

[235] L. Hao, C.A. ßmann, J.C. Gallop, D. Cox, F. Ruede, O. Kazakova, P. Josephs-Franks, D. Drung, Th Schurig, Detection of single magnetic nanobead with a nano-superconducting quantum interference device, Appl. Phys. Lett., 98 (2011) 092504.

[236] L. Hao, D. Cox, P. See, J. Gallop, O Kazakova, Magnetic nanoparticle detection using nano-SQUID sensors, J. Phys. D: Appl. Phys., 43 (2010) 474004.

[237] M. Faucher, P. O. Jubert, O. Fruchart, W. Wernsdorfer, V. Bouchiat, Optimizing the flux coupling between a nanoSQUID and a magnetic particle using atomic force microscope nanolithography, Supercond. Sci. Technol., 22 (2009) 064010.

[238] M. Martin, L. Roschier, P. Hakonen, U. Parts, M. Paalanen, B. Schleicher, and E. I. Kauppinen, Manipulation of Ag nanoparticles utilizing noncontact atomic force microscopy, Appl. Phys. Lett., 73 (1998) 1505.

[239] D. M. Schaefer, R. Reifenberger, A. Patil, and R. P. Andres, Fabrication of two dimensional arrays of nanometer size clusters with the atomic force microscope, Appl. Phys. Lett., 66 (1995) 1012.

[240] M. Jamet, W. Wernsdorfer, C. Thirion, D. Mailly, V. Dupuis, P. Melinon, A. Perez 2001 Magnetic anisotropy of individual cobalt clusters, Phys. Rev. Lett., 86 (2001) 4676.

[241] M. Jamet, W. Wernsdorfer, C. Thirion, V. Dupuis, P. Melinon, A. Perez, D. Mailly Magnetic anisotropy in single clusters, Phys. Rev. B, 69 (2004) 024401

[242] S. K. H. Lam, Wenrong Yang, H. T. R. Wiogo, C. P. Foley, Attachment of magnetic molecules on a nanoSQUID, Nanotechnology, 19 (2008) 285303.

[243] L. Neel, Theorie du trainage magnétique des ferromagnetiques en grains fins avec application aux terres cuites, Ann. Geophys., 5 (1949) 99-136.

[244] I. Volkov, M. Chukharkin, O. Snigirev, A. Volkov, S. Tanaka, and C. Fourie, Determination of the anisotropy constant and saturation magnetization of magnetic nanoparticles from magnetization relaxation curves, J. Nanopart. Res., 10 (2008) 487-497

[245] N. L. Adolphi, et al., Characterization of single-core magnetite nanoparticles for magnetic imaging by SQUID relaxometry, Phys. Med. Biol., 55 (2010) 5985-6003.

[246] E. C. Stoner, E. P., Wohlfarth, A mechanism of magnetic hysteresis in heterogeneous allows., Phil. Trans. R. Soc. Lond., A 240 (1948) 599–608.

[247] J. Clarke, A. N. Cleland, M. H. Devoret, D. Esteve, J. M. Martinis, Quantum mechanics of a macroscopic variable: the phase difference of a Josephson junction, Science 239 (1988) 992

[248] B. Ruggiero, V. Corato, C. Granata, L. Longobardi, S. Rombetto, Measurement of the effective dissipation in an rf SQUID system, Phys. Rev. B, 67 (2003) 132504.

[249] W. Wernsdorfer, K. Hasselbach, D. Mailly, B. Barbara, A. Benoit, L. Thomas, G. Swan, DC-SQUID magnetization measurements of single magnetic particles, J. Magn. Magn. Mat., 145 (1995) 33-39.

[250] W. Wernsdorfer et al, High sensitivity magnetization measurements of nanoscale cobalt clusters, J. Appl. Phys., 78 (1995) 7192.

[251] W. Wernsdorfer, B. Doudin, D. Mailly, K. Hasselbach, A. Benoit, J. Meier, J.-Ph. Ansermet, B. Barbara, Nucleation of Magnetization Reversal in Individual Nanosized Nickel Wires, Phys. Rev. Lett., 77 (1996) 1873.

[252] W. Wernsdorfer et al, Experimental Evidence of the Néel-Brown Model of Magnetization Reversal, Phys. Rev. Lett., 78 (1997) 1791.

[253] W. F. Brown, Thermal Fluctuations of a Single-Domain Particle, Phys. Rev., 130 (1963) 1677.

[254] W. Wernsdorfer et al, Macroscopic Quantum Tunneling of Magnetization of Single Ferrimagnetic Nanoparticles of Barium Ferrite, Phys. Rev. Lett., 79 (1997) 4014.

[255] E. Bonet et al., Three-Dimensional Magnetization Reversal Measurements in Nanoparticles, Phys. Rev. Lett., 83 (1999) 4188.

[256] C. Thirion, W. Wernsdorfer, D. Mailly, Switching of magnetization by nonlinear resonance studied in single nanoparticles, Nat. Mater., 2 (2003) 524-527

[257] W. Wernsdorfer, R. Sessoli, Quantum phase interference and parity effects in magnetic molecular clusters, Science, 284 (1999) 133

[258] O. Fruchart, J-P. Nozieres, W. Wernsdorfer, D. Givord, Enhanced coercivity in submicrometer-sized ultrathin epitaxial dots with in-plane magnetization, Phys. Rev. Lett., 82 (1999) 1305.

[259] M. J. Martınez-Perez, Alternating current magnetic susceptibility of a molecular magnet submonolayer directly patterned onto a micro superconducting quantum interference device, Appl. Phys. Lett., 99 (2011) 032504

[260] R. D. Piner, J. Zhu, F. Xu, S. Hong, C. A. Mirkin, "Dip-Pen" nanolithography, Science, 283 (1999), 661-663

[261] T. Schwarz et al, Low-Noise $YBa_2Cu_3O_7$ NanoSQUIDs for Performing Magnetization-Reversal Measurements on Magnetic Nanoparticles, arXiv:1503.06090v1 (2015).

[262] E. F. Fleet, S. Chatraphorn, F. C. Wellstood, C. Eylem, "Determination of magnetic properties using a scanning SQUID microscope", IEEE Trans. Appl. Supercond. 11, (2011) 1180 – 1183.

[263] S. Bechstein, A. Kirste, D. Drung, M. Regin, O. Kazakova, J. Gallop, L. Hao, D. Cox, and T. Schurig, Investigation of Material effects with micro-sized SQUID sensors, IEEE Trans. Appl. Supercond., 23 (2013) 1602004.

[264] T. Shuring, Making SQUIDs a practical tool for quantum detection and material characterization in the micro- and nanoscale, Journal of Physics: Conference Series, 568 (2014) 032015

[265] J. Nagel et al., Nanoscale multifunctional sensor formed by a Ni nanotube and a scanning Nb nanoSQUID, Phys. Rev. B, 88 (2013) 064425.

[266] S. Etaki, M. Poot, I. Mahboob, K. Onomitsu, H. Yamaguchi, H. S. J. van der Zant, Motion detection of a micromechanical resonator embedded in a d.c. SQUID, Nat. Phys., 4 (2008) 785-788.

[267] S. Pugnetti, Y. M. Blanter, and R. Fazio, Resonant coupling of a SQUID to a mechanical resonator, Euro. Phys. Lett., 90 (2010) 48007.

[268] C. Ke, H. D. Espinosa "Nanoelectromechanical Systems and Modeling", in Handbook of Theoretical and Computational Nanotechnology, Edited by Michael Rieth and Wolfram Schommers Volume 1, (2005) 1–38.

[269] L. Hao, D. C. Cox, J. C. Gallop, J. Chen, Member, IEEE, S. Rozhko, A. Blois, E. J. Romans, Coupled, NanoSQUIDs and Nano-Electromechanical Systems (NEMS) Resonators, IEEE trans. Appl. Supercond., 23 (2013) 1800304

[270] S. Bechstein, et al., Design and fabrication of coupled nanoSQUIDs and NEMS, IEEE Trans. Appl. Supercond., 25 (2015) 1602604.

[271] L. Hao, J. C. Gallop, D.C. Cox, J. Chen, Fabrication and Analogue Applications of NanoSQUIDs Using Dayem Bridge Junctions, EEE J. Sel. Top. Quantum Electron., 21 (2015) 9100108.

[272] G. C. Hilton, J. M. Martinis, K. D. Irwin, N. F. Bergren, D. A. Wollman, M. E. Huber, S. Deiker, S. W. Nam, Microfabricated transition-edge X-ray detectors, IEEE Trans. Appl. Supercond., 11 (2001) 739–42.

[273] A. J. Kreisler, A. Gaugue, Recent progress in high-temperature superconductor bolometric detectors: from the mid-infrared to the far-infrared (THz) range, Supercond. Sci. Technol., 13 (2000) 1235–1245.

[274] L. Hao, J. C. Gallop, C. Gardiner, P. Josephs-Franks, J. C. Macfarlane, S. K. H. Lam, C. Foley, Inductive superconducting transition-edge detector for single-photon and macro-molecule detection, Supercond. Sci. Technol., 16 (2003) 1479–1482.

[275] L. Hao, J. C. Macfarlane, P. Josephs-Franks, J. C. Gallop, Inductive Superconducting Transition-Edge Photon and Particle Detector, IEEE Trans. Appl. Supercond., 13 (2003) 622-625

[276] A. Watson, Measurement and the Single Particle, Science, 306 (2004) 1308-1309.

[277] L. Hao, J. C. Macfarlane, S. K. H. Lam, C. P Foley, P. Josephs-Franks, and J. C. Gallop, Inductive Sensor Based on Nano-Scale SQUIDs, IEEE Trans. Appl. Supercond., 15 (2005) 514-517.

[278] J. Tejada, E. M. Chudnovsky, E. del Barco, J. M. Hernandez, T. P. Spiller, Magnetic qubits as hardware for quantum computers, Nanotechnology, 12 (2001) 181–186.

[279] S. K. H. Lam, Flux transistor made from a single-layer niobium thin-film superconducting quantum interference device, Supercond. Sci. Technol., 19 (2006) 27–31.

[280] S. K. H. Lam, S. Gnanarajan, Hysteretic behaviour of nanoSQUIDs-prospective application as trapped-vortex memory, Supercond. Sci. Technol., 22 (2009) 064005.

[281] A. F. Hebard, A, T, Fiory, A memory device utilizing the storage of Abrikosov vortices at an array of pinning sites in a superconducting film, AIP Conf. Proc., 44 (1978) 465.

[282] S. Uehara, K. Nagata, Trapped vortex memory cells, Appl. Phys. Lett., 39 (1981) 992.

[283] J. Nygard, D. H. Cobden, P. E.Lindelof, Kondo physics in carbon nanotubes, Nature, 408 (2000) 342–346.

[284] C. Bena, S. Vishveshwara, L. Balents, M. P. A. Fisher, Quantum entanglement in carbon nanotubes. Phys. Rev. Lett. 89, 037901 (2002).